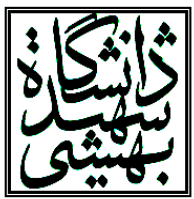

دانشگاه شهید بهشتی

دانشکده علوم

گروه فیزیک

پایان نامه کارشناسی ارشد

عنوان:

**بررسی حل های دقیق نا همسانگرد کسنری (Kasner)**

**در گرایش های مرتبه بالاتر**

استاد راهنما:

دکتر مهرداد فرهودی

استاد مشاور:

دکتر حمید رضا سپنجی

نگارش:

حمید شعبانی

مرداد ۸۸

# تقدیم به

پدرم و مادرم و تمام کسانی که دوستشان دارم

## تشکر و قدردانی

مراتب سپاس و قدردانی خود را نسبت به جناب آقای دکتر فرهودی استاد راهنما و جناب آقای دکتر سپنجی استاد مشاور اینجانب جهت راهنمایی های بسیارشان در طول انجام این پایان نامه ابراز می دارم و کمال تشکر را دارم.

همچنین از دوستان بسیار عزیزم آقایان حمید رضا شجاعی مهیب و جلال نوروز علیایی به دلیل کمک هایشان در این پایان نامه قدردانی می کنم.

# چکیده

این پایان نامه به تحقیق در وجود جواب های دقیق ناهمسانگرد از نوع کسنر (Kasner) در نظریه گرانش های مرتبه بالاتر در خلاء پرداخته است. مدل های بررسی شده در حالت کلی تابعی از سه کمیت اسکالر $R$، $R_{\mu\nu}R^{\mu\nu}$ و $R_{\alpha\beta\mu\nu}R^{\alpha\beta\mu\nu}$ هستند. در نزدیکی تکینگی، جملات غالب بسط این توابع در واقع بر حسب توان هایی از سه کمیت فوق به شکل $R^n$ (عملاً به صورت $R^{1+\delta}$، جهت نشان دادن مقدار انحراف از کنش اینشتین–هیلبرت)، $(R_{\mu\nu}R^{\mu\nu})^n$ یا $(R_{\alpha\beta\mu\nu}R^{\alpha\beta\mu\nu})^n$ می باشند. در بررسی انجام شده نشان داده شده است که در مدل هایی از نوع $R^n$ و $(R_{\mu\nu}R^{\mu\nu})^n$ همواره جواب های کسنری یافت می شود. اما در مدل هایی از نوع سوم جواب های نا همسانگرد از نوع کسنر یافت نشده است. همچنین رفتار این مدل ها در حضور ماده مورد بررسی قرار گرفته و نشان داده شده است که همواره این حل ها در شرایطی که ماده نسبیتی باشد معتبر هستند، ولی چنین اعتباری برای برخی جواب ها در حضور ماده غیر نسبیتی وجود ندارد. ضمناً شرایط انرژی برای مدل $R^{1+\delta}$ در حالت خلاء بررسی شده و نشان داده شده است که اعمال چهار شرط انرژی (ضعیف، تهی، قوی و غالب) قیدی روی $\delta$ اعمال می کنند. معمولاً در یک انبساط شتاب دار شرط انرژی قوی نقض می شود. لذا این جواب غیر فیزیکی ایجاب می کند که شرط انرژی قوی در این مدل باید ارضاء شود، که این یعنی انبساط ناهمسانگرد با یک شتاب کند شونده همراه است. همچنین در حالت خلاء با تعریف یک چگالی انرژی و فشار برای هندسه، معادله حالتی که بدست می آید از نوع تابش است. نتیجه این کار به همراه بررسی شرایط انرژی نشان می دهد که متریک کسنر می تواند حتی با وجود ماده تقریباً نسبیتی و بسیار نسبیتی یک حل تقریبی خوب باشد.

واژه های کلیدی: گرانش های مرتبه بالاتر، تکینگی، جواب های کسنری، جواب های ناهمسانگرد، ماده نسبیتی، ماده غیر نسبیتی، ماده بسیار نسبیتی، شرط انرژی، انبساط شتاب دار، شتاب کند شونده

فهرست ................................................... صفحه

فصل پنجم

# مطلع

نظریه گرانش های مرتبه بالاتر یکی از پر کاربرد ترین نظریه هایی است که امروزه در برنامه کاری فیزیک دانان نظری قرار گرفته است. در این نظریه ها معمولاً با اضافه کردن جملاتی به شکل توابعی از اسکالر ریچی و یا ناوردا هایی از تانسور ریچی و تانسور ریمان به کنش اینشتین–هیلبرت سعی بر ارائه مدل هایی می شود که بتوانند مشکلات موجود در کیهان شناسی استاندارد را تا حد امکان بر طرف کنند. مثلاً یکی از کاربرد های این نظریه استفاده از مدل های متنوع در توجیه انرژی تاریک و ماده تاریک است. اولین گزینه ای که برای انرژی تاریک معرفی شد ثابت کیهان شناسی در معادله اینشتین بود که در عین حال مشکلاتی چند را در پی داشت. مشکل اول بسیار کوچک بودن چگالی اندازه گیری شده این ثابت است. مشکل بعدی این است که چرا این چگالی از مرتبه چگالی اندازه گیری شده ماده در عالم امروز است. از مشکلات مهمی که باید نضریه معرفی شده به آن جواب دهد مسأله ثابت کیهان شناسی است. این مسأله می گوید چرا مقدار پیش بینی شده برای این ثابت در کیهان شناسی با مقدار محاسبه شده آن از نظریه میدان های کوانتومی اختلاف قابل ملاحظه ای دارد. نظریه گرانش های مرتبه بالاتر برای حل این مشکلات هم مورد استفاده قرار گرفتند.

فصل اول این پایان نامه مقدمه ای کوتاه در معرفی گرانش های مرتبه بالاتر را ارائه می کند. در این فصل به صورت خیلی گذرا به تاریخچه شکل گیری این نظریه می پردازیم و تا حدی از ضعف ها و نقاط قوت این نظریه صحبت می کنیم.در تمام نظریاتی که فرمول بندی لاگرانژی دارند، اولین مساله ای که مورد بررسی قرار می گیرد استخراج معادلات حرکت است، و در نهایت از این معادلات می توان پاسخ های مدل را بدست آورد. از این رو با توجه به رهیافت های متفاوت در استخراج معادلات میدان حاصل از کنشی معین در مساله نیاز به مطالعه روی مرسوم ترین این روش ها یعنی روش های وردشی[1] احساس می شود. روش های مختلفی برای استخراج معادلات میدان وجود دارد که پر استفاده ترین آنها در نظریه گرانش های مرتبه بالاتر به ترتیب روش وردش متریکی، و وردش پالاتینی است روش دیگری نیز وجود دارد که معادل با روش متریکی است با این تفاوت که الگوی مشخصی ارائه می کند. در فصل دوم به طور خلاصه اصول این سه روش وردشی با ذکر مثال هایی توضیح داده خواهد شد.

---

[1] variation principles

در نظریه گرانش های مرتبه بالا همواره بعد از بدست آوردن معادلات میدان به منظور برآوردی از میزان توانمندی مدل در تطابق با اطلاعات بدست آمده از مشاهدات رصدی، یا نیاز به برآورده شدن بعضی الزامات فیزیکی، و یا ارائه روشی برای جایگزین کردن این گونه مدل ها بجای مدل هایی با نا هم خوانی ها و مشکلات بیشتر، این مدل ها در بوته آزمایش قرار می گیرند. اصطلاحاً گفته می شود معادلات میدان توسط برخی روش ها مقید می شوند. در واقع روش های مختلف مقید کردن، مدل های جامع تر را محدود تر می کنند. معمولاً این مدل ها دارای یک یا چند متغیر آزاد هستند، و این روشها در واقع بازه هایی مجاز را برای این متغیر ها بدست می دهند و به عبارتی مدل ها محدودتر می شوند. در فصل سوم و چهارم روشهایی برای مقید سازی معادلات ارائه می گردد. یکی از این روش ها استفاده از شرایط انرژی است. در فصل سوم توضیح داده خواهد شد که مدل هایی که بتوانند هر یک از این شرایط انرژی را ارضاء کنند چگونه می توانند قید هایی را روی متغیر ها تحمیل کنند. در این فصل در مورد این قید ها صحبت می کنیم.

از روش های دیگر مقید سازی معادلات میدان مقایسه نتایج بدست آمده از پیش بینی پارامتر های مجهول در آزمایش های رصدی[1] با نتایج بدست آمده از مدل است. مثلاً یکی از پارامتر هایی که رسم منحنی مربوط به آن جزء برنامه آزمایش های رصدی است منحنی چرخش کهکشان ها است. در حقیقت اهمیت منحنی چرخش کهکشانی[2] در آن است که می توان اثرات ماده تاریک[3] را در آن ها مشاهده کرد. با این توضیح، اینکه منحنی چرخش به متغیر آزاد مدل وابسته شود به این معنی است که ماده تاریک می تواند به اثرات هندسی وابسته شود. انتهای فصل چهارم مروری کوتاه به این موضوع دارد. معادلات میدان را می توان با بدست آوردن روابطی برای عمر عالم، درخشندگی و یا مثلاً پارامتر شتاب کاهشی[4] مقید نمود. به این موضوعات در اواسط فصل چهارم پرداخته می شود.

یکی از کاربرد های نظریه گرانش های مرتبه بالاتر استفاده از آن به عنوان مدلی جایگزین برای مدل هایی است که دارای مشکلات و نقایص بیشتری در توضیح برخی مساله ها دارند. مثلاً در برخی مدل ها برای حل مساله انرژی تاریک از ثابت کیهان شناسی در معادله اینیشتن استفاده می کنند. با این وجود در

---

گرانش های مرتبه بالاتر انرژی تاریک را می توان به خود هندسه نسبت داد. در ابتدای فصل چهارم به این موضوع پرداخته ایم.

در سال ۱۹۲۵ کسنر (E. Kasner) یکی از حل های دقیق معادلات اینشتین را که موسوم به یک متریک ناهمسانگرد از نوع متریک بیانکی نوع اول است بدست آورد. این حل با داشتن دو قید روی متغیر های متریک دارای یک درجه آزادی است. یک ویژگی مهم این متریک این است که این متریک دارای یک تکینگی زمان گونه در $t = 0$ است. در فصل آخر این مساله در نظریه گرانش های مرتبه بالاتر بررسی شده است و در ادامه پایان نامه به عنوان کاربردی از فصول سوم و چهارم یکی از روش های مقید سازی یعنی روش شرایط انرژی را برای مدل $R^{1+\delta}$ بکار برده ایم.

قرارداد نوشتاری در تمام فصل های پایان نامه به این صورت است که تمامی شاخص ها اعم از لاتین و یونانی مقادیر صفر تا سه را می پذیرند و نشانگان متریک به صورت (+ + + −) است . بقیه روابط طبق قواعد نوشتاری کتاب d`Inverno می باشد.

# فصل اول

# مقدمه ای بر نظریه گرانش های مرتبه بالاتر[1]

دراین فصل به بررسی اجمالی نظریه گرانش های مرتبه بالاتر می پردازیم. با مطالعه این نظریه از ابتدای شکل گیری آن و آگاهی از برخی از نقاط ضعف و نقاط قوت این نظریه می توان به سودمندی آن پی برد .

علاوه بر موفقیت های نظریه گرانشی اینشتین، به این نظریه می توان تنها به منزله یک گام به طرف یک نظریه ساختاری جامع و کامل تر، به دلیل مشاهده ضعف هایی در این نظریه نگریست. حتی موفقیت های قبلی این نظریه دارای مشکلاتی عمیق و اساسی می باشد.

مثلاً این نظریه نتوانسته است نظریه وحدت یافته طبیعت را بدست دهد[۱]. شکست در بنا نهادن یک نظریه کوانتومی از گرانش، پرسش هایی اساسی در ساختار خود این نظریه مطرح می کند و از این رو  تلاش

---

برای یافتن نظریه های جایگزین را القاء می کند[۲]. گرایش فعالیت های اخیر به سمت این موضوع عمدتاً با امید به شکل دهی تحقیقاتی منسجم در اساس و بنیاد این نظریه صورت گرفته که در میان آنها می توان به تغییراتی در ابعاد و نیز در دینامیک فضا−زمان اشاره کرد. به طور خلاصه به توضیح هر یک از تغییرات فوق می پردازیم[۳]. اگر اصل ناوردایی عمومی [۱] در مورد گرانش بکار برده شود دیده می شود که انتگرال کنش باید کمیتی ناوردا باشد که در واقع قیدی است که توسط این اصل روی کنش تحمیل می شود. مفاهیم عمیق در این باره مرهون تلاش های هیلبرت[۲] است، او اولین کسی بود که لزوم ناوردایی عمومی کلیه قوانین فیزیک را در اصل وردشی[۳] بکار گرفت. استفاده از اصل کنش و اصل ناوردایی عمومی، یک ارتباط مستقیم بین اصول تقارن و قوانین بقاء را به عنوان اتحاد های ذاتی، مجاز می داند. تغییر در دینامیک عمدتاً با تغییر لاگرانژی یک سیستم بدست می آید. با این وجود، در نسبیت عام هنوز ارتباط بین سازوکار لاگرانژی برهمکنش ها و الزامات فیزیکی، موضوعی باز در مباحثات است.

در بین لاگرانژی اسکالر، معادلات میدان مبتنی بر لاگرانژی های ساخته شده از جملات درجه دوم تانسور خمش، همواره در نظریه گرانش از محبوبیت خاصی برخوردار بوده اند. این نظریه ها به دلایل گوناگون در حوزه های مختلف فیزیک نظری وارد شده اند[۳]. این نظریه ها در دو دهه اخیر با استقبال فراوانی مواجه شدند، زمانی آن ها عمدتاً به جهت حل مسائلی چند در کوانتیزه کردن گرانش پیشنهاد شدند. در این مورد، اولین ایده ها به روز های اولیه تولد نسبیت عام در تلاش های ادینگتون[۴][۴] و وایل[۵][۵] در راستای متحد کردن گرانش با الکترومغناطیس بر می گردد. با این همه، این تلاش ها به ثمر نرسیدند[۶,۷].

باخ[۶][۸] و لنچس[۷][۹] کنش هایی ساخته شده فقط با جملات مربعی از خمش در چهار بعد به عنوان اسکالر های ناوردا را بررسی کردند. در واقع این رهیافت از ایده وایل اتخاذ شده بود، که این ایده اصل ناوردایی پیمانه ای[۸] تحت تبدیلات همدیس[۹] با ضریب همدیس ثابت است. آنها از کنش $\int R^2 \sqrt{-g} d^4 x$ (صرف

نظر از یک ثابت) استفاده کردن. با این وجود این رهیافت به طور جدی مورد انتقاد قرار گرفت[۱۰]. دو ایراد عمده علیه این لاگرانژی ها عبارت اند از اینکه اولاً متریک مبتنی بر این لاگرانژی ها به طور مجانبی در فواصل دور به متریک تخت تبدیل نمی شود، ثانیاً عدم توافق با مشاهدات بعد از اضافه شدن ماده به مساله. در نتیجه اصلاحاتی در این رهیافت لازم شمرده شد. پیشنهادات اصلاحی در این حوزه از فعالیت ها با وجود بدست آمدن برخی نتایج مثبت تا قبل از دهه هفتاد خیلی فعال و پیشرو نبودند. در واقع پس از اقداماتی در پی کوانتیزه کردن نسبیت عام در دهه شصت، در این نظریات پیشرفت هایی حاصل شد. شاید انگیزه های مجاز ریاضی جهت آزمودن نظریه های گرانشی مبتنی بر لاگرانژی های غیر خطی، مشخصه پدیده شناسی نظریه اینشتین بوده که مجال یک چنین اصلاحاتی را باقی می گذارد، یعنی وابستگی تانسور اینشتین و لاگرانژی اینشتین به مشتقات متریک و ملاحظاتی در راستای افزایش ابعاد.

در حقیقت ملاحظات ابعادی و وارد کردن جملات شامل مشتقات متریک در لاگرانژی اینشتین، سر نخ هایی هستند که می توان در راستای تعمیم معادلات اینشتین بکار برد. لاگرانژی اینشتین عمومی ترین لاگرانژی مرتبه دوم مجاز از دید اصل ناوردایی عمومی نیست، از منظر این اصل تعمیمات بعدی می تواند تا حد اکثر هر مرتبه ای انجام شود و یک لاگرانژی اسکالر عام یک لاگرانژی متشکل ازمشتقات بالاتر متریک است[۳].

نقطه عطف نیاز به تعمیم لاگرانژی اینشتین اینجاست، با اینکه نشان داد شده است که نظریه اینشتین در حوزه پدیده های فیزیکی عادی مثلاً در میدان های گرانشی ضعیف موفق ظاهر می شود (یعنی این نظریه زمانی قابل کاربرد است که خمش فضا زمان خیلی زیاد نباشد)، ولی اشکال اصلی کار آنجا ناشی می شود که خمش قابل توجه می شود، یعنی وقتی که مقیاس ها بسیار کوچک باشند، و این به ویژه به زمان های اولیه تحول عالم بر می گردد. به بیان دیگر زمانی که مقیاسها از مرتبه طول پلانک باشند، و در چنین مقیاس هایی ساختار توپولوژیکی اقلیدسی بسیار نا محتمل است. در چنین فواصلی حتی افت و خیز های کوانتومی گرانش آنقدر شدید است که احتمالاً یک دینامیک متغیر را در ساختار توپولوژیکی عالم تحمیل می کند[۱۱]. بنا بر این احتمالاً در نظر گرفتن لاگرانژی های تعمیم یافته، فراسوی لاگرانژی اینشتین–هیلبرت به عنوان نظریه های جایگزین، کوششی بی فایده نباشد.

موضوعی دیگر این است که یک پارچه بودن فیزیک باید طبق اصل تطابق حفظ شود، یعنی در هر نظریه جدید، نظریه قبلی باید به عنوان یک حالت حدی از نظریه جامع تر جدید حفظ شود. همان طور که ذکر شد نظریه های گرانشی مبتنی بر تنها جملات مربعی خالص با انتقاد های شدید مواجه شدند[۱۰]. یک نظریه گرانشی نه تنها باید رفتار صحیحی از دینامیک کل عالم را بدست دهد، بلکه باید معادلات صحیحی از تحول ستاره ای را نشان دهد. بنا براین شخص باید انتظار داشته باشد که نسبیت عام به عنوان حالت حدی از نظریات گرانشی غیر خطی حفظ شود. در مورد بازه مجاز انرژی برای حوزه کاری نظریات مرتبه بالاتر لاگرانژی باید گفت که در نظریه وحدت[^1] برای ذرات بنیادی به نظر می رسد که انرژی مورد نیاز برای آزمودن این نظریه حداقل از مرتبه های $10^{15}\text{GeV}$ باشد. واضح است که زمان های اولیه تولد عالم تنها مجال مناسب برای آزمایش این نظریه ها می باشد. در واقع پیش داوری های استاندارد، مقدار حداقلی برای انرژی برای مشاهده اثرات کوانتومی، پیش گویی می کند.

قدرت اثرات کوانتومی برهم کنش های گرانشی نباید محسوس باشند مگر اینکه فواصل مورد بررسی در مسئله از مرتبه پلانک باشند. پس پرسش در مورد گرانش کوانتومی طبیعتاً با کیهان شناسی نخستین لحظات تولد عالم در ارتباط است[۱۲]. در ارتباط با اصل تطابق باید گفته شود که اثرات در حالت حدی لاگرانژی های غیر خطی باید به طور کامل ناچیز باشند یعنی زمانی که انرژی عالم کمتر از انرژی پلانک باشد.

در مورد نظریات با لاگرانژی مرتبه بالاتر باید گفت که این نظریات نقشی اساسی در رهیافت ابر گرانش[^2] ایفا می کنند[۱۳]. امروزه این موضوع به خوبی شناخته شده است که وقتی از گرانش اینشتین به عنوان شروعی برای گرانش کوانتومی استفاده شود، این نظریه به نظریه ای نا باز بهنجارش پذیر[^3] منجر می شود، هرچند این اشکال موقعی سر باز می زند که خمش فضا–زمان غیرقابل صرف نظر کردن باشد. برای حل این مشکل، یعنی برای اینکه بتوان واگرایی ها را باز بهنجار[^4] کرد، نشان داده شده است که کنش اینشتین – هیلبرت باید شامل جملات از مرتبه بالاتر خمش باشد[۱۴]. در واقع نشان داده شده است که لاگرانژی

$$L \propto (R + aR^2 + bR_{\alpha\beta}R^{\alpha\beta})$$

[^1]: unified theory
[^2]: supergravity
[^3]: non-normalizable
[^4]: renormalization

که حد نیوتونی مورد نیاز را دارد، و نیز قضیه گاوس-بونت[1] را ارضا می کند، عمومی ترین لاگرانژی مرتبه دوم در چهار بعد است که مساله بهنجارش را حل می کند[۱۵].

وجود این اشکال در نظریه اینشتین، لزوم استفاده از گرانش های مرتبه بالاتر را بازگو می کند. مکان دیگری که کنش اینشتین-هیلبرت رنگ می بازد، در نظریه ریسمان[2] است. این نظریه با هدف کاربرد فیزیک گرانش کلاسیک جملاتی از مرتبه بالاتر را وارد کرده و یک کنش موثر با انرژی کم که گرانش را در سطح کلاسیکی بیان می کند تشکیل می دهد.

از دیگر حوزه های گرانش های مرتبه بالا می توان به کیهان شناسی نسبیتی اشاره کرد. در اینجا از چنین لاگرانژی هایی برای اجتناب از ظهور تکینگی[3] که در برخی از حل های نسبیت عام وجود دارد استفاده می شود[۳]. از نقاط ضعف این نظریه (گرانش های مرتبه بالاتر) می توان به این نکته اشاره کرد، به علت حضور جملات مرتبه بالاتر خمش در لاگرانژی که منجر به معادلات میدان غیر خطی از درجات بالاتر می شود، یافتن جواب های غیر بدیهی بسیار مشکل می شود و این به نوبه خود استخراج تعبیر های فیزیکی مناسب را دچار مشکل می کند. به ویژه اینکه تکنولوژی حال حاضر ممکن است به سختی بتواند این تعابیر را مورد آزمایش قرار دهد.

نکته مهمی که شاید خیلی به آن دقت نشود این است که جملات مرتبه بالاتر اغلب به عنوان تصحیحی به لاگرانژی اولیه اینشتین-هیلبرت اضافه می شود، ولی این جملات نه تنها به معنی اختلال در نظریه اولیه نیستند بلکه واقعاً حضور آنها به عنوان جملات نا مقید و مستقل حتی با ضرایب کوچک، نظریه ای کاملا متفاوت با نظریه اولیه بدست می دهد[۱۷].

به عنوان جمع بندی می توان گفت با توجه به کاربرد های نظریه گرانش های مرتبه بالاتر مثلا در نظریه ریسمان یا کیهان شناسی نسبیتی لزوم مطالعه این نظریه به شکلی بنیادی احساس می شود که البته باید نسبت به مشکلاتی که ممکن است در کاربرد این نظریه ایجاد شود محتاط بود.

---

# فصل دوم

# اصول روش های وردشی در گرانش های مرتبه بالاتر

اصول وردشی نقشی تعیین کننده در فیزیک بازی می کنند. در طول این صده این موضوع مورد پذیرش همگان قرار گرفته که هر نظریه اساسی فیزیکی را می توان به بیان یک کنش به شکل یک هامیلتونی و یا یک لاگرانژی فرمول بندی کرد، که می توان از طریق این دو شکل به معادلات حرکت از طریق یک اصل وردشی دست یافت [۱۸].

در طی چند دهه اخیر این موضع به شکلی فراتر از یک قانون طبیعت پیشرفت کرده است. امروزه ساخت یک نظریه به این معنی است که فرد شروع به نوشتن یک کنش ساخته شده از میدان های موجود در نظریه کند، حتی اگر دانش در مورد شکل دقیق و واقعی برهم کنش ها، و ویژگی های تقارنی به شکلی قطعه-قطعه باشد. تعیین تابع لاگرانژی توسط ابزار های ریاضی و فیزیکی مثل ناوردایی پیمانه ای، باز بهنجارپذیری، ساده سازی و موارد دیگر انجام می شود.

از این رو در گرانش با سه رهیافت در اصول وردشی در مواجهه با چگالی لاگرانژی مساله رو به رو می شویم، وردش متریکی، وردش پالاتینی و ساختار مقید مرتبه اول.

## ۲-۱ وردش متریکی

در این روش که به آن وردش هیلبرت نیز گفته می شود فرض می شود همبستار[۱] ها از نوع لوی سیویتا[۲] هستند یعنی داریم: $\nabla_\mu\, g_{\alpha\beta} = 0$.

پس فضا-زمان ما یک خمینه[۳] $M$ با متریک لورنسی[۴] $g_{\alpha\beta}$ با همبستار های لوی سیویتاست. در روش وردش هیلبرت فرض می شود چگالی لاگرانژی در حالت کلی تابعی از متریک و مشتقات متریک است:

$$\mathcal{L} = \mathcal{L}(g, \partial g, \partial^2 g, \dots) \qquad (۲-۱)$$

که چگالی لاگرانژی عبارت است:

$$\mathcal{L} = \sqrt{-g}\, L$$

که $L$ لاگرانژی مساله و $g$ دترمینان ماتریس تشکیل شده از مولفه های $g_{\alpha\beta}$ است. بنا بر این انتگرال کنش بدین شکل خواهد شد:

$$S[g] = \int \mathrm{d}^4\Omega \mathcal{L} \qquad (۲-۲)$$

عنصر انتگرال گیری عبارت است از:

$$d^4\Omega = dx^0 \wedge dx^1 \wedge dx^2 \wedge dx^3$$

و این انتگرال روی کل ناحیه $u$ روی خمینه $M$ گرفته می شود. معادلات میدان از این شرط بدست می آیند که کنش فوق تحت وردش های اختیاری پایا باشد، با این شرط مرزی که متریک و مشتقات اول

---

آن روی مرز $\partial u$ ثابت هستند. این وردش، مشتق تابعی[1] $\mathcal{L}_{\alpha\beta}$ از چگالی لاگرانژی $\mathcal{L}$ را به شکل زیر تعریف می کند:

$$\delta S[\mathrm{g}] = \int \mathrm{d}^4\Omega\, \mathcal{L}_{\alpha\beta}\delta \mathrm{g}^{\alpha\beta}$$

$$\mathcal{L}_{\alpha\beta} \equiv \frac{\delta \mathcal{L}}{\delta \mathrm{g}^{\alpha\beta}}$$

$\mathcal{L}_{\alpha\beta}$ را همچنین مشتق اولر–لاگرانژ چگالی لاگرانژی می نامند. معادلات میدان از صفر بودن مشتق اولر–لاگرانژ بدست می آید یعنی:

$$\mathcal{L}_{\alpha\beta} = 0 \qquad\qquad (۲-۳)$$

به عنوان مثال می توانیم چگالی لاگرانژی اینشتین–هیلبرت به شکل $\mathcal{L}_{EH} = \sqrt{-g}\,R$ را نام ببریم. وقتی از این روش وردش گیری استفاده کنیم، معادلات خلاء اینشتین بدست می آید. جزئیات این محاسبه در بسیاری از کتاب ها آمده است. اگر در این چگالی لاگرانژی، چگالی لاگرانژی مربوط به ماده هم اضافه کنیم به معادله $R_{\alpha\beta} - \frac{1}{2}\mathrm{g}_{\alpha\beta}R = \frac{\chi}{2}\mathrm{T}_{\alpha\beta}$ خواهیم رسید.(با فرض صفر بودن ثابت کیهان شناسی)

مثال بعدی کنش زیر است:

$$S = \int \mathrm{d}^4\Omega[\mathcal{L}_{EH} + c_1\mathcal{L}_1 + c_2\mathcal{L}_2 + c_3\mathcal{L}_3] \qquad\qquad (۲-٤)$$

در این حالت چگالی لاگرانژی علاوه بر قسمت مربوط به اینشتین–هیلبرت شامل سه قسمت دیگر نیز می باشد. هر یک از $\mathcal{L}_i$ ها ناوردادهای مربعی از اسکالر ریچی[2]، تانسور ریچی[3] و تانسور خمش[4] به شکل زیر هستند:

$$\mathcal{L}_1 \equiv \sqrt{-g}\,R^2 \qquad\qquad (الف-۵-۲)$$

$$\mathcal{L}_2 \equiv \sqrt{-g}\,R_{\alpha\beta}R^{\alpha\beta} \qquad\qquad (ب-۵-۲)$$

---

[1] functional derivative
[2] Ricci scalar
[3] Ricci tensor
[4] curvature tensor

$$\mathcal{L}_3 \equiv \sqrt{-g} R_{\alpha\beta\mu\nu} R^{\alpha\beta\mu\nu}$$

<div dir="rtl">

(پ-۵-۲)

مثلاً ترکیبی معروف از این سه جمله همان ترکیب گاوس-بونت با ضرایب $c_1 = 1$، $c_2 = -4$ و $c_3 = 1$ است.

مشتق اولر-لاگرانژ سه چگالی لاگرانژی فوق به ترتیب عبارت است از:

</div>

$$\mathcal{L}_1{}^{\alpha\beta} = \sqrt{-g}[\tfrac{1}{2}g^{\alpha\beta}R^2 - 2RR^{\alpha\beta} + 2\nabla^\alpha\nabla^\beta R - 2g^{\alpha\beta}g_{\mu\nu}\nabla^\mu\nabla^\nu R]$$

$$\mathcal{L}_2{}^{\alpha\beta} = \sqrt{-g}[\tfrac{1}{2}g^{\alpha\beta}R_{\mu\nu}R^{\mu\nu} - 2R^{\beta\mu\alpha\nu}R_{\mu\nu} + \nabla^\alpha\nabla^\beta R - g_{\mu\nu}\nabla^\mu\nabla^\nu R^{\alpha\beta} - \tfrac{1}{2}g^{\alpha\beta}g_{\mu\nu}\nabla^\mu\nabla^\nu R]$$

$$\mathcal{L}_2{}^{\alpha\beta} = \sqrt{-g}[\tfrac{1}{2}g^{\alpha\beta}R_{\mu\nu\delta\gamma}R^{\mu\nu\delta\gamma} - 2R_{\mu\nu\delta}{}^\alpha R^{\mu\nu\delta\beta} - 4g_{\mu\nu}\nabla^\mu\nabla^\nu R^{\alpha\beta} + 2\nabla^\alpha\nabla^\beta R - 4R_{\mu\nu}R^{\beta\mu\alpha\nu} + 4R^\beta{}_\mu R^{\mu\alpha}]$$

<div dir="rtl">

که این معادلات مشتقاتی درجه چهارم از مولفه های متریک اند. هر یک از این جملات به تنهایی در بسیاری از مقاله های کار شده به چشم می خورند. در واقع گفته می شود که چگالی لاگرانژی مربعی از تانسور خمش، معادلات میدانی از درجه چهار بدست می دهند (وردش مربوط به چگالی لاگرانژی دوم در ضمیمه ب به طور مفصل توضیح داده شده است).

به عنوان مثال آخر می توان چگالی لاگرانژی تشکیل شده از تابع غیر خطی و عمومی به شکل زیر را بررسی کرد:

</div>

$$\mathcal{L} = \sqrt{-g}f(R)$$

<div dir="rtl">

با شرط $0 \neq f''(R)$. مشتق اولر-لاگرانژ عبارت اند از:

</div>

$$f'R_{\alpha\beta} - \tfrac{1}{2}fg_{\alpha\beta} - \nabla_\alpha\nabla_\beta f' + g_{\alpha\beta}g_{\mu\nu}\nabla^\mu\nabla^\nu f' = 0$$

که این مشتق اولین بار توسط بوچدال[19] بدست آمد (وردش مربوطه در ضمیمه پ توضیح داده شده است).

## ۲-۲ وردش پالاتینی

در این روش که به آن روش وردشی متریک آفین[2] نیز می گویند خمینه مثل قبل عبارت است از یک فضا-زمان چهار بعدی با یک متریک لورنسی. اما هموستارهای $\nabla_\alpha$ هموستار های اختیاری و متقارن هستند یعنی شرط سازگاری متریک[3] $\nabla_\alpha g_{\rho\sigma} = 0$ وجود ندارد. یعنی در این روش هموستار و متریک مستقل از هم فرض می شوند و هیچ قید پیش فرضی بین آن ها وجود ندارد. در اینجا پارامتر پیچش[4] صفر است.

در این روش، چگالی لاگرانژی تشکیل شده است از متریک و مشتقات همموردا[5] از متریک تا مرتبه معینی، و ضرایب هموستار و مشتقات ضرایب هموستار حداکثر تا مرتبه معینی. نکته مهم در این چگالی لاگرانژی این است که در این حالت چون هموستارها و متریک مستقل از هم هستند یعنی شرط $\nabla_\alpha g_{\rho\sigma} = 0$ وجود ندارد چگالی لاگرانژی بجای اینکه تابعی از $\partial g$ باشد تابعی از مشتقات همموردا از متریک است. پس در این روش چگالی لاگرانژی به شکل عمومی زیر است:

$$\mathcal{L} = \mathcal{L}(g, \nabla g, \nabla\nabla g, \dots; \Gamma, \partial\Gamma, \partial^2\Gamma, \dots)$$  (۲-۶)

که $\Gamma$ هموستار است. انتگرال کنش عبارت است از:

$$S[g, \Gamma] = \int d^4\Omega \mathcal{L}$$  (۲-۷)

و این انتگرال روی خمینه مساله گرفته می شود. وردش کنش نسبت به متغیر های مستقل هموستار و متریک که روی مرز مساله صفر در نظر گرفته می شود عبارت است از:

$$\delta S[g, \Gamma] = \int d^4\Omega (\mathcal{R}_\alpha{}^{\beta\gamma}\delta\Gamma_{\beta\gamma}{}^\alpha + \mathcal{B}_{\mu\nu}\delta g^{\mu\nu})$$  (۲-۸)

---

که در آن $\mathfrak{R}_\alpha{}^{\beta\gamma}$ و $\mathcal{B}_{\mu\nu}$ مشتقات اول لاگرانژ $\mathcal{L}$ نسبت به همبستار و متریک است. معادلات میدان عبارتند از:

$$\mathfrak{R}_\alpha{}^{\beta\gamma} = 0 \quad , \quad \mathcal{B}_{\mu\nu} = 0 \qquad (۲-۹)$$

به عنوان مثال کنش زیر را در نظر بگیرید:

$$\mathcal{L} = \sqrt{-g}\, g^{\alpha\beta} R_{\alpha\beta}(\Gamma, \partial\Gamma)$$

که در آن تانسور ریچی تنها تابعی ازضرایب همبستار و مشتق اول این ضرایب است. وردش این چگالی لاگرانژی نسبت به متریک بدست می دهد:

$$\delta S = \int d^4\Omega \sqrt{-g}\left(R_{\alpha\beta} - \frac{1}{2} R g_{\alpha\beta}\right)\delta g^{\alpha\beta} = 0$$

$$R_{\alpha\beta} - \frac{1}{2} R g_{\alpha\beta} = 0$$

که این معادله همان معادله اینشتین در خلاء است. در حالی که جواب نهایی وردش نسبت به همبستار عبارت است از:

$$\delta S = \int d^4\Omega (\delta_\alpha{}^\beta \nabla_\mu \mathscr{g}^{\gamma\mu} - \nabla_\alpha \mathscr{g}^{\gamma\beta})\delta\Gamma^\alpha{}_{\gamma\beta} = 0$$

که تعریف کرده ایم: $\mathscr{g}^{\gamma\beta} \equiv \sqrt{-g}\, g^{\gamma\beta}$. از آنجایی که $\Gamma^\alpha{}_{\gamma\beta}$ نسبت به دو شاخص پایین متقارن است در نتیجه قسمت متقارن عبارت داخل پرانتز در بالا باید صفر شود یعنی :

$$\delta_\alpha{}^{(\beta} \nabla_\mu \mathscr{g}^{\gamma)\mu} - \nabla_\alpha \mathscr{g}^{\gamma\beta} = 0$$

این معادله شرط متریک سازگاری را به صورت زیر بدست می دهد:

$$\nabla_\alpha \mathscr{g}^{\gamma\beta} = 0 = \nabla_\alpha g^{\gamma\beta}$$

و از این رو همبستار ها لزوما نمادهای کریستوفل [۱] هستند: $\Gamma^\alpha{}_{\gamma\beta} = \begin{Bmatrix} \alpha \\ \gamma\beta \end{Bmatrix}$

---

[۱] Christofel symbol

با استفاده از روش وردش اینشتین پالاتینی توانستیم معادله اینشتین همراه با شرط متریک سازگاری را بدست آوریم. این در حالی است که در روش وردش متریکی با فرض وجود شرط متریک سازگاری معادله اینشتین بدست آمد. یعنی شاید تصور شود این دو روش معادلند. ولی بدون وارد شدن در جزئیات باید گفت که در حالت کلی این دو روش معادل نمی باشند. مثلاً وقتی چگالی لاگرانژی، شامل جمله مربوط به ماده می شود روش اینشتین پالاتینی با شکست مواجه می شود. اگر کنش کلی مساله، شامل جمله مربوط به ماده به شکل زیر باشد:

$$S = \int d^4\Omega \left[ \sqrt{-g}\, g^{\alpha\beta} R_{\alpha\beta}(\Gamma, \partial\Gamma) + \mathcal{L}_M \right]$$ (۲-۱۰)

مشتقات اول لاگرانژ به شکل زیر خواهند بود:

$$\sqrt{-g}\left( R_{\alpha\beta} - \frac{1}{2} R g_{\alpha\beta} \right) = -2 \frac{\delta \mathcal{L}_M}{\delta g^{\alpha\beta}}$$

$$\delta_\alpha{}^{(\beta} \nabla_\mu \mathscr{g}^{\gamma)\mu} - \nabla_\alpha \mathscr{g}^{\gamma\beta} = \frac{\delta \mathcal{L}_M}{\delta \Gamma^\alpha{}_{\gamma\beta}}$$

واضح است که معادله اول، معادله میدان اینشتین را در حضور ماده بیان می کند. ولی در رابطه پایین می بینیم که هموستار های بدست آمده دیگر نمادهای کریستوفل نیستند و در حالت کلی به چگالی لاگرانژی ماده وابسته اند. در نتیجه این دو معادله، معادله ای شبیه به معادله کامل میدان اینشتین در یک خمینه غیر ریمانی را نمایش می دهند. با این وجود در نظریاتی که لاگرانژی ماده صریحا به ضرایب هموستار وابسته نیست، سمت راست معادله پایین صفر شده و در نتیجه هموستارها ی بدست آمده نماد های کریستوفل خواهند بود. در نتیجه در شرایطی که $\frac{\delta \mathcal{L}_M}{\delta g^{\alpha\beta}} = 0$ و $\frac{\delta \mathcal{L}_M}{\delta \Gamma^\alpha{}_{\gamma\beta}} = 0$ باشد ما معادله کامل اینشتین را در یک خمینه ریمانی با هموستار های کریستوفل خواهیم داشت[۱۸].

مثال بعدی لاگرانژی های مربعی است که قبلا معادلات آنها را در روش ورش متریکی بدست آوردیم، در اینجا معادلات میدان را در روش اینشتین-هیلبرت بدست می آوریم[۲۰]:

$$\delta \mathcal{L}_1 \rightarrow R\left( R_{(\alpha\beta)} - \frac{1}{4} R g_{\alpha\beta} \right) = 0$$

$$\nabla_\mu (R \sqrt{-g}\, g^{\alpha\beta}) = 0$$

$$\delta \mathcal{L}_2 \rightarrow R_\alpha{}^\mu R_{\beta\mu} + R^\mu{}_\alpha R_{\mu\beta} - \frac{1}{2} R_{\rho\sigma} R^{\rho\sigma} g_{\alpha\beta} = 0$$

$$\nabla_\mu\left(\sqrt{-g}R^{\alpha\beta}\right) = 0$$

$$\delta\mathcal{L}_3 \rightarrow 2R_{\rho\sigma\alpha}{}^\mu R^{\rho\sigma}{}_{\beta\mu} - R_{\alpha\rho\sigma\mu}R_\beta{}^{\rho\sigma\mu} + R^\mu{}_{\alpha\rho\sigma}R_{\mu\beta}{}^{\rho\sigma} - R_{\rho\sigma\mu\delta}R^{\rho\sigma\mu\delta}g_{\alpha\beta} = 0$$

$$\nabla_\sigma\left(\sqrt{-g}R_\mu{}^{(\alpha\beta)\sigma}\right) = 0$$

در هر یک از سه لاگرانژی، برای بدست آوردن معادله دوم (معادلات شامل مشتقات همموردا) از انتگرال گیری به جزء استفاده کرده ایم (ضمیمه ث و ج). جنبه های مهم این معادلات عبارتند از:

الف) این معادلات تحت تبدیلات همدیس، یعنی تحت تبدیلاتی به شکل $\tilde{g}_{\alpha\beta} = \Omega^2(x)g_{\alpha\beta}$ که $\Omega^2(x)$ یک میدان نرده ای مثبت و اختیاری است، ناوردا هستند. می توان این موضوع را از اینجا فهمید که هموستار ها در اینجا مستقل از متریک هستند و چگالی های لاگرانژی ها نسبت به خمش به درجه دوم است. یک چگالی لاگرانژی وجود دارد که از تانسور وایل ساخته می شود. در روش متریکی معادلات میدان این چگالی لاگرانژی نسبت به تبدیلات همدیس ناوردا است. این چگالی لاگرانژی به شکل زیر است:

$$\mathcal{L} = \sqrt{-g}C_{\alpha\beta\gamma\delta}C^{\alpha\beta\gamma\delta}$$

یکی از کاربرد های این موضوع این است که مثلاً حل های مربوط به دو لاگرانژی $\mathcal{L}_1$ و $\mathcal{L}_2$ تحت تبدیلات همدیس به معادله اینشتین به علاوه یک ثابت اختیاری تبدیل می شوند[21].

ب) این معادلات، معادلات دیفرانسیل مرتبه دوم نسبت به مشتق متریک اند. خصوصا اینکه اینها با معادلات مرتبه چهارم که با روش وردش متریکی بدست می آیند معادل نیستند[18].

ج) این معادلات هموستارهای لوی سیویتا را در حالت کلی بدست نمی دهند، یعنی فضای ریمان را القاء نمی کنند. فضای القاء شده توسط مساله در حالت کلی یک فضای ویل است که در حالت خاص می توانیم فضای اینشتین را با یک سری تبدیلات همدیس بدست آوریم. در واقع فضای القایی، فضای ویل نیست بلکه فضای ریمان با یک پیمانه نامعین است. در زیر این موضوع را نشان می دهیم[18].

با اعمال روش وردشی اینشتین-پالاتینی، به چگالی لاگرانژی زیر:

$$\mathcal{L} = \sqrt{-g}f(g^{\alpha\beta}R_{\alpha\beta}(\Gamma)) \tag{٢-١١}$$

جفت معادله زیر را بدست می آوریم:

$$f'R_{(\alpha\beta)} - \frac{1}{2}fg_{\alpha\beta} = 0 \qquad (\text{۱۲-۲})$$

$$\nabla_\mu\left(f'\sqrt{-g}g^{\alpha\beta}\right) = 0 \qquad (\text{۱۳-۲})$$

در روش وردش متریکی معادله دیگری بدست آوردیم که با جفت معادله بالا متفاوت است. حال معادلات بالا را بررسی می کنیم.

معادله پایینی را به شکل زیر می نویسیم:

$$f'\nabla_\mu\left(\sqrt{-g}g^{\alpha\beta}\right) + f''\left(\sqrt{-g}g^{\alpha\beta}\right)\nabla_\mu R = 0$$

با استفاده از دو فرمول $\nabla_\mu\sqrt{-g} = \partial_\mu - \sqrt{-g}\Gamma_\mu$ و $\nabla_\mu R = \partial_\mu R$ و با فرض $f' \neq 0$ معادله را به شکل زیر ساده می کنیم:

$$(\text{۱٤-۲})$$

$$\left[\partial_\mu\left(\ln\sqrt{-g}\right) + (\ln f')' \partial_\mu R - \Gamma^\rho{}_{\mu\rho}\right]g^{\alpha\beta} + \partial_\mu g^{\alpha\beta} + \Gamma^\alpha{}_{\mu\rho}g^{\rho\beta} + \Gamma^\beta{}_{\mu\rho}g^{\alpha\rho} = 0$$

در نتیجه بعد از عمل رد[1] گیری از معادله بالا خواهیم داشت:

$$\Gamma_\mu = \partial_\mu\left(\ln\sqrt{-g}\right) + 2(\ln f')'\partial_\mu R$$

از آنجایی که داریم $R_{[\alpha\beta]} = \partial_{[\alpha}\Gamma_{\beta]}$ این معادله به این معناست که تانسور ریچی متقارن است. اگر مقدار بدست آمده برای $\Gamma_\mu$ را در معادله (۱۲-۲) قرار دهیم خواهیم داشت:

$$\partial_\mu\left(f'g_{\alpha\beta}\right) = f'(\Gamma^\rho{}_{\mu\alpha}g_{\beta\rho} + \Gamma^\rho{}_{\mu\beta}g_{\alpha\rho})$$

اگر تعریف کنیم $\tilde{g}_{\alpha\beta} = f'g_{\alpha\beta}$ معادله بالا شکل مطلوب زیر را به خود خواهد گرفت:

$$\partial_\mu\left(\breve{g}_{\alpha\beta}\right) = \Gamma^\rho{}_{\mu\alpha}\tilde{g}_{\beta\rho} + \Gamma^\rho{}_{\mu\beta}\tilde{g}_{\alpha\rho} \qquad (\text{۱٥-۲})$$

---

نتیجه بالا این را القا می کند که مشتق همموردای متریک جدید $\tilde{g}_{\alpha\rho}$ نسبت به هموستار $\nabla_\mu$ صفر است. از این رو ضرایب هموستار $\Gamma^\rho{}_{\alpha\beta}$، هموستار های لوی سیویتای متریک جدید است. از سویی دیگر می توان معادله (۲-۱۴) را به شکل زیر نوشت:

$$\nabla_\mu g^{\alpha\beta} = \nabla_\mu (\ln f') g^{\alpha\beta}$$

یعنی به نظر می رسد روش وردشی اینشتین-پالاتینی یک فضای ویل را انتخاب کرد با $Q_\mu = \nabla_\mu (\ln f')$. با وجود این، این موضوع درست نیست یعنی فضای مساله یک فضای ریمانی با یک پیمانه نا معلوم است[۲۲]. در یک فضای ریمانی با هموستار متقارن یک فرمی [1] $Q_\mu$ صفر است. حال اگر در یک فضای ویل از تبدیل همدیس $\tilde{g}_{\alpha\beta} = \Omega^2 g_{\alpha\beta}$ استفاده کنیم یک فرمی به شکل $\tilde{Q}_\mu = Q_\mu - 2\nabla_\mu (\ln\Omega)$ تبدیل می یابد. حال اگر $Q_\mu$ به شکل گرادیان کامل باشد، آنگاه داریم $\tilde{Q}_\mu = 0$ یعنی با فرض اینکه یک فرمی $Q_\mu$ گرادیان کامل باشد مساله ما تحت یک تبدیل همدیس فضای ریمان را انتخاب می کند. و این حالتی است که در اینجا وجود دارد زیرا در اینجا داریم $Q_\mu = \nabla_\mu (\ln f')$ و در نتیجه در مساله ما $\tilde{Q}_\mu = 0$ است.

در نتیجه ما همواره با انتخاب $f' = \Omega^2$ می توانیم مساله ای با فضای ریمان داشته باشیم. از طرفی از (۲-۱۲) خواهیم داشت:

$$f'(R)R = 2f(R)$$

یعنی تا حداکثر یک ضریب انتگرال گیری داریم $f(R) = R^2$. که البته این نتیجه انتظار می رفت زیرا در نکته الف گفتیم که معادلات ناشی از چگالی های لاگرانژی جملات مربعی تحت تبدیلات همدیس ناوردا است. و ما در اینجا این ناوردایی را در حالت کلی نشان دادیم. پس انتظار داریم تابع $f(R)$ یک تابع مربعی باشد. حال اگر نتیجه بالا را در (۲-۱۰) جاگذاری کنیم داریم:

$$R_{\alpha\beta} - \frac{1}{4} R g_{\alpha\beta} = 0$$

حال این معادله تحت تبدیل $\tilde{g}_{\alpha\beta} = k g_{\alpha\beta}$ برابر خواهد شد با:

―――――――――
[1] one form

$$\tilde{R}_{\alpha\beta} - \frac{1}{4}k^{-1}R\tilde{g}_{\alpha\beta} = 0$$

در واقع به این نتیجه می رسیم که خمینه مساله یک فضای اینشتین  با یک ثابت اختیاری است.

۲-۳ ساختار مقید مرتبه اول

هدف این روش تعمیم روش های وردشی بالا به چگالی لاگرانژی هایی است که عمومیت بیشتری دارند. استفاده از روش ضرایب لاگرانژ در یک روش وردشی زمانی قابل دفاع است که هر گاه فرد با برخی قید های تحمیل شده روی پارامتر های مستقل مساله که در حالت کلی ممکن است با روش وردشی مورد استفاده سازگاری نداشته باشد، مواجه شود. در هندسه ریمانی این روش به صورت اضافه کردن شرط متریک سازگاری به شکل یک قید با یک ضریب لاگرانژ به چگالی لاگرانژی مساله انجام می شود. سپس شخص کنش حاصل را نسبت به متریک، هموستار و ضریب لاگرانژ به عنوان سه پارامتر مستقل وردش می دهد.

وقتی این روش به چگالی لاگرانژی هیلبرت-پالاتینی به شکل $\mathcal{L}_{HP} = \sqrt{-g}\,g^{\alpha\beta}R_{\alpha\beta}(\Gamma, \partial\Gamma)$ که به دو روش قبل آن را بررسی کردیم، اعمال می شود، ضریب لاگرانژ به عنوان نتیجه ای از معادلات میدان صفر می شود. این روش تعبیر هم ارز بودن دو روش متریکی خاص و اینشتین پالاتینی ، در خلاء را به صورت فقط یک اتفاق تقویت می کند. این روش وقتی به لاگرانژی های مرتبه بالاتر اعمال می شود مراحل انجام محاسبه کمتر شده و دید بهتری در تفسیر معادلات به دست می آید[۱۸].

برای شروع فضای ریمانی ، با تعریف معمول خود را به عنوان خمینه مساله تعریف می کنیم و در مرحله بعد قید متریک سازگاری یعنی $\nabla_\mu g^{\alpha\beta} = 0$ را رها کرده و فرض می کنیم که $\nabla_\mu g^{\alpha\beta} \neq 0$ به عنوان یک قید مستقل وجود دارد. برای اینکه بعد از محاسبات معادلات، فضای اولیه را القاء کنند، (در اینجا فضای اولیه ریمان فرض شده است ولی در حالت کلی هر فضایی را می توان تعریف کرد) یک قید به لاگرانژی اولیه اضافه می شود که به شکل زیر است:

$$\mathcal{L}_c(g, \Gamma, \Lambda) = \sqrt{-g}\,\Lambda_\mu{}^{\alpha\beta}(\Gamma^\mu{}_{\alpha\beta} - \begin{Bmatrix} \mu \\ \alpha\beta \end{Bmatrix} - C^\mu{}_{\alpha\beta}) \qquad (\text{۲-۱۶})$$

که $\Lambda_\mu{}^{\alpha\beta}$ ضرایب نامعین لاگرانژ هستند و $C^\mu{}_{\alpha\beta}$ تانسور اختلاف بین همبستار اختیاری $\Gamma^\mu{}_{\alpha\beta}$ و نماد های کریستوفل هستند. مثلاً در هندسه ریمانی. و قید بالا به صورت زیر تغییر می کند: $C^\mu{}_{\alpha\beta} = 0$

$$\mathcal{L}_c(g, \Gamma, \Lambda) = \sqrt{-g}\,\Lambda_\mu{}^{\alpha\beta}(\Gamma^\mu{}_{\alpha\beta} - \begin{Bmatrix} \mu \\ \alpha\beta \end{Bmatrix}) \qquad (۲-۱۷)$$

در نتیجه کنش نهایی مساله عبارت است از:

$$S = \int d^4\Omega\, [\mathcal{L}(g, \Gamma, \psi) + \mathcal{L}_c(g, \Gamma, \Lambda)] \qquad (۲-۱۸)$$

فرض کرده ایم لاگرانژی اصلی مساله شامل یک میدان اسکالر $\psi$ است. حال این کنش باید نسبت به $g^{\alpha\beta}$، $\Lambda_\mu{}^{\alpha\beta}$ و $\Gamma^\mu{}_{\alpha\beta}$ و $\psi$ وردش داده شود.

وردش نسبت به متریک به معادلاتی موسوم به معادلات **g** منجر می شوند که به شکل زیر هستند:

$$\frac{\delta\mathcal{L}}{\delta g^{\alpha\beta}}\Big|_\Gamma + \sqrt{-g}\,B_{\alpha\beta} = 0 \qquad (\text{الف} -۲-۱۹)$$

$$\sqrt{-g}\,B_{\alpha\beta} \equiv -\frac{1}{2}\sqrt{-g}\,\nabla^\mu(\Lambda_{\beta\alpha\mu} + \Lambda_{\alpha\mu\beta} - \Lambda_{\mu\alpha\beta}) \qquad (\text{ب} -۲-۱۹)$$

وردش نسبت به همبستار به معادلات **Γ** به صورت زیر منجر می شوند:

$$\frac{\delta\mathcal{L}}{\delta\Gamma^\mu{}_{\alpha\beta}}\Big|_g + \Lambda_\mu{}^{\alpha\beta} = 0 \qquad (۲-۲۰)$$

وردش نسبت به میدان اسکالر به معادلات حرکت مربوط منجر می شود. سرانجام وردش نسبت به ضرایب لاگرانژ قید (۲-۱۶) را بدست می دهد.

اگر این روش را روی فضای ریمان روی چگالی هیلبرت پالاتینی اعمال کنیم همان معادلاتی بدست خواهد آمد که در روش متریکی خالص بدست می آید. به عنوان مثال برای چگالی لاگرانژی های (۲-۵) داریم:

$$\frac{1}{2}g^{\alpha\beta}R^2 - 2RR^{\alpha\beta} + B^{\alpha\beta} = 0 \qquad (۲-۲۱)$$

$$\Lambda_\mu{}^{\alpha\beta} = (2g^{\alpha\beta}\delta_\mu{}^\rho - g^{\alpha\rho}\delta_\mu{}^\beta - g^{\rho\beta}\delta_\mu{}^\alpha)\nabla_\rho R$$

$$B^{\alpha\beta} = -2g^{\alpha\beta}\Box R + 2\nabla^\beta\nabla^\alpha R$$

برای $\mathcal{L}_1$ و

$$\frac{1}{2}g^{\alpha\beta}R_{\rho\sigma}R^{\rho\sigma} - R_\gamma{}^\beta R^{\alpha\gamma} - R_\gamma{}^\alpha R^{\beta\gamma} + B^{\alpha\beta} = 0 \qquad (\text{۲۲-۲})$$

$$\Lambda_\mu{}^{\alpha\beta} = 2\nabla_\mu R^{\alpha\beta} - \delta_\mu{}^\alpha \nabla_\rho R^{\rho\beta} - \delta_\mu{}^\beta \nabla_\sigma R^{\alpha\sigma}$$

$$B^{\alpha\beta} = -\Box R^{\alpha\beta} + 2\nabla_\sigma \nabla^\beta R^{\alpha\sigma} - g^{\alpha\beta}\nabla_\sigma\nabla_\rho R^{\rho\sigma}$$

برای $\mathcal{L}_2$ و

$$\frac{1}{2}g^{\alpha\beta}R_{\rho\sigma\gamma\delta}R^{\rho\sigma\gamma\delta} - 2R_{\rho\sigma\gamma}{}^\beta R^{\alpha\rho\sigma\gamma} + B^{\alpha\beta} = 0 \qquad (\text{۲۳-۲})$$

$$\Lambda_\mu{}^{\alpha\beta} = 2\nabla_\rho R_\mu{}^{\alpha\beta\rho} + 2\nabla_\rho R_\mu{}^{\beta\alpha\rho} - \delta_\mu{}^\beta \nabla_\sigma R^{\alpha\sigma}$$

$$B^{\alpha\beta} = 4\nabla_\sigma\nabla_\rho R^{\alpha\rho\beta\sigma}$$

برای $\mathcal{L}_3$ برای مدل کلی غیر خطی خواهیم داشت

$$\frac{1}{2}fg^{\alpha\beta} - f'R^{(\alpha\beta)} + B^{\alpha\beta} = 0 \qquad (\text{۲٤-۲})$$

$$\Lambda_\mu{}^{\alpha\beta} = \frac{1}{2}(2g^{\alpha\beta}\delta_\mu{}^\rho - g^{\alpha\rho}\delta_\mu{}^\beta - g^{\rho\beta}\delta_\mu{}^\alpha)\nabla_\rho f'$$

$$B^{\alpha\beta} = -g^{\alpha\beta}\Box f' + \nabla^\beta\nabla^\alpha f'$$

حال اگر در هر یک از چهار مدل بالا $B^{\alpha\beta}$ را که از معادلات $\Gamma$ بدست آمده را در معادله اول قرار دهیم، مجددا معادلات مشتقات اول- لاگرانژ را که در روش متریکی خالص بدست آمد، بدست می آیند.

به بیان دیگر در مقایسه با روش هیلبرت عادی، در همه مثال های بررسی شده تا به حال، با استفاده از سازوکار مقید شده مرتبه اول، شخص از یک چگالی لاگرانژی که در فضای تابعی (یعنی چگالی لاگرانژی تابعی از توابع مختلف است) تعریف شده، و از یک روش وردشی متفاوت استفاده می کند به یک مجموعه معادلات، معادل با معادلاتی که در روش متریکی بدست می آید منتهی می شود. در واقع سودمندی این روش از اینجاست که با استفاده از یک الگوی مشخص به جواب نهایی می رسیم.

این الگو به صورت زیر است:

۲۱

الف) تشکیل کنش مساله با استفاده از قید (۲-۱۶)

ب) وردش نسبت به $\boldsymbol{\Gamma}$ و بدست آوردن ضرایب لاگرانژ

ج) وردش نسبت به متریک و بدست آوردن معادله $\mathbf{g}$.

د) محاسبه $B^{\alpha\beta}$ با استفاده از ضرایب لاگران بدست آمده از مرحله ب) با استفاده از رابطه (۲-۱۹ ب) و اگذاری آن در معادله $\mathbf{g}$.

# فصل سوم

# شرایط انرژی در گرانش های مرتبه بالا

در این فصل در مورد شرایط انرژی در گرانش های مرتبه بالاتر صحبت می کنیم[۲۳،۲۴].

مشاهدات برای عالم کنونی، انبساط شتاب دار با مقدار مثبت را نشان می دهد[۲۵]. مدل های بسیاری برای توجیه این مورد ارائه شده اند، که همگی به دو گروه تقسیم می شوند. گروه اول مدل هایی بر پایه تصحیحاتی روی تانسور انرژی-تکانه در معادله اینشتین است، که در برخی از آن ها به وجود برخی میدان های جدید در این تانسور اشاره شده است و بیشتر در فیزیک انرژی بالا مورد بررسی قرار می گیرند. و یا در برخی دیگر به یک شاره ایده آل اشاره می کند و در هر دو مورد این ماده باید شرط انرژی قوی[۱] را نقض کنند تا بتوانند عالمی شتاب دار را توصیف نمایند.

در گروه دوم پژوهش ها ماده از نوعی معمولی انتخاب می شود ولی در اینجا تصحیحاتی روی ساختار هندسی نظریه نسبیت عام انجام می گیرد. نظریات گرانش های مرتبه بالاتر یعنی نظریاتی که شامل توان هایی از مرتبه بیشتر از یک، از اسکالر ریچی هستند (از این حیث نسبیت عام اینشتین نظریه ای خطی به

---

حساب می آید زیرا در کنش آن، توان یک از اسکالر ریچی ظاهر می شود)، و نظریات با ابعاد بیشتر از چهار بعد از این گروه اند[٢٤].

نظریات مختلف گرایش های مرتبه بالا در زمینه های مختلفی بررسی شده اند. در بررسی شرایط پایداری، سیستم خورشیدی، بررسی روی کهکشان ها، تحولات اولیه عالم و دوره تورمی[1] و مواردی دیگر. این نظریات روش جدیدی در بررسی سرعت فزاینده تحول عالمی، بدون داخل کردن انرژی تاریک در مساله ارائه نموده اند[٢٣]. اختیاری که در انتخاب تابع تحلیلی $f(R)$ وجود دارد مشکلاتی چند را به وجود آورده است. مثلاً معادلات حرکت منتج آنها عموماً به گونه ای غیر خطی اند که حل ساده ترین آن ها در بسیاری موارد امکان پذیر نیست. مثلاً  وقتی از متریک $FRW$ استفاده می کنیم به معادلاتی مرتبه چهار برای ضریب مقیاس[2] برمی خوریم که به سختی قابل بررسی اند.

مشکل بعدی که وجود دارد این است که چگونه این $f(R)$ های مختلف را با استفاده از مشاهدات مقید کنیم، یعنی فضای پارامتر های مجهول مساله را کاهش دهیم. قید ها از معادلات حاصل از گرایش های مرتبه بالا که با استفاده از برخی داده ها ساده تر و قابل بررسی شده اند. این قید ها از روش های زیر بدست می آیند:

١) داده های اختر فیزیکی و عالمی
٢) آزمایش هایی روی سیستم خورشیدی
٣) نیروی پنجم؛ اطلاعات حاصل از سنتز هسته ی انفجار بزرگ،BBN.
٤) ارائه توالی صحیحی از فازهای شتاب کند شونده–تند شونده در تحول عالمی

اکثر آزمایش های کیهانی یا شامل تبدیلاتی است که نظریه را به یک مدل معادل به علاوه یک میدان اسکالر تبدیل می کند و یا شامل بررسی نظریه در چارچوب های مختلف و یا شامل فرضیاتی روی وابستگی ثابت هابل به پارامتر انتقال قرمز است.

---

با این وجود می توان قید ها را از طریق بررسی هایی روی انرژی نیز بدست آورد. که به آنها شرایط انرژی گفته می شود. این شیوه مقید سازی مستقل از هر گونه داده است، بدون حل معادلات حرکت[1]، بدون فرضیاتی روی ثابت هابل و یا بدون بکار بردن تبدیلاتی روی نظریه.

تنها فرض این است که شتاب عالم حاصل از یک مدل تعمیم یافته و یک ماده عادی است. این شرایط در موقعیت های مختلف اعمال شده اند و نتیجه عمومی این بوده است که در هر موقعیت یک یاچند شرط انرژی باید همزمان ارضاء شوند. مثلاً در تکینگی پنروز–هاوکینگ[2] شرط ضعیف و شرط قوی باید ارضاء شوند، در حالی که اثبات قانون دوم ترمودینامیک سیاه چاله ها به ارضاء شرط تهی وابسته است[۲٤].

ذکر این نکته بد نیست که شرایط انرژی ابتدا در نسبیت عام فرمول بندی شده اند، به عبارت دیگر شخص باید در بکار بردن شرایط انرژی در نظریات عمومی گرانش محتاط باشد. نظریات مرتبه بالاتر گرانشی ویژگی جالبی دارند؛ با شروع از چارچوب جوردان[3] (در این چارچوب کنش شامل تابع $f(R)$ است)، با اعمال یک تبدیل مناسب روی متریک می توان نشان داد که هر مدل $f(R)$ به طور ریاضی معادل است با نظریه اینشتین (که در آن کنش شامل اسکالر ریچی است) به علاوه یک میدان اسکالر که به طور مینیمال جفت شده است[۲٦].

بنابراین در اصل شخص باید از یک شرایط انرژی انتقالی، مستقیما از نسبیت عام و اعمال آن روی فشار و چگالی انرژی موثره در چارچوب جوردان تعریف می شود صحبت کند.

## ۳–۱ شرایط انرژی در نسبیت عام

بیشتر روش هایی که برای استخراج شرایط انرژی در نظریه های گرانش های مرتبه بالا استفاده می شود، ریشه در نظریه نسبیت عام اینشتین دارند. برای کامل شدن مطلب مروری کوتاه بر شرایط انرژی در نسبیت عام خواهیم داشت. عموما چهار شرط انرژی به جهت ارضاء شدن برخی نیازهای فیزیکی تعریف می شوند که به طور خلاصه عبارتند از[۲۷]:

---

شرط انرژی ضعیف[1]

این شرط بیان می کند که چگالی انرژی هر توزیع ماده، از دید هر ناظری باید مقداری نامنفی داشتـه باشـد. چون چگالی انرژی برای هر ناظر با چهار بردار سرعت $u^{\alpha}$ برابر $T_{\alpha\beta}u^{\alpha}u^{\beta}$ است برای ارضاء شدن ایـن شرط باید داشته باشیم:

$$T_{\alpha\beta}u^{\alpha}u^{\beta} \geq 0$$

برای هر بردار زمان گونه. که نتیجه این شرط عبارت است از:

$$\rho \geq 0 \quad , \quad \rho + P_i \geq 0$$

که $P_i$ مولفه اختیاری از فشار است.

شرط انرژی تهی[2]

این شرط، شرط انرژی ضعیف را برای بردار های تهی $k^{\alpha}$ بیان می کند یعنی برای هر بردار اختیاری تهی داریم:

$$T_{\alpha\beta}k^{\alpha}k^{\beta} \geq 0$$

که در نتیجه داریم:

$$\rho + P_i \geq 0$$

شرط انرژی قوی

منشاء این شرط از معادله رایچودری[3] است:

$$\frac{d\theta}{d\tau} = -\frac{1}{3}\theta^2 - \sigma_{\alpha\beta}\sigma^{\alpha\beta} + \omega_{\alpha\beta}\omega^{\alpha\beta} - R_{\alpha\beta}u^{\alpha}u^{\beta}$$

که $\theta$، $\sigma_{\alpha\beta}$ و $\omega_{\alpha\beta}$ به ترتیب پارامتر انبساط، تنش[4] و پارامتر چرخش است. این ها تعریف صرفا هندسی برای یک دسته ژئودزیک در یک خمینه معین است. میدان گرانش یک میدان جاذب است، که به بیان

---

هندسی به این معنا است که دسته منحنی ژئودزیک های زمان گونه باید همگرا باشند. برای این نوع دسته ژئودرزیک باید داشته باشیم $\frac{d\theta}{d\tau} \leq 0$. جمله اول سمت راست معادله بالا همواره منفی است جمله دوم نیز همینطور زیرا $\sigma^2 = \sigma_{\alpha\beta}\sigma^{\alpha\beta} \geq 0$. پس برای یک دسته ژئودزیک بدون خاصیت چرخش یعنی $\omega_{\alpha\beta} = 0$ باید داشته باشیم:

$$R_{\alpha\beta}u^\alpha u^\beta = (T_{\alpha\beta} - \frac{T}{2}g_{\alpha\beta})u^\alpha u^\beta \geq 0$$

که نتایج این شرط عبارت است از:

$$\rho + \sum_{i=1}^{3} P_i \geq 0 \quad , \quad \rho + P_i \geq$$

توجه کنید که برای بردار های پوچ معادله رایچوداری به شکل زیر تغییر می کند:

$$\frac{d\theta}{d\tau} = -\frac{1}{2}\theta^2 - \sigma_{\alpha\beta}\sigma^{\alpha\beta} + \omega_{\alpha\beta}\omega^{\alpha\beta} - R_{\alpha\beta}k^\alpha k^\beta$$

شرط انرژی غالب[1]

این شرط تاکید می کند که ماده باید در راستای جهان خط های پوچ یا زمان گونه شارش یابد. این به بیان ریاضی یعنی:

$$-T_\beta{}^\alpha u^\beta \quad \text{یک بردار پوچ یا زمان گونه در جهت به سمت آینده است}$$

از آن جایی که برای هر ناظر با چهار بردار $u^\beta$ کمیت فوق، چگالی چهار جریان انرژی–تکانه است که توسط ناظر دیده می شود، این شرط به این معنی است که سرعت شارش انرژی ماده کمتر از سرعت نور است.

شرایط انرژی EC، وقتی برای یک ماده کامل می روند به انواع زیر تقسیم می شوند:

(۳-۱) شرط انرژی تهی (یا پوچ) $\qquad\qquad \rho + P \geq 0$

$$\rho \geq 0, \quad (\rho + P) \geq 0 \qquad \text{(۲-۳) شرط انرژی ضعیف}$$

$$\rho + 3P \geq 0, \quad \rho + P \geq 0 \qquad \text{(۳-۳) شرط انرژی قوی}$$

$$\rho \geq 0, \quad \rho \pm P \geq 0 \qquad \text{(٤-۳) شرط انرژی غالب}$$

نشان داده شده است که EC در زمینه تکینه های کیهان شناسی و مدل های bouncing کاربرد دارد.

۳-۲ شرایط انرژی در گرانش های مرتبه بالاتر

به دلیل اینکه تعریف ها و اصول هندسی نسبیت عام مستقل از نظریه های سوار بر آن هستند، از این رو معادله رایچوداری و همتای ژئودزیک های تهی آن، برای حد اقل نظریه هندسی گرانش های مرتبه بالاتر برقرار است و در واقع این موضوع محرک اصلی بر قراری اصول هندسی تعریف شده در نسبیت عام در هر نظریه بعدی است. بنابراین ما می توانیم دوباره از معادله رایچوداری برای مدل های مختلف $f(R)$ استفاده کنیم.

کنش در گرانش های مرتبه بالا به صورت زیر تعریف می شود:

$$S = \int f(R)\sqrt{-g}\, d^4x + S_m \qquad \text{(٥-۳)}$$

که $g$ رد ماتریس متریک است، $R$ اسکالر ریچی است و $S_m$ کنش استاندارد برای میدان های ماده است. با وردش متریکی این کنش، معادلات میدان به صورت زیر بدست می آید[۲۳]:

$$f'(R)R_{\alpha\beta} - \frac{f(R)}{2}g_{\alpha\beta} - (\nabla_\alpha\nabla_\beta - g_{\alpha\beta}g^{\alpha\beta}\nabla_\alpha\nabla_\beta)f'(R) = T_{\alpha\beta} \qquad \text{(٦-۳)}$$

در نظریات $f(R)$ می توان تانسور انرژی- تکانه موثر به شکل زیر تعریف کرد:

$$T_{\alpha\beta}{}^{eff} = \frac{1}{f'}[T_{\alpha\beta} + \frac{1}{2}(f - Rf')g_{\alpha\beta} + (\nabla_\alpha\nabla_\beta - g_{\alpha\beta}g^{\alpha\beta}\nabla_\alpha\nabla_\beta)f'] \qquad \text{(٧-۳)}$$

که از آن می توان فشار و چگالی انرژی موثر را تعریف کرد:

$$\rho_{eff} = \frac{1}{f'}[\rho + \frac{1}{2}(f - Rf') - 3\dot{R}Hf'']$$

$$P_{eff} = \frac{1}{f'}[P - \frac{1}{2}(f - Rf') + (\ddot{R} + 2\dot{R}H)f'' + \dot{R}^2f''']$$

حال اگر شرایط انرژی را برای چگال انرژی موثر و فشار موثر که در بالا معرفی شد اعمال کنیم برای $P_{eff} + \rho_{eff} \geq 0$ که همان شرط در حالت جدید است، خواهیم داشت:

(۳-۸)

شرط انرژی تهی: $\quad \rho + P + (\ddot{R} - \dot{R}H)f'' + \dot{R}^2 f''' \geq 0$

به طور مشابه برای $\rho_{eff} + 3P_{eff} \geq 0$ داریم:

$$\rho + 3P - f + Rf' + 3(\ddot{R} + \dot{R}H)f'' + \dot{R}^2 f''' \geq 0$$

شرط بالا علاوه بر شرط (۳-۸) شرط انرژی قوی را بدست می دهد.

(۳-۹)

شرط انرژی قوی: $\quad \rho + P + (\ddot{R} - \dot{R}H)f'' + \dot{R}^2 f''' \geq 0$

$$\rho + 3P - f + Rf' + 3(\ddot{R} + \dot{R}H)f'' + \dot{R}^2 f''' \geq 0$$

برای $\rho_{eff} \geq 0$ داریم:

$$\rho + \frac{1}{2}(f - Rf') - 3\dot{R}Hf'' \geq 0$$

شرط بالا علاوه بر شرط (۸-۳-۲) شرط انرژی ضعیف را بدست می دهد:

(۳-۱۰)

شرط انرژی ضعیف: $\quad \rho + P + (\ddot{R} - \dot{R}H)f'' + \dot{R}^2 f''' \geq 0$

$$\rho + \frac{1}{2}(f - Rf') - 3\dot{R}Hf'' \geq 0$$

و برای $\rho_{eff} - P_{eff} \geq 0$ داریم:

$$\rho - P + f - Rf' - (\ddot{R} + 5\dot{R}H)f'' - \dot{R}^2 f''' \geq 0$$

۲۹

از این رو رابطه بالا علاوه بر روابط (۲-۳-۸) و (۲-۳-۱۰) شرایط انرژی غالب را به صورت زیر بدست می دهند:

(۳-۱۱)

$$\rho + P + (\ddot{R} - \dot{R}H)f'' + \dot{R}^2 f''' \geq 0$$ شرط های انرژی غالب:

$$\rho + \frac{1}{2}(f - Rf') - 3\dot{R}Hf'' \geq 0$$

$$\rho - P + f - Rf' - (\ddot{R} + 5\dot{R}H)f'' - \dot{R}^2 f''' \geq 0$$

## ۳-۳ مقید کردن معادلات

با استفاده از شرایط انرژی و نیز اطلاعات بر گرفته از مشاهدات عالمی می توانیم حدودی را برای ضرایب ثابت بکار رفته در تابع $f(R)$ بدست آوریم. توجه کنید که در تمام محاسبات متریک را فریدمان-رابرتسون-واکر تخت انتخاب کردیم. پارامتر هایی که به منظور تعیین ویژگی های دینامیک عالمی مورد استفاده قرار می گیرد طبق تعریف عبارت اند از:

$$H = \frac{\dot{a}}{a}$$

$$q = -\frac{1}{H^2}\frac{\ddot{a}}{a}$$

$$j = \frac{1}{H^3}\frac{\dddot{a}}{a}$$

$$s = \frac{1}{H^4}\frac{\frac{d^4 a}{dt^4}}{a}$$

که به ترتیب ثابت هابل، پارامتر شتاب کاهشی،جرک[۱] و اسنپ[۲] نامیده می شوند. مقادیر اندازه گیری شده فعلی آن ها طبق توافق انجام شده عبارت است از: $H_0 = 72 \pm 8 \ \text{km/sMpc}$، $q_0 = -0.81 \pm 0.14$ و $j_0 = 2.16^{+0.81}_{-0.75}$ با وجود این مقادیر هنوز مقداری با خطای قابل قبول برای $s_0$ هنوز گزارش نشده. اگر اسکالر ریچی و مشتق های آن را بر حسب این پارامتر ها باز نویسی کنیم خواهیم داشت:

---

[۱] jerk
[۲] snap

$$R = -6H^2(1-q) \qquad \text{(الف -۱۲-۳)}$$

$$\dot{R} = -6H^3(j - q - 2) \qquad \text{(ب -۱۲-۳)}$$

$$\ddot{R} = -6H^4(s + q^2 + 8q + 6) \qquad \text{(ج -۱۲-۳)}$$

برای بدست آوردن روابط مربوط به $\dot{R}$ و $\ddot{R}$ کافی است که از $R$ به ترتیب یک و دو بار نسبت به زمان مشتق بگیریم. حال اگر تمام پارامتر ها را بر حسب مقادیر زمان حال آن ها در معادلات مربوط به چهار شرط انرژی قرار دهیم خواهیم داشت: (در اینجا فقط روابط غیر تکراری ذکر می گردد)

$$-[s_0 - j_0 + (q_0+1)(q_0+8)]f_0'' + 6[H_0(j_0 - q_0 - 2)]f_0''' \geq 0 \qquad \text{(۱۳-۳)}$$

برای شرط انرژی تهی.

$$2\rho_0 + f_0 + 6H_0{}^2(1-q_0)f_0' + 36H_0{}^4(j_0 - q_0 - 2)f_0'' \geq 0 \qquad \text{(۱٤-۳)}$$

برای شرط انرژی ضعیف.

$$\text{(۱٥-۳)}$$

$$\rho_0 + 3P_0 + f_0 - 6H_0{}^2(1-q_0)f_0' - 6H_0{}^4(s_0 + j_0 + q_0{}^2 + 7q_0 + 4)f_0'' +$$
$$3\left[6H_0{}^3(j_0 - q_0 - 2)\right]^2 f_0''' \geq 0$$

برای شرط انرژی قوی و

$$\text{(۱٦-۳)}$$

$$\rho_0 - P_0 + 6H_0{}^2(1-q_0)f_0' - 6H_0{}^4(s_0 + (q_0-1)(q_0+4) + 5j_0)f_0'' -$$
$$\left[6H_0{}^3(j_0 - q_0 - 2)\right]^2 f_0''' \geq 0$$

برای شرط انرژی غالب. با روشی دیگر نیز می توان به این قضیه نگاه کرد. می توان بدون معرفی فشار و چگالی انرژی موثر از معادله میدان بدست آمده از کنش مساله، فشار و چگالی انرژی خود ماده را برحسب جملات هندسی ناشی مدل گرانش های مرتبه بالا محاسبه کرد[٢٤]. یعنی با فرض اینکه ماده کامل داریم و تانسور انرژی — تکانه قطری است می توانیم از معادله میدان:

$$f'(R)R_{\alpha\beta} - \frac{f(R)}{2}g_{\alpha\beta} - (\nabla_\alpha\nabla_\beta - g_{\alpha\beta}g^{\alpha\beta}\nabla_\alpha\nabla_\beta)f'(R) = T_{\alpha\beta} \qquad (٦-٣)$$

فشار را به صورت زیر بدست آوریم:

$$\rho = 3f'(R)qH^2 - \frac{f}{2} + 3f''(R)H\dot{R} \qquad (١٧-٣)$$

$$P = -\frac{f'(R)}{3}(3qH^2 + R) + \frac{f(R)}{2} - f''(R)(\ddot{R} - \frac{2\dot{a}\dot{R}}{a}) - f'''(R)\dot{R}^2 \qquad (١٨-٣)$$

که برای ماده کامل $T_{tt} = \rho$ و $T_{ii} = P$ هستند.

مشاهدات اخیر نشان داده اند فشار ماده پرکننده عالم (با فرض اینکه ماده، معمولی است) بسیار کم و در حد صفر است. با این فرض تمام چهار شرط انرژی به شرط $\rho \geq 0$ کاهش می یابند. از (٣-١٧) استفاده کرده و بدست می آوریم:

$$3q_0H_0{}^2f_0' - \frac{f_0}{2} - 18H_0{}^4f_0''(j_0 - q_0 - 2) \geq 0 \qquad (١٩-٣)$$

نکته مهمی که باید یاد آوری شود این است که برای پارامتر اسنپ هنوز هیچ مقدار مورد توافق عام گزارش نشده است. به این دلیل ما از روابطی که در آن ها پارامتر اسنپ وجود دارد استفاده نمی کنیم. برای $P = 0$ از (٣-١٨) استفاده کرده و عبارتی به شکل زیر برای اسنپ بدست می آوریم:

$$(٢٠-٣)$$

$$s_0 = \frac{f_0'}{6H_0{}^2f_0''}(q_0 - 2) + 6H_0{}^2\frac{f_0'''}{f_0''}(-q_0 + j_0 - 2) - [q_0(q_0 + 6) + 2(1 + j_0)] - \frac{f_0}{12H^4f_0''}$$

ذکر چند مثال

در اینجا توابع مختلفی از اسکالر ریچی را معرفی می کنیم و روابط قیدی بالا را در مورد این توابع تحقیق می کنیم[٢٣،٢٤].

تابع اول به شکل زیر مورد توجه است:

$$f(R) = \alpha R^{-n} \quad , n \in N \qquad (\text{۲۱–۳})$$

با جاگذاری این تابع در(۳-۱۹) برای $\alpha > 0$ و $n$ های زوج خواهیم داشت:

$$-3q_0 H_0{}^2 n R_0 - \frac{1}{2} R_0{}^2 - 18 H_0{}^4 n(n+1)(j_0 - q_0 - 2) \geq 0$$

با جاگذاری مقادیر فعلی پارامتر ها ی شتاب کاهشی، ثابت هابل و جرک قید زیر برای مقادیر زوج $n$ بدست می آید:

$$-17.64 n^2 - 44.50 n - 59.62 \geq 0 \qquad (\text{۲۲–۳})$$

چون این نامساوی برای هیچ $n$ حقیقی ارضاء نمی شود نتیجه می گیریم که $n$ نمی تواند برای $\alpha > 0$ زوج باشد. برای $n$ های فرد علامت نامساوی فوق برعکس می شود. این یعنی برای $\alpha > 0$ فقط n های فرد مجاز هستند. با تحلیلی مشابه در می یابیم که برای $\alpha < 0$ فقط n های زوج مجاز هستند.

اگر ما مقدار فعلی اسنپ را می دانستیم با استفاده از (۲۰–۳) قید دیگری روی n بست می آوردیم. با وجود این هنوز می توانیم از این رابطه به عنوان رابطه ای که مقدار اسنپ را بر حسب تابعی از n بدست می دهد استفاده کنیم.

اگر تعریف کنیم:

$$\emptyset \equiv -17.64 n^2 - 44.50 n - 59.62$$

مقادیر این تابع در نمودار زیر رسم شده است:

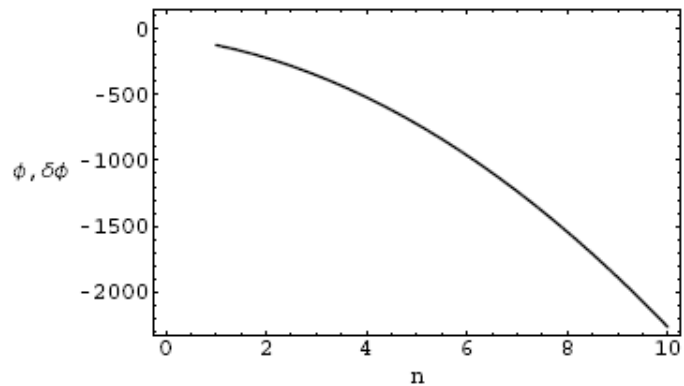

نمودار۳-۱ منحنی مربوط به $\emptyset$

همانطور که می بینیم این عبارت همواره منفی است. این واقیت نکات توضیح داده شده در بالا را تاکید می کند. نموداری که برای اسنپ بر حسب مقادیر $n$ می توان رسم نمود عبارت است از:

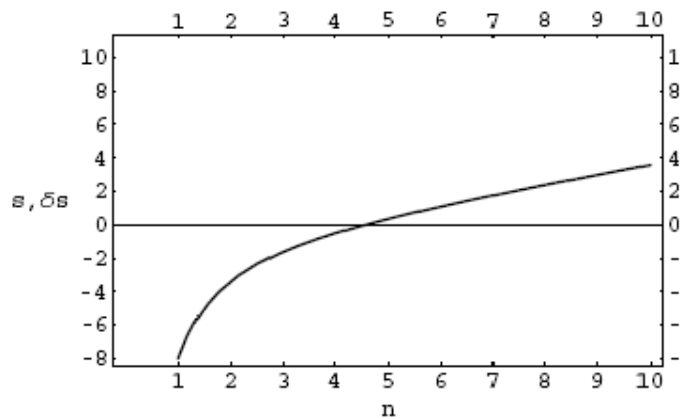

نمودار۳-۲ منحنی مربوط به s

تابع دوم به شکل زیر است:

$$f(R) = R + \alpha ln \frac{R}{\mu} \qquad (\text{۳-۲۳})$$

با توجه به اینکه $R = 6H^2(q-1) < 0$ پس برای جواب دار بودن تابع فوق باید داشته باشیم $\mu < 0$. استفاده از (۳-۱۹) برای این تابع و نیز برای همه $\alpha < 0$ به عبارت زیر می انجامد:

$$0 < \frac{\mu}{R_0} < e^{-g(\beta)}$$

که تعریف کرده ایم:

$$\beta \equiv \frac{\alpha}{R_0}$$

$$g(\beta) = \frac{1}{\beta}\left[-6q_0(1+\beta)\frac{H_0^2}{R_0} + 1 - 36\frac{H_0^4}{R_0^2}\beta(j_0 - q_0 - 2)\right]$$

منحنی زیر مقادیر مجاز برای $\frac{\mu}{R_0}$ در حالت $\alpha < 0$ را نشان می دهد. در منحنی زیر تمام حالات بین خط افقی و منحنی رسم شده مقادیر مجاز را نشان می دهند. نمودار نشان می دهد، $\frac{\mu}{R_0}$ برای $\beta$ بزرگ به مقدار ۲.۱ میل می کند.

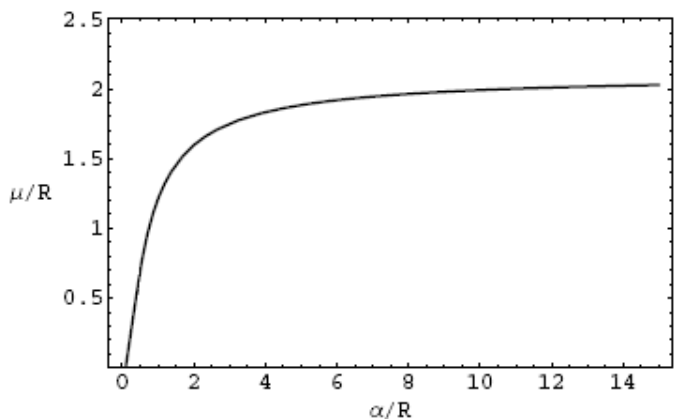

نمودار۳-۳ منحنی مربوط به $\frac{\mu}{R_0}$ برحسب $\frac{\alpha}{R_0}$

با محاسبه $R_0$ این بدین معناست که داریم $|\mu| \lesssim 1.2 \times 10^{-41} m^2$. یعنی با استفاده از بررسی شرایط انرژی  توانستیم برای دو ثابت مساله حد مجازی را بدست آوریم. برای $\alpha < 0$ مقادیر زیر منحنی مجاز نیستند.

مدل بعدی که بررسی خواهیم کرد عبارت است از:

$$f(R) = R + \alpha R^n \tag{۳-۲٤}$$

$n$ عدد صحیح و $\alpha$ مثبت یا منفی می تواند باشد. می خواهیم از روابطی استفاده کنیم که نیازی به پارامتر اسنپ نداشته باشند یعنی از (۳-۱٤) برای شرط انرژی ضعیف استفاده می کنیم. بعد از قرار دادن تابع فوق در (۳-۱٤) داریم:

$$\tag{۳-۲٥}$$

$$\alpha(-1)^n(An^2 - (A+1)n + 1) \geq 0$$

که تعریف کرده ایم:

$$A \equiv \frac{j_0 - q_0 - 2}{(1 - q_0)^2}$$

شرط دیگری که وجود دارد این است که باید داشته باشیم $f'(R) > 0$ این شرط به معادله زیر می
انجامد:

$$\alpha(-1)^n n(3.3H_0)^{2n-2} < 1 \qquad (\text{۳-۲۶})$$

از رابطه اول می بینیم که رابطه مربوط به شرط انرژی ضعیف به طور آشکار به علامت $\alpha$ بستگی دارد. در
نتیجه می توان جواب هایی برای $n$ و $\alpha$ های مثبت و منفی بدست آورد (جزئیات محاسبات (۳-۲۵) و
(۳-۲۶) در ضمیمه خ ارائه شده است).

الف) به ازاء $\alpha > 0$ مقادیر مجاز برای $n$ عبارت است از: $n = \{3, 1, -2, -6, \dots\}$ و
$n = \{4, 6, 8, \dots\}$ مقادیرمجازی برای $0 < \alpha < [n(3.3H_0)^{2n-2}]^{-1}$ هستند.

ب) به ازاء $\alpha > 0$ مقادیر مجاز عبارت اند از $n = \{2, -1, -3, -5, \dots\}$ و $n = \{1, 5, 7, 9, \dots\}$
مقادیر مجازی برای $0 < \alpha < -[n(3.3H_0)^{2n-2}]^{-1}$ هستند.

مثلاً یک کاربرد این تحلیل این است که برای گرانش با تابع $f(R) = R + \alpha R^2$، برای اینکه شرط
انرژی ضعیف برآورده شود $\alpha$ باید منفی باشد. یا مثلاً مدل $f(R) = R - \frac{\mu^4}{R}$ شرط ضعیف را برآورده
می کند زیرا برای $\alpha = -\mu^4$، $n = -1$ مجاز است. ولی مدل $f(R) = R - \frac{\mu^6}{R^2}$ برای $\alpha = -\mu^6$،
$n = -2$ مجاز نیست در نتیجه این مدل شرط انرژی ضعیف را نقض می کند.

مدل آخر عبارت است از:

$$f(R) = \alpha R^n \qquad (\text{۳-۲۷})$$

که $n$ عددی صحیح است. برای این مدل شرط انرژی ضعیف عبارت است از:

$$\alpha(-1)^n(An^2 - (A+1)n + 1) \geq 0 \qquad (\text{۳-۲۸})$$

که در مدل قبل نیز همین نامعادله برای شرط انرژی ضعیف بدست آمد. ولی برای شرط $f'(R) > 0$
داریم:

$$\alpha(-1)^n n(3.3H_0)^{2n-2} < 0 \qquad (۲۹-۳)$$

مثل مدل قبل دو دسته جواب داریم:

الف) برای $\alpha > 0$ مقادیر مجاز برای $n$ عبارت است از: $n = \{3, 1, -2, -4, -6, \ldots\}$

ب) برای $\alpha < 0$ مقادیر مجاز برای $n$ عبارت است از: $n = \{2, -1, -3, -5, \ldots\}$

# فصل چهارم

## روش های مقید سازی معادلات با استفاده از کمیت های بدست آمده از مشاهدات

به منظور همخوان کردن نتایج نسبیت عام با مشاهدات اخیر در کیهان شناسی، مثلاً شتاب کیهانی که مرهون نتایج آزمایشگاهی رسیده از آزمایش های مختلف است، وجود انرژی تاریک مورد نیاز است. از سویی دیگر، برخی از نظریه های جایگزین نسبیت عام به طور مثال گرانش های مرتبه بالاتر شتاب جهان را بدون نیاز به حضور انرژی تاریک توجیه می کنند. هر نظریه ای که با نتایج مشاهداتی همخوانی دارد باید نسبیت عام را در مقیاس سیستم خورشیدی بدست دهد. در مقام توضیح باید گفت انرژی تاریک ماده ای با فشار منفی، تقریباً همگن است که حدود ۷۰ درصد انرژی کل را تشکیل داده است.

رهیافت های متفاوتی در مواجهه با انرژی تاریک وجود دارد که یکی از آنها مدل $\Lambda CDM$ است. این مدل ثابت کیهان شناسی را مسئول انرژی تاریک می داند که به رغم موفقیت های چشم گیر در مقایسه با مشاهدات، مشکلات مختلف در مقیاس های مختلف را داراست. در برخی نظریه ها به آن به چشم انرژی

خلاء نگاه شده که در این صورت مقدار ثابت کیهان شناسی $\Lambda$، حدود ۱۲۰ مرتبه اندازه کوچکتر از مقدار پیش بینی شده است.

علاوه بر این شخص باید مساله انطباق[1] را حل کند، این مساله به چرایی این موضوع می پردازد که اگر انبساط شتابدار از زمان اولیه تحول عالمی آغاز شده باشد، ساختار هایی شبیه کهکشان ها فرصت کافی برای شکل گیری نمی داشته اند. مساله بعدی توجیه سهم $\Lambda$ در انرژی کل است. برای حل مساله انطباق مدل هایی quintessence معرفی شد ه اند. این مدل ها شامل یک میدان اسکالر هستند که این میدان اسکالر با حرکت بهمنی انرژی پتانسیل خود برهم کنشی مربوطه، اجازه می دهند که انرژی خلاء در دوره عالمی حاضر غالب شود.

امکان دیگر این است که شتاب کیهانی، ناشی از ظهور جزء دیگری در توده عالمی نباشد و تنها اولین علائم شکست دانش ما از قوانین گرانش باشد. در این راستا نظریه های مختلف، مشکلات و موفقیت هایی دارند که موثر ترین آنها نظریه گرانش های مرتبه بالا است. نشان داده شده است که در حد میدان ضعیف پتانسیل نیوتونی با یک جمله اضافی تصحیح می شود. وقتی منحنی های چرخشی راه شیری بدست آمده از طریق مشاهدات را با منحنی های حاصل از پتانسیل نیوتونی تصحیح شده مقایسه کردند، همخوانی قابل قبولی مشاهده شد. این همخوانی بدون وارد کردن ماده تاریک به مساله مشاهده شد[۲۸].

نتایج نشان می دهد که بررسی گرانش های مرتبه بالا می تواند راهی برای توجیه ماده تاریک و انرژی تاریک باشد.

یک نظریه جایگزین باید قادر به توضیح پدیده های زیر باشد[۲۹]:

الف) اطلاعات رسیده از منحنی چرخشی کهکشانی

ب) اطلاعات دریافتی با پرتو ایکس، مربوط به mass profile از خوشه های کهکشانی

ج) اطلاعات مربوط به اثر لنز گرانشی برای کهکشان ها و خوشه های کهکشانی

د) شکل گیری کهکشان های قدیمی در عالم نخستین و رشد کهکشان ها

---

[1] coincidence problem

و) تابش کیهانی زمینه که شامل اطلاعات در مورد افت و خیز های آکوستیکی طیفی اند.

ه) شبیه سازی N جسمی برای پژهش های کهکشانی

ی) شتاب انبساط کیهانی

در ادامه به بررسی کیفی منحنی چرخش کهکشانی و شتاب انبساط عالمی در گرایش های مرتبه بالا جهت درک بهتر رویارویی این نظریه در مواجهه با این دو مساله می پردازیم.

## ۴-۱ شتاب انبساط عالمی-خمش عنصر پنجم (quintessence)

بدین منظور از کنش زیر استفاده می کنیم:

$$S = \int d^4 \sqrt{-g}(f(R) + \mathcal{L}_M) \qquad (۴-۱)$$

برای نشان دادن اینکه می توان انبساط کیهانی را صرفا توسط خصوصیت هندسی ماده توضیح داد از اینجا به بعد فرض می کنیم ماده وجود ندارد یعنی $\mathcal{L}_M = 0$. بعد از وردش کنش فوق معادلات میدان عبارت خواهند بود از:

$$(۴-۲)$$

$$f'(R)R_{\alpha\beta} - \frac{1}{2}f(R)g_{\alpha\beta} - \nabla^\rho\nabla^\sigma f'(R)\big(g_{\alpha\rho}g_{\beta\sigma} - g_{\alpha\beta}g_{\rho\sigma}\big) = 0$$

اگر معادله بالا را به صورت معادله اینشتین مرتب کنیم یعنی به شکل زیر:

$$G_{\alpha\beta} = R_{\alpha\beta} - \frac{1}{2}g_{\alpha\beta}R = T_{\alpha\beta}^{\text{curve}} \qquad (۴-۳)$$

آنگاه خواهیم داشت:

$$G_{\alpha\beta} = \frac{1}{f'(R)}\left\{\frac{1}{2}\big[f(R) - Rf'(R)\big]g_{\alpha\beta} + \nabla^\rho\nabla^\sigma f'(R)\big(g_{\alpha\rho}g_{\beta\sigma} - g_{\alpha\beta}g_{\rho\sigma}\big)\right\} = T_{\alpha\beta}^{\text{curve}}$$

که با مقایسه با (۴-۳) می توانیم نتیجه بگیریم:

$$T_{\alpha\beta}^{curve} \equiv \frac{1}{f'(R)}\left\{\frac{1}{2}[f(R) - Rf'(R)]g_{\alpha\beta} + \nabla^\rho\nabla^\sigma f'(R)\left(g_{\alpha\rho}g_{\beta\sigma} - g_{\alpha\beta}g_{\rho\sigma}\right)\right\}$$

که در واقع اثر تابع f(R)، به شکل یک تانسور انرژی-تکانه ناشی از صرفا هندسه که به معادله اینشتین اضافه شده است دیده می شود. در مرحله بعد با استفاده از روش کانونی کردن کنش (٤-١)، آن را به صورت زیر باز نویسی می کنیم [٣٠]:

$$A_{(curv)} = \int dt\, \mathcal{L}\left(a, \dot{a}; R, \dot{R}\right) \qquad (٦-٤)$$

در بالا فرض کرده ایم متریک ما فرید مان-رابرتسون-واکر است. که در آن ضریب مقیاس و اسکالر ریچی متغیر های کانونیکال [1] هستند داریم:

$$\mathcal{L}\left(a, \dot{a}; R, \dot{R}\right) = a^3\left(f(R) - Rf'(R)\right) + 6a\dot{a}^2 f'(R) \qquad (٧-٤)$$

$$+6a^2\dot{a}\dot{R}f''(R) - 6af'(R)k$$

که سهم ماده به صورت یک جمله فشار ظاهر می شود. مشتقات اولر – لاگرانژ عبارت است از:

$$2\left(\frac{\ddot{a}}{a}\right) + \left(\frac{\dot{a}}{a}\right)^2 + \frac{k}{a^2} = -p_{(curv)} \qquad (٨-٤)$$

$$f''(R)\left\{R + 6\left[\frac{\ddot{a}}{a} + \left(\frac{\dot{a}}{a}\right)^2 + \frac{k}{a^2}\right]\right\} = 0 \qquad (٩-٤)$$

$$\left(\frac{\dot{a}}{a}\right)^2 + \frac{k}{a^2} = \frac{1}{3}\rho_{(curv)} \qquad (١٠-٤)$$

از معادله اول و سوم می توانیم نتیجه زیر را بگیریم:

$$\frac{\ddot{a}}{a} = -\frac{1}{6}[\rho_{(curv)} + 3p_{(curv)}] \qquad (١١-٤)$$

رفتار با شتاب مثبت ضریب مقیاس ایجاب می کند که داشته باشیم:

$$\rho_{(curv)} + 3p_{(curv)} < 0 \qquad (١٢-٤)$$

---

[1] canonical variable

که برای فشار و چگالی انرژی خمش داریم:

$$p_{(curv)} = -\frac{1}{f'(R)}\{2\left(\frac{\dot{a}}{a}\right)\dot{R}f''(R) + \ddot{R}f''(R)\} \qquad (13-4)$$

$$\rho_{(curv)} = -\frac{1}{f'(R)}\{\frac{1}{2}[f(R) - Rf'(R)] - 3\left(\frac{\dot{a}}{a}\right)\dot{R}f''(R)\} \qquad (14-4)$$

(جزئیات استخراج روابط بالا و صفحه قبل در ضمیمه د ارائه شده است) با فرض وجود رابطه های به شکل $p_{(curv)} = w_{(curv)}\rho_{(curv)}$ برای چگالی خمش و فشار متناظر برای خمش، با استفاده از رابطه (12-4) برای خلاء با $\rho_{(matter)} = 0$ خواهیم داشت:

$$w_{curv} < -\frac{1}{3} \qquad (15-4)$$

با استفاده از روابط (13-4) و (14-4) می توانیم بنویسیم:

$$w_{curv} = \frac{p_{curv}}{\rho_{curv}} = -1 + \frac{\ddot{R}f''(R)+\dot{R}[\dot{R}f'''(R)-\left(\frac{\dot{a}}{a}\right)f''(R)]}{\frac{1}{2}[f(R)-Rf'(R)]-3\left(\frac{\dot{a}}{a}\right)\dot{R}f''(R)} \qquad (16-4)$$

حال روابط فوق را می خواهیم برای نظریه $f(R) = f_0 R^n$ بکار ببریم. از آنجایی که این نوع تابع پیش بینی های خوبی از کمیت های رصدی مثل منحنی های چرخش کهکشانی داشته است همواره مورد توجه خاصی قرار گرفته است. برای سهولت فرض می کنیم ضریب مقیاس به شکل توانی باشد یعنی به شکل $a(t) = a_0 \left(\frac{t}{t_0}\right)^\alpha$ باشد. برای اینکه انبساط شتاب دار داشته باشیم باید $\alpha$ مثبت باشد. برای نشان دادن رفتار خود خمش بدون حضور ماده، فرض کرده ایم $\rho_{(matter)} = 0$ باشد و نیز فرض می کنیم فضا-زمانی تخت داریم. با این مفروضات به دو معادله قیدی برای دو ضریب موجود در مساله می رسیم:

$$\alpha[\alpha(n-2) + 2n^2 - 3n + 1] = 0$$

$$\alpha[n^2 + \alpha(n-2-n-1)] = n(n-1)(2n-1)$$

که مقادیر مجاز عبارتند از:

$$\alpha = 0 \quad , \quad n = 0, \frac{1}{2}, 1$$

$$\alpha = \frac{2n^2 - 3n + 1}{2 - n} \quad , \quad n \neq 2$$

حل های دسته اول مورد توجه نیستند زیرا عالم ایستا را بدست می دهند. در نتیجه برای معادله (٤-١٦) داریم:

$$w_{curv} = -\left(\frac{6n^2 - 7n - 1}{6n^2 - 9n + 3}\right)$$

در واقع روابط فوق بیان گر این حقیقت هستند که توانستیم بدون استفاده از انرژی تاریک و صرفا با استفاده از خود هندسه، انبساط شتاب دار عالم را توجیه کنیم. حضور خمش از مراتب بالا متناظر است با نوعی ماده با فشار منفی.

در مرحله بعد می توان با استفاده از اطلاعات رصدی و با استفاده از روابط دامنه−درخشندگی[1]، مدل داده شده را مقایسه کرد. بدین منظور با داشتن منحنی های درخشندگی $d_L$ بر حسب پارامتر جابه جایی قرمز $Z$ حاصل از اطلاعات رصدی، و منحنی هایی که مدل ما با استفاده از رابطه زیر بدست می دهد می توان تخمینی از مدل معرفی شده بدست آورد.

$$\int_0^r \frac{dr'}{(1 - kr'^2)^{\frac{1}{2}}} = \int_t^{t_0} \frac{cdt'}{a(t')} = \int_a^{a_0} \frac{cda'}{a'\dot{a}'}$$

که در آن $c$ سرعت نور و $a_0$ ضریب مقیاس در زمان حال حاضر است. با توجه به رابطه بین ضریب مقیاس و پارامتر جابه جایی قرمز به شکل $a = \frac{a_0}{1+Z}$ و فرمول ضریب مقیاس در مدل ما یعنی $a(t) = a_0 \left(\frac{t}{t_0}\right)^\alpha$ برای درخشندگی داریم:

$$d_L(Z, H_0, n) = \frac{c}{H_0}\left(\frac{\alpha}{\alpha - 1}\right)(1 + Z)[(1 + Z)^{\frac{\alpha}{\alpha - 1}} - 1]$$

که در آن $H_0$ ثابت هابل در زمان حال حاضر است. در این مرحله می توان با یک روش برازش منحنی وابسته به این رابطه و منحنی های بدست آمده از مشاهدات محدوده مناسبی برای ضریب $\alpha$ بدست آورد. می بینیم که در این حالت درخشندگی به پارامتر هندسی وابسته شده است ($\alpha = \alpha(n)$). سن عالم را نیز می توان محاسبه کرد:

---

[1] amplitude-luminosity

$$t = \left(\frac{3n^2 - 3n + 1}{2 - n}\right) H_0^{-1}$$

که البته مقادیر عددی دو رابطه فوق را با استفاده از روش های مرسوم در فیزیک نجومی می توان تخمین زد و با مقادیر رسیده از مشاهدات عددی مقایسه کرد که در اینجا مورد نظر ما نیست. ملاک بعدی کنترل پارامتر شتاب کند شونده $q_0 = -(\frac{\ddot{a}a}{\dot{a}^2})$ است. با جاگذاری $a(t) = a_0(\frac{t}{t_0})^\alpha$ در این رابطه و استفاده از رابطه $\alpha$ برحسب $n$ می توان و با توجه به این نکته که برای انبساط شتاب دار عالم باید $q_0 < 0$ باشد می توان بازه های جالب برای $n$ که شتاب کند شونده را بدست می دهند را بدست آورد.

فقط به عنوان جمع بندی می توان گفت که بعد از مقایسه اطلاعات داده شده توسط مشاهدات رصدی و مقادیر تخمینی با استفاده از مدل پیشنهادی، بازه هایی از $n$ را می توان به عنوان مناسب ترین مقادیر از جهت بیشترین شباهت حاصل از مقایسه برگزید. که بدین معناست که مدل پیشنهادی خمش ما محدود به حالت های خاصی می شود که بهترین تخمین را برآورده می سازد. یعنی با این شیوه می توان ثابت های مساله را مقید به مقادیر خاصی کرد.

٤-٢ هندسه به عنوان گزینه ای برای ماده تاریک

در حد انرژی های پایین، گرانش های مرتبه بالا یک پتانسیل گرانشی تصحیح شده را القاء می کند. با شروع از یک جرم نقطه ای $M$ و حل معادلات خلاء برای یک متریک شبه شوارزشیلد، برای مدل $f(R) = f_0 R^n$ می توان پتانسیل تصحیح شده زیر را بدست آورد[۳۱]:

$$\phi(r) = -\frac{GM}{2r}\left[1 + \left(\frac{r}{r_c}\right)^\beta\right] \qquad (۴-۱۷)$$

$$\beta = \frac{12n^2 - 7n - 1 - \sqrt{36n^4 + 12n^3 - 83n^2 + 50n + 1}}{6n^2 - 4n + 2}$$

در واقع پتانسیل نیوتونی با یک جمله با قانون توانی تصحیح می شود:

$$\phi(r) = -\frac{GM}{2r} - \frac{GM}{2r_c}\left(\frac{r}{r_c}\right)^{\beta-1}$$

ثابت $r_c$ به جرم سیستم بستگی دارد یعنی برای هر کهکشان متفاوت است. از آنجایی که $\beta$ مربوط به ساختار هندسی مساله باید برای کلیه نمونه های مورد بررسی (کلیه کهکشان ها) مقداری ثابت باشد. با

تثبیت β با استفاده از یک نمونه می توان $r_c$ را برای هر نمونه دیگری در مساله محاسبه کرد. برای $n = 1$ داریم $β = 0$ و در نتیجه پتانسیل نیوتونی حاصل می شود. در واقع به ازاء $n = 1$ ما کنش نسبیت عام را بدست می آوریم. این یک پتانسیل نقطه ای است، برای یک توزیع غیر نقطه ای جرمی می توان سیستم را به بینهایت جرم نقطه ای تقسیم کرده و روی پتانسیل همه این نقاط انتگرال بگیریم. با استفاده از اینکه شتاب حرکت دورانی یک جرم نقطه ای در حال دوران $m\frac{v^2}{r}$ است، و اینکه در حرکت دورانی یک جرم نقطه ای حول یک توزیع جرم (در اینجا توزیع جرم، یک جرم نقطه ای به جرم M فرض می شود) نیروی تامین کننده حرکت دورانی نیروی جاذبه گرانشی است، در حالت تعادل خواهیم داشت:

$$\vec{F}_r = -m\frac{v^2}{r}\vec{e}_r$$

که $\vec{F}_r$ بردار نیروی تامین کننده حرکت دورانی در راستای شعاعی، $\vec{e}_r$ بردار یکه شعاعی که طبق قرار داد به سمت بیرون است و $v$ اندازه سرعت جسم است.

در اینجا داریم:

$$\vec{F}_r = -\vec{\nabla}_r \varphi(r) = -M\vec{\nabla}_r \phi(r)$$

که $\varphi$ انرژی پتانسیل گرانشی است. در نتیجه منحنی های چرخش که همان اندازه سرعت دورانی هستند عبارت خواهند شد از:

$$v_c^2 = \frac{GM}{2r}\left[1 + (1-β)\left(\frac{r}{r_c}\right)^β\right] \qquad (٤-١٨)$$

باز هم در قیاس با مشابه این رابطه در حد نیوتونی نسبیت عام $v_c^2 = \frac{GM}{r}$، می بینیم که تصحیحی به شکل جمله توانی از فاصله ظاهرشده است. به ازاء $n = 1$ رابطه $v_c^2 = \frac{GM}{r}$ حاصل می شود. به ازاء $0 < β < 1$ منحنی چرخش تصحیح شده بزرگ تر از جواب نیوتونی است. از آنجا که اندازه گیری منحنی های چرخش کهکشان های مارپیچ، نسبت به پیش گویی های مبتنی بر اندازه گیری جرم درخشان[1] و پتانسیل نیوتونی، منحنی چرخش بزرگتری را نشان می دهد، نتیجه بالا پیشنهاد می کند که پتانسیل

---

[1] luminous mass

گرانشی تصحیح شده ما ممکن است بتواند این اختلاف مشاهده شده حاصل از اندازه گیری و نتایج معادلات نیوتونی را بدون نیاز به معرفی ماده تاریک توجیه کند.

همان طور که دیده می شود منحنی چرخشی تصحیح شده در بازه مورد نظر $1 > \beta > 0$، در فواصل زیاد به طور مجانبی به سمت صفر میل می کند (همین اتفاق در حالت نیوتونی اتفاق می افتد). با وجود این گفته می شود که منحنی چرخشی حاصل از داده های مشاهداتی تخت هستند. ولی این موضوع احتمال اینکه منحنی چرخشی در فواصل زیاد صفر شود را مردود نمی سازد. زیرا هیچ کدام از آزمایش های مشاهداتی مقادیر $v_c$ را تا فواصل بسیار زیاد بدست نمی آورند و معمولاً این منحنی ها با احتساب عدم قطعیت های مبتنی بر آخرین نقاط اندازه گیری تخت هستند.

# فصل پنجم

# حل های دقیق تعمیم یافته نا همسانگرد از نوع حل های کسنری

هدف این فصل بررسی شرایط وجود تعمیم هایی از حل متریک نا همسانگرد کسنر در گرانش های مرتبه بالاتر است. در قسمت اول به طور خلاصه حل کسنر در شرایط خلاء و نیز در زمانی که ماده به مساله اضافه می شود در نسبیت عام اینشتین، مورد بررسی قرار می گیرد. در قسمت دوم به جستجو برای یافتن حل های مشابه در گرانش های مرتبه بالاتر با معرفی سه چگالی لاگرانژی و بررسی معادلات میدان آنها به هدف یافتن جواب های تعمیم یافته کسنر می پردازیم. منظور از تعمیم یافته این است که جواب هایی مورد نظر ما هستند که در حالاتی خاص جواب های کسنر در نسبیت عام اینشتین را بدست دهند.

۵-۱ معرفی حل های کسری به عنوان حل های دقیق معادلات میدان اینشتین در شرایط خلاء

همانطور که می دانیم در کیهان شناسی حد اقل دو ویژگی همواره اصل فرض مـی شـوند. ایـن دو ویژگـی همسانگردی و همگنی را در ابعاد وسیع عالم به همه ناظرها تحمیل می کند. طبق نتـایج بدسـت آمـده از تحقیقات و مشاهدات، معلوم شده است که این دو اصل توصیف نسبتا دقیقی از عالم قابل مشـاهده را ارائـه می دهند. اگر این دو اصل در معادلات اینشتین بکار رود متریکـی بـرای فضـا زمـان بدسـت مـی آیـد کـه همسانگردی و همگنی را دارا بود این متریک که به F.R.W موسوم است، عبارت است از:

$$dS^2 = dt^2 - a^2(t)[\frac{dr^2}{1-kr^2} + r^2d\Omega^2] \qquad (۵-۱)$$

که در متریک فوق تمام کمیت ها همان تعریف معمولشان را دارند. با وجود کفایت مدل های همسانگرد در توصیف مراحل اخیر (منظور زمان های خیلی بعد تر از تکینگی اولیه است) تحول عالم، این همخـوانی بـه خودی خود نمی تواند دلیل مناسبی برای کارایی مشابه مدل هایی از نـوع همسـانگرد در مراحـل نخسـتین تحول عالم خیلی نزدیک به تکینگی زمانی اولیه باشد. چرا که معیاری کـه مـدل هـای همسـانگرد را بـرای توصیف جهان قابل مشاهده مناسب نشان می دهد مقایسه نتایج حاصل از این مدل هـا بـا نتـایج حاصـل از مشاهدات مستقیم از جهان قابل مشاهده است. ولی چنین ملاکی برای زمان های اولیه وجـود نـدارد، تنهـا کاری که می توان کرد این است که آزمایش هایی (روی زمین) ترتیب داد و نظریه ها را با نتایج حاصـل از آنها مقایسه کرد. مثالی از این آزمایش ها در شتاب دهنده ها انجام می شود. در نتیجـه هنـوز بررسـی مـدل های ناهمسانگرد می تواند اطلاعات مفیدی از تحولات نخستین عالم را بدست دهـد. نکتـه جـالبی کـه در بررسی مدل های ناهمسانگرد وجود دارد این است که بسیاری از این مدل ها در آینده خود، تبدیل به مدلی همسانگرد می شوند. از سویی دیگر یکی از اهداف مهم کیهان شناسی امروز توضیح درجه بالای همگنـی و همسانگردی مشاهده شده در عالم می باشد که نیل به این مقصود بررسی مدل هایی بـا عمومیـت کـافی را می طلبد.

ساده ترین مدل نا همسانگرد کیهانی مدلی است که با متریک بیانکی نوع یک به فرم زیر توصیف می شود:

$$dS^2 = -dt^2 + a^2(t)dx^2 + b^2(t)dy^2 + c^2(t)dz^2 \qquad (۵-۲)$$

در اینجا $a(t)$، $b(t)$ و $a(t)$  فاکتور مقیاس متریک فوق است که توسط معادلات میدان اینشتین تعیین می شوند. این سه تابع زمانی مستقل از هم را دارند.

ا. کسنر[1] اولین کسی بود که معادلات اینشتین را برای متریک عمومی فوق در حالت خلاء حل نمود[٣٢]. جواب نهایی با استخراج سه تابع فاکتور مقیاس به شکل زیر می باشد:

$$dS^2 = -dt^2 + t^{2p_1}dx^2 + t^{2p_2}dy^2 + t^{2p_3}dz^2 \qquad (٣-٥)$$

سه ضریب کسنر $p_1$، $p_2$ و $p_3$ که در اینجا عدد می باشند و به ضرایب کسنر موسوم هستند دو قید زیر را ارضاء می کنند:

$$\sum_{i=1}^{3} p_i = 1$$

$$\sum_{i=1}^{3} {p_i}^2 = 1$$

سه ضریب کسنر واضح است که نمی توانند یکسان باشند. حالاتی که فقط دو تا ضرایب یکسان باشند فقط برای سه تایی $(\frac{2}{3}, \frac{2}{3}, -\frac{1}{3})$ و $(0,0,1)$ اتفاق می افتد. در بقیه حالات سه ضریب با هم متفاوتند، یکی از آنها منفی و بقیه مثبت خواهند بود.

بدون از دست دادن عمومیت مساله فرض می کنیم $p_1 < p_2 < p_3$. در این صورت ضرایب متریک در بازه های زیر معتبر خواهند بود:

$$-\frac{1}{3} < p_1 < 0$$

$$0 < p_2 < \frac{2}{3}$$

$$\frac{2}{3} < p_3 < 1$$

به طور کلی این حل ها یا یک تکینگی سیگار[2] شکل در $t = 0$ دارند، بدین معنی که یک ناحیه کوچک فضایی که در زمانی معین تقارن کروی داشته است در زمانهای نزدیک به صفر یعنی در $t \to 0$ این ناحیه بینهایت دراز و باریک می شود. حالت دیگر این است که در آن متریک در دو بعد تحول داشته باشد، در این صورت ناحیه با تقارن کروی در $t \to 0$ یک ناحیه دیسکی بی نهایت نازک خواهد شد که به آن

---

تکینگی کلوچه شکل [1] می گویند[۳۳]. در این مدل ها اگر ما ماده را به مساله اضافه کنیم، حل هـای بیانکی تکینگی های فیزیکی خواهند داشت، از این حیث که چگالی ماده $\rho$ بی نهایت بـزرگ مـی شـود یعنی در $t \to 0$ داریم $\rho \to 0$.

متریک کسنر یک حل دقیق معادلات میدان اینشتین برای فضای تهی است. اما نزدیک نقطـه تکیـن $t = 0$ یک حل تقریبی خواهد بود زیرا جملات معادلات میدان خلاء که با $t^{-2}$ متناسب انـد در $t = 0$ واگرا مـی شوند (در بخش بعدی این موضوع به تفصیل توضیح داده خواهـد شـد)، حتی بـرای حـالتی کـه توزیـع یکنواختی از ماده وجود داشته باشد. زیرا جملات ماده نیز در نقطه تکین واگرا می شوند[۳٤]. بـا توجـه بـه تحلیلی که در بخش بعد ارائه خواهد شد، وابستگی چگالی ماده و مولفه هـای چهـار بـردار سـرعت بـرای حالت خاص $P = \frac{\rho}{3}$ عبارت است از:

$$\rho \sim t^{-2(1-p_3)} \tag{٤-٥}$$

$$u_a \sim t^{(1-p_3)/2}$$

از آنجا که در نسبیت عام همواره $p_3 \le 1$ است می بینیم که چگالی ماده با کاهش پـارامتر زمان بـه سـمت صفر طبیعتی واگرا را از خود نشان می دهد. ولی چهار بردار سرعت به سمت صفر مـی رود. بـرای حالـت خاص $p_3 = 1$ ظاهرا در $t = 0$ چگالی و مولفه های چهاربردار مقداری مبهم می اختیار می کنند. ولی می توان نشان داد برای ضرایب کسنر به شکل $(0,0,1)$ متریک با تبدیلی همدیس به متریک مینکوسکی تبدیل می شود، یعنی برای ضرایب فوق تکینگی $t = 0$ یک تکینگی فیزیکـی نیست و فقـط حاصـل انتخـاب مختصاتی خاص است.

در تمام این صحبت ها فرض بر این بود که ماده اختلال قابل توجهی روی تحول متریک ندارد زیرا در غیر این صورت ما مساله را می بایست برای تانسوری خاص از ماده حل مـی کـردیم و در نتیجـه متریکـی کـه بدست می آوردیم دیگر متریک کسنر نمی بود. ما به هنگام اضافه کردن ماده به مساله به طور ضمنی فـرض کردیم که متریک ما متریک کسنر است. در قبل گفتیم که متریک کسنر یک حل خلاء است. از آنجایی ایـن حل در خلاء دارای یک تکینگی است حضور ماده نباید طبیعت این تکینگی را تغییـر دهـد. یعنی بایـد در حضور ماده تکینگی همچنان وابسته به هندسه باشد نه حضور ماده. همانطور که بیان شد تانسـور انـرژی در

---

[1] pancake singularity

زمان صفر نیز دارای یک تکینگی است. حال اگر سرعت رشد اندازه عددی جملات تانسور انرژی کمتـر از رشد جملات میدان باشد، حضور ماده را می توان بی تاثیر تلقی کرد. در بخش بعد این موضـوع را بررسـی می کنیم.

حال جملات غالب در مولفه های تانسور انرژی تکانه را بدست می آوریم:

$$T_0^0 \sim \rho u_0{}^2 \sim t^{-(1+p_3)} \qquad (5-5)$$

$$T_1^1 \sim \rho \sim t^{-2(1-p_3)}$$

$$T_2^2 \sim \rho u_2 u^2 \sim t^{-(1+2p_2-p_3)}$$

$$T_3^3 \sim \rho u_3 u^3 \sim t^{-(1+p_3)}$$

از آنجایی که همواره $p_3 \leq 1$ می باشد، می بینیم برای تمامی مولفه هـای تانسور انـرژی تکانـه داریـم $T_\alpha^\alpha \leq t^{-2}$ و این بدین معنی است که در حضور ماده در $t \to 0$ فقط جملات میدان هستند که مهم مـی شوند و بقیه (جملات تانسور انرژی تکانه) در مقابل جملات میدان قابل صرف نظر کردن خواهنـد بـود و این به نوبه خود بدین معنی است که در نسبیت عام در $t \to 0$ متریک کسنر که حـل خـلاء بـود حتـی بـا وجود ماده توصیف کننده خوبی برای فضا زمان خواهد بود. توجه کنید که شرط $T_\alpha^\alpha \leq t^{-2}$ خـود، زمانی را به شکل:

$$T_\alpha^\alpha \leq t_0^{-2}$$

نشان می دهد که از تکینگی تا این زمان اثر ماده را می توان نادیده گرفت و به خلاصه گفت : متریک کسنر تا زمان $t_0$ مستقل از اینکه ماده باشد یا نباشد توصیف کننده خوبی از فضا زمان است و این خوب بودن هر چه به سمت تکینگی به عقب برگردیم دقیق تر خواهد بود.

۵-۲ حل های تعمیم یافته کسنری در گرانش های مرتبه بالاتر

با توجه به توضیحات داده شده در قسمت قبل برای حل های کسنری در نسبیت عام، اکنون مـی خـواهیم مساله مشابه را در گرانش های مرتبه بالا بررسی کنیم و مشخص کنیم در چـه شـرایطی چنـین حـل هـایی وجود دارند[۳۵].

برای یک نظریه واقع گرایانه، انتظار داریم که جملات غالب در بسط تابع تحلیلی معرفی شده جـایگزین $R$ در کنش اینشتین هیلبرت، یعنی بسط تحلیلی $f(R)$ در مدل هایی با کنش:

$$S_{HG} = \frac{1}{\chi} \int \sqrt{-g} f(R) d^4 x \qquad (۶-۵)$$

در حد های کلاسیکی یعنی در گرانش نیوتونی به شکل $R \approx R$ در $f(R)$ در آیند. که انتخاب یک تابع تحلیلـی با این ویژگی در واقع به ما این امکان را می دهد که در حالات بـا گرانش ضـعیف نظریه عـامتر (۶-۵)، نظریه نسبیت عام اینشتین با کنش اینشتین–هیلبرت به شکل زیر را بدست دهد:

$$S_{EH} = \frac{1}{\chi} \int \sqrt{-g} R d^4 x \qquad (۷-۵)$$

یعنی به طور خلاصه نظریه مطلوب در گرانش ضعیف رفتاری به شکل زیر را خواهد داشت:

$$S_{HG} \xrightarrow{\text{گرانش ضعیف}} S_{EH}$$

(HG در (۶-۵) مخفف Higher Order Gravity و EH در (۷.۵) مخفف Einstein- Hilbert مـی باشد). با وجود این، چنین ویژگی را نمی توان از نظریه در گرانش های قوی انتظار داشت، زیرا در گرانش های قوی ما اطلاعی درستی از شکل تابع $f(R)$ نداریم. در واقع در چنین موقعیت هـایی انتظـار داریم تصحیحات کوانتومی جای خود را باز کرده و نظریات ما دستخوش تغییراتی شوند. با توجه به اینکـه کـنش نسبیت عام بدون ثابت کیهان شناسی شامل جمله ای ناوردا از اسکالر ریچی است ما برای یک بررسی جامع ترجیح می دهیم توابعی تحلیلی از $R_{\alpha\beta} R^{\alpha\beta}$ و $R_{\alpha\beta\mu\nu} R^{\alpha\beta\mu\nu}$ را مطالعه کنیم. یعنی تـابعی بـه شـکل $f(X, Y, Z)$ که $X$، $Y$ و $Z$ به ترتیب $R$، $R_{\alpha\beta} R^{\alpha\beta}$ و $R_{\alpha\beta\mu\nu} R^{\alpha\beta\mu\nu}$ هستند را مطالعه می نماییم. بـا فرض تحلیلی بودن این تابع در تمام نقاط دامنه آن، جملات غالب در بسط چنین تابعی می تواند به شـکل، $R^n$، $(R_{\alpha\beta} R^{\alpha\beta})^n$ یا $(R_{\alpha\beta\mu\nu} R^{\alpha\beta\mu\nu})^n$ باشد. پس تمام کاری که انجام می دهیم این است کـه سـه

تابع فوق را بررسی کرده و جواب های تعمیم یافته کسری را در صورت وجود بدست آوریـم. از سـوی دیگر معلوم می شود که آیا این جواب ها در شرایط خاص منجر به جواب های نسبیت عـام مـی شـوند یـا خیر. پاسخ تعمیم یافته بدین معنی است که دو قید موجود در (۵-۳) تغییر کرده و در قسمت بعدی خواهیم دید که تابعی از متغیر آزاد موجود در مساله یعنی $n$ خواهد شد.

در قسمت بعدی در ابتدا تابع به شکل $R^{1+\delta}$ را مطالعه می کنیم. در آن بجای $n$ از $\delta + 1$ اسـتفاده کـرده ایم، با این کار در واقع انحراف آن از $R$ در کنش اینشتین-هیلبـرت (۵-۷) را بـا ضـریب $\delta$ کـه محـدوده اعتبار آن بعد از بررسی معادلات میدان منتج از مدل مشخص می شود، نشان داده ایم. در هر رابطـه ای کـه در آن $\delta$ وجود داشته باشد با اتخاذ مقدار صفر بجای آن، روابط مشابه در نسبیت عام بدسـت مـی آیـد. در دو تابع دیگر همچنان ضریب $n$ را داریم زیرا در کنش اینشتین-هیلبرت جمله های به شکل $R_{\alpha\beta}R^{\alpha\beta}$ و $R_{\alpha\beta\mu\nu}R^{\alpha\beta\mu\nu}$ وجود ندارد تا ما در این بررسی میزان انحراف را آن ها را بررسی کنیم.

۵-۳ در اینجا به مدل اول با چگالی لاگرانژی زیر می پردازیم[۳٦]:

$$L_{HG} = \frac{1}{\chi}\sqrt{-g}\,R^{1+\delta} \qquad (٨-۵)$$

که $\delta$ و $\chi$ هر دو مقادیر ثابتی هستند. در حد $\delta \to 0$ چگالی لاگرانژی اینشـتن هیلبـرت در نسبیت عـام معمولی بدست می آید. در اینجا فرم $f(R)$ به صورت نمایی انتخاب شده است، و آن هم به این دلیل کـه اولاً تحلیل معادلات منتج ساده تر خواهد شد و نیز در معادلات با گذاشتن $\delta = 0$ به معـادلات در نسبیت عام معمولی می رسیم که در نتیجه می توان به ازای مقادیر مختلف $\delta$ مقدار انحـراف از معـادلات معمـولی نسبیت عام را مورد بررسی قرار داد.

کـنش کلـی، حاصـل از کـنش ناشـی از مـاده و کـنش مربـوط بـه جملـه هندسـی اضـافه شـده یعنـی $f(R) = R^{1+\delta}$ است:

$$S_{HG} = \int L_{HG}\,d^4x + S_m \qquad (٩-۵)$$

که $S_m$ کنش ماده است. طبق معمول از اثرات کنش در بینهایت صرف نظر می کنیم که این شرط مرزی ما در بینهایت است. برای بدست آوردن معادلات میدان کنش کلی را وردش می دهیم. حاصل به شکل زیر است:

$$\bar{P}_{\alpha\beta} = \kappa T_{\alpha\beta}$$

(۵-۱۰-الف)

که $\bar{P}_{\alpha\beta}$ تانسور اینشتین تعمیم یافته، $G_{\alpha\beta}$ است:

(۵-۱۰-ب)

$$\bar{P}_{\alpha\beta} = \delta(1-\delta^2)R^\delta \frac{R_{,\alpha}R_{,\beta}}{R^2} - \delta(1+\delta)R^\delta \frac{R_{;\alpha\beta}}{R} + (1+\delta)R^\delta R_{\alpha\beta} - \frac{1}{2}g_{\alpha\beta}RR^\delta - g_{\alpha\beta}\delta(1-\delta^2)R^\delta \frac{R_{,c}R^{,c}}{R^2} + \delta(1+\delta)g_{\alpha\beta}R^\delta \frac{\Box R}{R}$$

که $T_{\alpha\beta}$ تانسور انرژی-تکانه مربوط به ماده است و به طور معمول تعریف می شود(ضمیمه ذ).

حال در اینجا به دنبال شرایطی هستیم که تحت آن شرایط یا معـادلات قیـدی، متریک کسنرحل معـادلات میدان (ب-۵-۱۰) باشد. متریک کسنر عبارت است از:

$$dS^2 = -dt^2 + t^{2p_1}dx^2 + t^{2p_2}dy^2 + t^{2p_3}dz^2$$

(۵-۱۱)

که $p_i$ ها ضرایب کاسنر هستند. در نسبیت عام معمولی روابط زیر بین این ضرایب بر قرار است:

$$\sum_{i=1}^{3} p_i = 1$$

$$\sum_{i=1}^{3} p_i^2 = 1$$

و$P_i$ ها در بازه های زیر قرار می گیرند:

$$-\frac{1}{3} \leq p_1 \leq 0$$

$$0 \leq p_2 \leq \frac{2}{3}$$

$$\frac{2}{3} \leq p_3 \leq 1$$

برای یافتن حل های خلاء سمت راست رابطه (۵-۱۰-الف) صفر خواهد بود:

$$\bar{P}_{\alpha\beta} = 0 \qquad (۵-۱۲)$$

اگر مؤلفه های متریک کسنر را در معادله میدان فوق قرار دهیم، به دو معادله قیدی زیر می رسیم:

$$(1 - \delta)(p_1 p_2 + p_1 p_3 + p_2 p_3) + 3\delta^2(1 + 2\delta) = \delta(2 + \delta)(p_1 + p_2 + p_3) \qquad (۵-۱۳)$$

$$(1 - \delta)(p_1{}^2 + p_2{}^2 + p_3{}^2) + 3\delta(1 + 2\delta) = (1 + 2\delta^2)(p_1 + p_2 + p_3) \qquad (۵-۱٤)$$

در واقع متریک کسنر حل خلاء معادلات میدان برای مدل $f(R) = R^{1+\delta}$ است اگر و فقط اگر دو معادله بالا ارضاء شوند.

حل معادلات بالا دو دسته جواب زیر را ارائه می کند. دسته اول عبارت است از:

$$\sum_{i=1}^{3} p_i = \frac{3\delta(1+2\delta)}{1-\delta} \qquad (۵-۱۵)$$

$$\sum_{i=1}^{3} p_i{}^2 = \frac{3\delta^2(1+2\delta)^2}{(1-\delta)^2} \qquad (۵-۱٦)$$

این دسته فقط یک حل دارد که به شکل زیر است:

$$p_1 = p_2 = p_3 = \frac{\delta(1+2\delta)}{1-\delta} \qquad (۵-۱۷)$$

در واقع اگر معادلات (۵-۱۵) و(۵-۱٦) را با هم حل کنیم هر سه ضریب کسنر با هم برابر می شوند. چون این جواب برای $p_i$ ها جواب یکسانی را بدست می دهد و نیز این جواب قبلا توسط بلیر[1] و اشمیت[2] [۳۷،۳۸] بدست آمده و مورد بحث قرار گرفته در اینجا به آن نمی پردازیم.

دسته دوم جواب عبارت است از:

---

$$\sum_{i=1}^{3} p_i = 1 + 2\delta \qquad (\text{۱۸–۵})$$

$$\sum_{i=1}^{3} p_i{}^2 = 1 - 4\delta^2 \qquad (\text{۱۹–۵})$$

که حالتی جدید و در واقع تعمیمی است از عالمی که با متریک کستر توصیف می شود[۳۲]. همان طور که می بینید اگر قرار دهیم $\delta = 0$ این معادلات، حل های استاندارد کسر در نسبیت عام عادی خواهند بود و همان قید های ذکر شده در بالا را خواهند داد.

قبل از ادامه نشان می دهیم که شرط اینکه با متریک کسنر یک عالم انبساط شونده را داشته باشیم این است که:

$$\sum_{i=1}^{3} p_i \geq 0 \qquad (\text{۲۰–۵})$$

برای اینکه عالم انبساط یابد باید در حد یک بعد در $t \to \infty$ افزایش بعد داشته باشیم یعنی حد اقل یکی از ضرایب متریک کسنر در $t \to \infty$ صفر نشود. شرط فوق بدین معنی است که اگر در بد ترین حالت دو تا از ضرایب متریک منفی باشد، ضریب سوم هم باید مثبت باشد وهم اندازه آن از اندازه مجموع دو ضریب دیگربزرگ تر باشد (یعنی سرعت انبساط از سرعت انقباض در دو بعد دیگر بیشتر باشد) یعنی:

$$p_1 \geq 0$$

$$p_1 \geq |p_2 + p_3|$$

$$|p_1| = p_1 \,, |p_2 + p_3| = -p_2 - p_3$$

$$p_1 \geq -p_2 - p_3 \ \to p_1 + p_2 + p_3 \geq 0$$

لازم به توضیح نیست که اگر یکی از ضرایب کسنر منفی باشد، مؤلفه متناظر با آن در متریک، زمانی که t به سمت بینهایت میل کند کوچک وکوچکتر (تقریبا صفر) می شود و کیهان در آن راستا دیگر انبساط نخواهد یافت.

این نتیجه را می توان از مثبت بودن متوسط ثابت هابل نیز بدست آورد:

$$H = \frac{1}{3}\left(\frac{\dot{a}}{a} + \frac{\dot{b}}{b} + \frac{\dot{c}}{c}\right) \geq 0$$

که $a$، $b$ و $c$ به ترتیب ضرایب مقیاس در سه بعد است. با جاگذاری این ضرایب از متریک کسنر شرط فوق بدست خواهد آمد.

در ادامه با استفاده از رابطه (۵-۱۸) و (۵-۲۰) خواهیم داشت:

$$1 + 2\delta \geq 0 \;\; \rightarrow \;\; \delta \geq -\frac{1}{2}$$

واز رابطه (۵-۱۹) داریم:

$$1 - 4\delta^2 > 0 \;\; \rightarrow \;\; -\frac{1}{2} < \delta < \frac{1}{2}$$

اگر $\delta$ خارج از این بازه باشد، حل های نا همسانگرد برای این نوع $f(R)$ وجود نخواهد داشت. (یعنی متریک کسنر دیگر جواب معادله میدان (۵-۱۰-ب) نیست). دقت می کنیم که اهمیت شرط اول که در شرط دوم نیز وجود دارد این است که با یک مفهوم فیزیکی یعنی انبساط بدست آمده است بر خلاف شرط دوم که تنها با یک ملاحظه منطقی ریاضی بدست آمده است.

برای تحلیل جواب های (۵-۱۹) و (۵-۱۸) بدون از دست دادن عمومیت مساله شرط اضافه زیر را وارد می کنیم:

$$p_1 < p_2 < p_3 \tag{۵-۲۱}$$

حال می خواهیم محدوده جواب هر یک از ضرایب کسنر را بدست آوریم. روش کار بدین صورت است که در (۵-۱۸) و (۵-۱۹) یک بار قرار می دهیم $p_1 = p_2$ بار دیگر قرار می دهیم $p_2 = p_3$ ولی در تحلیل نهایی حالات مساوی را در نظر نمی گیریم. با این کار دو دسته جواب به صورت زیر خواهیم داشت. دسته اول عبارت است از:

$$p_1 = p_2 = \frac{1 + 2\delta \pm \sqrt{(1+2\delta)(1-4\delta)}}{3} \tag{۵-۲۲- الف}$$

$$p_3 = \frac{1 + 2\delta \pm 2\sqrt{(1+2\delta)(1-4\delta)}}{3} \tag{۵-۲۲- ب}$$

ودسته دوم عبارت است از:

$$p_2 = p_3 = \frac{1+2\delta \pm \sqrt{(1+2\delta)(1-4\delta)}}{3} \qquad \text{(پ -۲۲-۵)}$$

$$p_1 = \frac{1+2\delta \pm 2\sqrt{(1+2\delta)(1-4\delta)}}{3} \qquad \text{(ت -۲۲-۵)}$$

توجه می کنیم که در رابطه (۲۱-۵) بین $p_1$ و $p_3$ به طور مستقیم رابطه ای وجود ندارد پس ما نمی توانیم $p_1 = p_3$ را بررسی کنیم. از دسته اول می توان حالت بالایی (۲۲-۵- ب) را برای $p_3$ انتخاب کرد. چون طبق (۲۱-۵) از همه بزرگتر است و از دسته دوم می توان حالت پایینی (۲۲-۵- ت) را برای $p_1$ انتخاب کرد چون از همه کوچکتر است. تا اینجا می توانیم بازه ها را بدین شکل بنویسیم:

$$\frac{1+2\delta - 2\sqrt{(1+2\delta)(1-4\delta)}}{3} \leq p_1 \leq \Sigma \qquad \text{(ث -۲۲-۵)}$$

$$\Sigma \leq p_2 \leq \Psi \qquad \text{(ج -۲۲-۵)}$$

$$\Psi \leq p_3 \leq \frac{1+2\delta + 2\sqrt{(1+2\delta)(1-4\delta)}}{3} \qquad \text{(چ -۲۲-۵)}$$

حال باید $\Sigma$ و $\Psi$ را با استفاده ازجواب های بالایی در دو دسته جواب تعیین کرد. با این علم که $\Sigma$ و $\Psi$ هر دو نمی توانند جوابهای یکسان (هر دو + یا هر دو -) را انتخاب کنند (زیرا بازه بدست آمده فقط شامل یک عدد خواهد شد) دوحالت می ماند:

$$p_2 = p_3 = \Psi = \frac{1+2\delta - \sqrt{(1+2\delta)(1-4\delta)}}{3} \qquad \text{(ح -۲۲-۵)}$$

$$p_1 = p_2 = \Sigma = \frac{1+2\delta + \sqrt{(1+2\delta)(1-4\delta)}}{3} \qquad \text{(خ -۲۲-۵)}$$

یا

$$p_2 = p_3 = \Psi = \frac{1+2\delta + \sqrt{(1+2\delta)(1-4\delta)}}{3} \qquad \text{(د -۲۲-۵)}$$

$$p_1 = p_2 = \Sigma = \frac{1+2\delta - \sqrt{(1+2\delta)(1-4\delta)}}{3} \qquad \text{(ر -۲۲-۵)}$$

به وضوح مشخص است نمی توان (۵-۲۲- ح) و (۵-۲۲- خ) را در(۵-۲۲- ج) قرار داد چون هیچ عددی در این بازه وجود ندارد ولی (۵-۲۲- د) و(۵-۲۲- ر) بازه ای با معنی را بدست می دهد پس با این توضیحات بازه های نهایی عبارت خواهند بوداز:

$$\frac{1+2\delta-2\sqrt{(1+2\delta)(1-4\delta)}}{3} \leq p_1 \leq \frac{1+2\delta-\sqrt{(1+2\delta)(1-4\delta)}}{3} \qquad (\text{ن} \ -\text{۲۲}-\text{۵})$$

$$\frac{1+2\delta-\sqrt{(1+2\delta)(1-4\delta)}}{3} \leq p_2 \leq \frac{1+2\delta+\sqrt{(1+2\delta)(1-4\delta)}}{3} \qquad (\text{و} \ -\text{۲۲}-\text{۵})$$

$$\frac{1+2\delta+\sqrt{(1+2\delta)(1-4\delta)}}{3} \leq p_3 \leq \frac{1+2\delta+2\sqrt{(1+2\delta)(1-4\delta)}}{3} \qquad (\text{ه} \ -\text{۲۲}-\text{۵})$$

این بازه ها را می توان به ازاء مقادیر مختلف $\delta$رسم نمود:

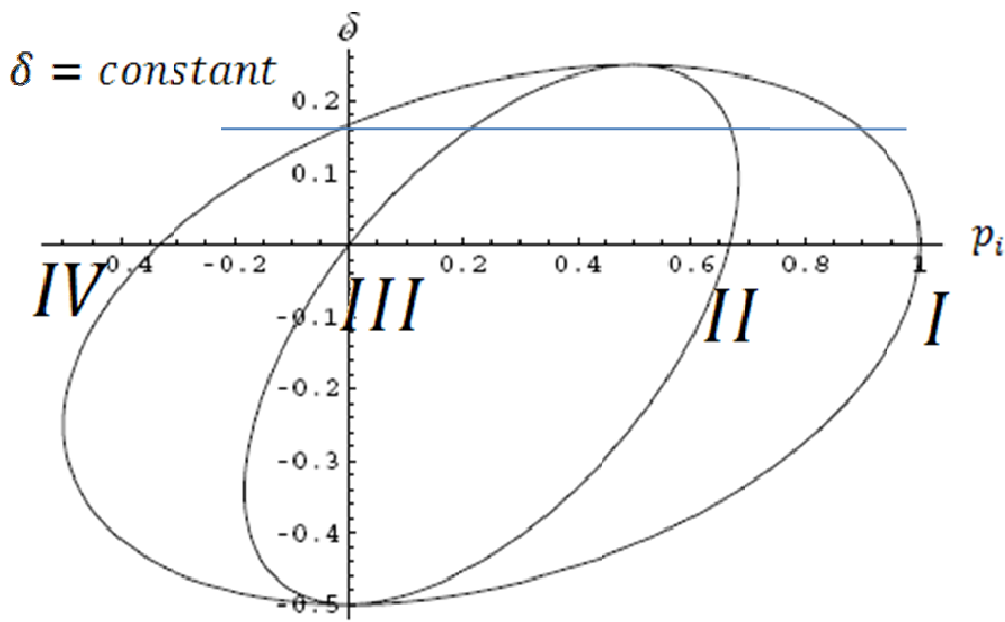

منحنی ۵-۱ منحنی جواب های مربوط به ضرایب کسر تعمیم یافته برای مدل $L = \frac{1}{\chi}R^{1+\delta}$

همانطور که می بینید نمودار قید بیشتری را روی $\delta$ نشان می دهد.

$$-\frac{1}{2} < \delta < \frac{1}{4} \qquad (\text{ی}-\text{۲۲}-\text{۵})$$

در واقع این تحلیل نسبتا طولانی نشان داد که بازه معتبر برای $\delta$ بازه (۵-۲۲-ی) است و نه بازه قبلی و به ازاء مقادیر داخل این بازه متریک کسر حل معادلات میدان در حالت خلا است. خط  ثابت $\delta=$ دو

منحنی را در چهار نقطه قطع می کند که سه بازه که به ازاء آن مقدار $\delta$ برای $p$ ها بدست می دهد. برای $\delta = 0$ خواهیم داشت:

$$-\frac{1}{3} \leq p_1 \leq 0$$

$$0 \leq p_2 \leq \frac{2}{3}$$

$$\frac{2}{3} \leq p_3 \leq 1$$

که همان بازه هایی اند که قبلا در حل اینشتنی معرفی کرده بودیم.

برای درک بهتری از نحوه ترسیم منحنی فوق تعریف کرده ایم:

$$I \equiv \frac{1+2\delta+2\sqrt{(1+2\delta)(1-4\delta)}}{3}$$

$$II \equiv \frac{1+2\delta+\sqrt{(1+2\delta)(1-4\delta)}}{3}$$

$$III \equiv \frac{1+2\delta-\sqrt{(1+2\delta)(1-4\delta)}}{3}$$

$$IV \equiv \frac{1+2\delta-2\sqrt{(1+2\delta)(1-4\delta)}}{3}$$

یعنی منحنی فوق از روی هم افتادن چهار منحنی که با چهار عدد یونانی برچسب خورده اند بدست آمده است.

بحث روی مقادی$\delta$

$\frac{1}{2} - > \delta$ مادامی که $\delta$ این شرط را ارضا کند در $0 \to t$ ما تکینگی انحنا از نوع ویل[1] خواهیم داشت. که در واقع همان تکینگی در زمان صفر است که با توجه به شکل ضریب مقیاس وابسته بـه زمـان در متریـک کسر واضح است. در واقع گفتیم که به ازاء $\delta$ در بازه (۵-۲۲-ی) ما این تکینگی را خـواهیم داشت زیـرا فقط این بازه است که به ازاء آن متریک کسر حل معادلات میدان خواهـد بـود و در نتیجـه تکینگـی آن در این بازه است که معنی پیدا می کند.

$\frac{1}{2} - > \delta$ در این حالت $0 < 2\delta + 1$ است و طبق توضیحات قبلی داریم:

$$\sum_{i=1}^{3} p_i < 0$$

یعنی یک یا چند ضریب کسر منفی خواهند بود و مولفه های متریک در$\infty \to t$ به سمت صفر میل می کنند، در این حالت رشد انقباطی بر رشد انبساطی (در یک یا هر سه بعد) غلبه یافته و در واقع در بینهایت، متریک ما که باید در واقع فضا-زمان را توصیف کند صفر می شود. این تکینگی شکاف بزرگ[2] نامیده می شود. ولی همان طور که قبلا متذکر شدیم این حالت در بازه مجاز (۵-۲۲-ی) قرار ندارد بدین معنی که حل ما تکینگی شکاف بزرگ را شامل نمی شود وفقط تکینگی انحنای ویل را دارد (تکینگی در $0 \to t$) که به طور پیش فرض در متریک کسر موجود است.

$0 < \delta$ در این حالت همانطور که از نمودار دیده می شود $p_2$ و $p_3$ همیشه مثبت اند ولی $p_1$ مـی توانـد مقدار منفی هم اختیار کند. حد بالای $p_1$ همواره مثبت است ولی حد پایین آن بین $\frac{1}{2}$ و $\frac{1}{3} -$ تغییر می کنـد، یعنی برای این حالت فقط $p_1$ ممکن است منفی شود ولی در هرصورت عالم انبساط می یابد.

$0 < \delta$ در این حالت $p_1$ همواره منفی است، $p_2$ می تواند منفی یا مثبت باشد ولی $p_3$ همواره مثبت است بدین معنی که عالم همواره در این حالت انبساط می یابد.

---

در نسبت عام متریک کسنر، حل خلاء معادلات اینشتن در حالت ناهمسانگرد است و دارای تکینگی انحنا از نوع ویل می باشد. ضرایب کسنر در دو قیدی که در ابتدا ذکر شد صدق می کند. این حل انبساط عالم را در پی دارد. در تعمیم مرتبه بالاتر لاگرانژی انیشتن-هیلبرت با نوع خاصی از تابع انحنا که معرفی گردید، این متریک باز هم حل این معادلات میدان تعمیم یافته ولی با معادلات قیدی جدید است. در این مدل تابع انحنا با یک ثابت (δ) معرفی گردید و نشان دادیم اگر متریک یاد شده جواب معادلات میدان باشد باید این ثابت در بازه معینی تغییر کند و نمی تواند هر مقداری باشد. و نیز نشان دادیم که حل معادلات تعمیم یافته با شرط تغییر این ثابت در بازه (۵-۲۲-ی)، منجر به مدلی می شود که تکینگی انحنای ویل را حفظ کرده و انبساط عالم را سبب می شود. در واقع ما مدل کلی تری را ارائه کردیم که همان نتایج معـادلات معمـولی را در پی دارد و در حالتی که ثابت صفر باشد معادلات معمولی بدست می آید.

در حل عادی، متریک کسنر نه تنها حل خلاء معادلات میدان است، بلکه در حضور مـاده و در مجـاورت تکینگی ویل می تواند حل معادلات غیر خلاء باشد. این در واقع نشان می دهد که این متریـک کسـنر حـل کلی معادلات اینشتن است به شرط اینکه نتیجه گیری ها در شرط $t \to 0$ صـورت گیـرد. از آنجـا کـه مـدل در خلاء دارای یک تکینگی است ($t = 0$) این بدین معنی است که ایـن تکینگـی فقـط ناشـی از هندسـه ای است که انتخاب کردیم یعنی ناشی از متریک تعمیم یافته کسنر است که به نوبه خود به این معنی اسـت کـه در حضور ماده این تکینگی نباید دسخوش تغییر شود. یعنی تکینگی در حضور ماده فقـط و فقـط ناشـی از هندسه انتخابی باشد. برای اینکه هندسه انتخابی این موضوع را نشان دهد باید با پیشـروی بـه سـمت نقطـه تکین جملات ناشی از حضور ماده خود به خود اندازه کوچکتری نسبت به جملات خلاء داشته باشند یعنی قابل صرف نظر کردن باشد. این موضوعی است که در بخش بعدی به آن می پردازیم.

۵-۳-۲ حل در مجاورت ماده خیلی نزدیک به تکینگی t → 0

دراین قسمت به دنبال شرایطی هستیم که متریک کسنر تحت آن شرایط در حضور ماده کامل هنـوز جـواب معادلات میدان باشد یعنی در آن صورت ما بتوانیم فضا زمان را با متریک خلاء کسنر توصیف کنیم. این بـه این معنی است که حضور ماده نقش مهمی را ایفا نمی کند البته در مجاورت تکینگی $t \to 0$. و هر چه بـه سمت این تکینگی پیش می رویم توصیفات ما دقیقتر خواهد بود. شرطی که باید ارضاء شود این اسـت کـه جملات معادله میدان (۵-۱۰-ب) سریع تراز جملات مربوط به تانسور انرژی-تکانه واگرا شـوند، یعنـی در

مجاورت $t \to 0$ این جملات بر جملات تانسور ماده غلبه کنند. به عبارتی بتـوان از جمـلات مربـوط بـه تانسور ماده در مقابل جملات معادله میدان صرف نظر کرد. این توضیح لازم است که در اینجا ما شرط هـای جدیدی برای ضرایب $p_i$ بدست نمی آوریم، در عوض بررسی می کنیم که آیا همین که قید هـای (۵-۲۲-ن) ، (۵-۲۲-و) و(۵-۲۲-ه) بر قرار باشند کافی است که ماده اثر ناچیز داشته باشد یا خیر.

برای این موضوع ابتدا نشان می دهیم که جملات میدان از مرتبـه $O(t^{-2(1+\delta)})$ هسـتند. در قـدم بعـدی مؤلفه تانسور انرژی-تکانه ماده را در دو حالت حرکت نسبیتی ماده و حرکت غیر نسبیتی ماده بدست می آوریـم و سپس مرتبه بزرگی جملات مربوط به تانسور انرژی-تکانه و جملات میدان را با هم مقایسه می کنیم.

برای نشان دادن مرتبه جملات میدان چون در حالت خلاء سمت راست معادله صفراسـت، دو طـرف را در $g^{\alpha\beta}$ ضرب می کنیم، به این دلیل ضرایب شامل مؤلفه های متریک که حاوی ضرایب کسـر هسـتند حـذف می شوند. برای نمونه جمله اول را تحلیل می کنیم:

$$\delta(1-\delta^2)g^{\alpha\beta}R^\delta \frac{R_\alpha R_\beta}{R^2} \propto R^\delta \frac{R_\alpha R^\alpha}{R^2} = R^\delta \frac{R_{,t}R^{,t}}{R^2} = R^\delta \frac{(R_{,t})^2}{R^2} = t^{-2(1+\delta)}$$

که از این نکته استفاده کردیم که $R \propto t^{-2}$ می توان نشان داد که بقیه جملات نیز به همین شیوه با $t^{-2(1+\delta)}$ متناسب خواهند بود.

## ۵-۳-۲  تانسور انرژی-تکانه برای ماده

وقتی معادلات میدان را در اطراف یک نقطه تکین بررسی می کنیم، یعنی نقطه ای که در آن فشار و چگـالی ماده مقادیر بسیار عظیمی را اتخاذ می کنند (به بیان ریاضی یعنی وقتی به بینهایت میل می کنند)، لازم است که برای ماده در اطراف آن نقطه رفتاری نسبیتی را در نظر بگیریم. یعنی فرض بر آن است که سـرعت مـاده در آن نقطه از مرتبه سرعت نور است. عموما ماده هایی که چنین رفتاری را می تـوان بـرای آنهـا در نظـر گرفت در بازه $\gamma < \frac{4}{3}$ قرار دارند (در نزدیکی های $\gamma < \frac{4}{3}$). ولی برای اینکه کلیت مطلـب از دسـت نـرود باید ماده هایی را نیز در نظر گرفت که حتی نزدیک نقطه تکین، سرعت هایی غیر نسبیتی دارند. این ماده ها در بازه $\gamma > \frac{4}{3}$ قرار دارند. یعنی برای مقادیر به اندازه کافی بزرگ $\gamma$ در $\gamma > \frac{4}{3}$ می توان فرض کـرد مـاده رفتاری غیر نسبیتی دارد. پس بررسی خود را برای دو نوع ماده نسبیتی و غیر نسبیتی انجام می دهیم.

برای اینکار فرض می کنیم متریک زمینه، متریک کسنر است و نیز محتویات عالم، ماده ای کامـل بـا معادلـه حالت $P = (\gamma - 1)\rho$ و $2 > \gamma \geq 1$ است.

از معادله پیوستگی $\nabla_\alpha T^{\alpha\beta} = 0$ به دو معادله

$$\frac{\partial}{\partial x^i}\left(t^{p_1+p_2+p_3}\rho^{\frac{1}{\gamma}}u^i\right) = 0$$

$$(\rho + P)u^k\left(u_{i,k} - \frac{1}{2}u^l g_{kl,i}\right) = -\frac{1}{3}\rho_{,i} - u_i u^k P_{,k} \qquad (\text{۲٤}-\text{۵})$$

می رسیم[۳۹]. $u^k$ ها چهار بردار ها با اتحاد $u_i u^i = -1$ هستند.

در این مرحله از یک تقریب با عنوان تقریب سرعت غالب[1] استفاده می کنیم[٤۰]. بدین شـکل کـه از تمـام مشتقات فضایی در مقابل مشتقات زمانی صرف نظر می کنیم. این تقریب به این معنی است که توجه مـا بـه مقیاس های بزرگتر از افق ذره[2] است[۳٦].

با توجه به این تقریب برای معادله (۲۳-۵) داریم:

$$\frac{\partial}{\partial x^0}\left(t^{p_1+p_2+p_3}\rho^{\frac{1}{\gamma}}u^0\right) + \frac{\partial}{\partial x^1}\left(t^{p_1+p_2+p_3}\rho^{\frac{1}{\gamma}}u^1\right) + \frac{\partial}{\partial x^2}\left(t^{p_1+p_2+p_3}\rho^{\frac{1}{\gamma}}u^2\right) +$$
$$\frac{\partial}{\partial x^3}\left(t^{p_1+p_2+p_3}\rho^{\frac{1}{\gamma}}u^3\right) = 0$$

$$\frac{\partial}{\partial x^t}\left(t^{p_1+p_2+p_3}\rho^{\frac{1}{\gamma}}u^0\right) = 0 \qquad (\text{۲۵}-\text{۵})$$

$$t^{p_1+p_2+p_3}\rho^{\frac{1}{\gamma}}u^0 = constant \equiv C$$

و برای معادله (۲٤-۵) داریم:

$$\gamma\rho u^k u_{i,k} - \frac{1}{2}\gamma\rho u^k u^l g_{kl,i} = -\frac{1}{3}\rho_{,i} - u_i u^k(\gamma-1)\rho_{,k} \qquad (\text{۲٦}-\text{۵})$$

$i = \alpha \rightarrow \alpha$ شاخص فضا گونه است

---

$$\gamma \rho u^k u_{i,k} = -u_i u^k (\gamma - 1) \rho_{,k}$$

$$\gamma \rho u_{\alpha,0} = -u_i (\gamma - 1) \rho_0$$

$$\frac{\gamma}{1-\gamma} \frac{\frac{\partial}{\partial t} u_\alpha}{u_\alpha} = \frac{\frac{\partial}{\partial t} \rho}{\rho} \;\rightarrow\; u_\alpha \rho^{\frac{(\gamma-1)}{\gamma}} = constant \equiv D$$

در این مرحله فرض می کنیم که سرعت در یکی از راستا ها بر راستاهای غلبه دارد. این راستا را راستای ۳ فرض می کنیم. با استفاده از این نکته و نیز با استفاده از اتحاد مربوط به چهاربردار سرعت می توان شرط حرکت نسبیتی در راستای سوم را به صورت زیر بدست آورد:

$$u_i u^i = -1 \;\rightarrow\; u_0 u^0 + u_1 u^1 + u_2 u^2 +$$

$$u_3 u^3 = -1$$

$$u^3 = u_3 t^{-2p_3} >> u^1 , u^2$$

$$u_0 u^0 + u_3 u^3 = -1 = -u_0{}^2 + u_3{}^2 t^{-2p_3}$$

$$u_0{}^2 = u_3{}^2 t^{-2p_3} + 1$$

$$u_0{}^2 \propto u_3{}^2 t^{-2p_3} \tag{۵-۲۷}$$

شرط فوق شرط حرکت نسبیتی ماده در راستای سوم است. همان طور که می بینیم با نزدیک شدن به نقطه تکین $u_0$ به سمت بینهایت میل می کند. با توجه به اینکه $u^0 = \frac{c}{\sqrt{1-\frac{v^2}{c^2}}}$، بینهایت شدن $u_0$ به معنای این است که $v \sim c$. یعنی هر چه به طرف نقطه تکین پیش می رویم سرعت ماده به سرعت نور نزدیک تر می شود. در واقع فرض شده است که ضریب سوم کسر از دیگر ضرایب کسر بزرگتر است.

با حل معادلات (۲۵-۵) و (۲۶-۵) و نیز استفاده از شرط (۲۷-۵)، با یک عملیات ساده ریاضی می تـوان چگالی ماده ومؤلفه های فضایی چهارسرعت را به شکل زیر به دست آورد:

$$\rho = \frac{D}{C} t^{\frac{-\gamma}{2-\gamma}(p_1+p_2)} \propto t^{\frac{-\gamma}{2-\gamma}(p_1+p_2)} \tag{۲۸-۵}$$

$$u_\alpha \propto t^{\frac{(p_1+p_2)(\gamma-1)}{2-\gamma}} \tag{۲۹-۵}$$

تانسور انرژی-تکانه برای $\gamma > \frac{4}{3}$

برای این گونه مواد فرض می‌شود سرعت‌ها نسبیتی هستند. حال مؤلفه‌های تانسور انرژی-تکانه را که برابر $T_\alpha{}^\beta = (\rho + P)u_\alpha u^\beta + P\delta_\alpha{}^\beta$ هستند را تشکیل می دهیم. این مؤلفه‌ها با توجه به بزرگ ترین جملات آنها برای $\gamma < \frac{4}{3}$ عبارت خواهند بود:

$$T_0{}^0 \propto \rho u_0{}^2 \propto t^{-1-2\delta-p_3}$$

(۵-۳۰-الف)

$$T_1{}^1 \propto \rho \propto t^{-\frac{\gamma}{2-\gamma}(1+2\delta-p_3)}$$

(۵-۳۰-ب)

$$T_2{}^2 \propto \rho u_2 u^2 \propto t^{-2p_2-(1+2\delta-p_3)}$$

(۵-۳۰-ج)

$$T_3{}^3 \propto \rho u_3 u^3 \propto t^{-1-2\delta-p_3}$$

(۵-۳۰-د)

نکته‌ای که باید به آن توجه کرد این است که برای $\gamma < \frac{4}{3}$ خواهیم داشت:

$$\frac{\gamma}{2-\gamma} < 2$$

در نتیجه برای چگالی ماده می توان نوشت:

$$\rho < \rho_{max} \propto t^{-2(p_1+p_2)} = t^{-2(1+2\delta-p_3)}$$

(۵-۳۱)

در تمام عبارات (۵-۳۰) از این نکته استفاده شده کـه بـه جـای $\rho \propto t^{-\frac{\gamma}{2-\gamma}(p_1+p_2)}$ از مقـدار بیشـینه آن یعنی از (۵-۳۱) استفاده کرده ایم.

عبارات بالا با توجه به رابطه (۵-۳۱) به صورت زیر بدست می آیند:

$$T_0{}^0 \propto \gamma t^{-1-2\delta-p_3} + (\gamma-1)t^{-2(1+2\delta-p_3)} = \gamma t^{-p_3}(t^{-1-2\delta}) + (\gamma-1)t^{2p_3}(t^{-2(1+2\delta)})$$

در $t \to 0$ عبارت جمله دوم به سمت صفر میل کرده و عبارت اول واگرا می شود. در نتیجه مولفه $T_0{}^0$ در مجاورت تکینگی با عبارت اول متناسب می شود.

$$T_1{}^1 \propto t^{-(1+2\delta-p_3)} = \gamma t^{-p_1}t^{-p_1} + (\gamma-1)t^{-p_1}t^{-p_2}$$

با توجه به شرط $p_1 < p_2 \rightarrow t^{-p_2} > t^{-p_1}$ در عبارت بالا جمله اول کوچک تر از جمله دوم خواهد شد و در نتیجه $T_1^{\ 1}$ در مجاورت تکینگی با جمله دوم متناسب خواهد بود. برای $T_2^{\ 2}$ توجیهی مشابه برقرار است با این تفاوت که $T_2^{\ 2}$ با جمله اول متناسب می شود.

با توجه به شرط نسبیتی (۲۷-۵) $T_3^{\ 3}$ مشابه $T_0^{\ 0}$ است.

حال باید بررسی کنیم که آیا همه مؤلفه های تانسور انرژی-تکانه که در سمت راست معادله میدان ما القاء شده است در مجاورت $t \rightarrow 0$ با سرعت خیلی کمتر ازجملات میدان (از مرتبه $O(t^{-2(1+\delta)})$ واگرا می شوند یا خیر.

دقت می کنیم که منظور از واگرا شدن با سرعت کمتر این است که در یک زمان t اندازه عددی جملات تانسور انرژی از اندازه عددی جملات میدان کمتر باشد.

در اینجا باید به این نکته دقت داشت، از آنجایی که حل ما متریک (۱۱-۵) با قید های (۲۲-۵) است، وقتی معادلات را در حضور ماده بررسی می کنیم باید بررسی کنیم که آیا این جواب به طور خود کار منجر به ناچیز ماندن ماده می شود یا خیر. یعنی به طور صریح اگر مقادیر ضرایب $p_i$ در قید های (۲۲-۵) را در معادلات در حضور ماده جاگذاری کنیم باید جملات تانسور انرژی- تکانه در مقابل جملات خلاء کوچک باشند.

از چهار مؤلفه (۳۰-۵) مؤلفه $T_3^{\ 3}$ (که برابر با مؤلفه $T_0^{\ 0}$ است) نسبت به بقیه واگرایی بیشتری دارد یعنی در هر لحظه اندازه بزرگتری دارد. پس کافیست نشان دهیم که $T_3^{\ 3}$ نسبت به $O(t^{-2(1+\delta)})$ واگرایی کمتری خواهد داشت. اگر این شرط برای این مؤلفه صادق باشد برای سایر مؤلفه ها نیز صادق خواهد بود.

$$T_3^{\ 3} \propto t^{-1-2\delta-p_3} < t^{-2(1+\delta)}$$

$$-1 - 2\delta - p_3 > -2 - 2\delta$$

$$p_3 < 1 \qquad\qquad (۳۲-۵)$$

اگر به نمودار ۵-۱ دقت کنیم می بینیم که این شرط همواره صادق است زیرا در آنجا $p_3 < 1$ است.

$$\gamma = \frac{4}{3} \ \text{تانسور انرژی-تکانه برای}$$

برای حل تابشی یعنی $\gamma = \frac{4}{3}$ به خاطر وجود عبارت (۳۱-۵) از دستگاه همراه[1] استفاده می کنیم. زیرا استفاده از دستگاه غیر همراه نتایج مربوط به $\gamma < \frac{4}{3}$ را بدست می دهد. با فرض اینکه تانسور را در دستگاه همراه با ماده با شرط $u_i = \delta_i{}^0$ بررسی می کنیم خواهیم داشت:

$$T_\alpha{}^\beta = (\rho + P)u_\alpha u^\beta - P\delta_\alpha{}^\beta = 0 + P \propto \rho$$

توجه می کنیم که در چارچوب همراه تمام مؤلفه های فضایی چهارسرعت صفر هستند. از این رو مؤلفه های فضایی تانسور انرژی-تکانه دارای جمله صفر هستند وهر سه مؤلفه برابر با فشار و در نتیجه متناسب با چگالی ماده خواهند بود. مولفه زمانی این تانسور دقیقا برابر فشار خواهد شد.

پس کافیست نشان دهیم که سرعت واگرایی $\rho$ از سرعت واگرایی جملات میدان کمتر است. با توجه به (۲۸-۵) به ازاء $\gamma = \frac{4}{3}$ داریم:

$$\rho \propto t^{-2(1+2\delta-p_3)} < t^{-2(1+\delta)}$$

$$-1 - 2\delta + p_3 > -1 - \delta$$

$$p_3 > \delta \qquad\qquad (۳۳-۵)$$

حال باید نشان دهیم که این نتیجه همواره صادق است. با توجه به (۲۲-۵-ه) می نویسیم:

$$p_3 > \frac{1+2\delta+\sqrt{(1+2\delta)(1-4\delta)}}{3} = \delta + \frac{1-\delta+\sqrt{(1+2\delta)(1-4\delta)}}{3}$$

با توجه به شرط (۲۲-۵-ی) می بینیم که که عبارت بعد از $\delta$ در قسمت دوم عبارت بالا همواره مثبت است. شرط (۳۳-۵) می گوید که برای اینکه جملات تانسور انرژی در مقابل جملات میدان خلاء ناچیز باشند کافی است که $\delta > p_3$. اگر به نمودار ۵-۱ نگاه کنیم می بینیم که تمام مقادیر $p_3$ در این بازه وجود دارند. به بیان دیگر تمام مقادیر $p_3$ در جواب خلاء می توانند شرایطی را به وجود آورند که ماده حضوری غیر مخرب روی میدان داشته باشد. در نتیجه ملاحظات بالا نشان می دهند که در حضور ماده با حرکت نسبیتی همواره متریک کسنر را می توان حل غیر خلاء معادلات میدان دانست.

---

[1] comoving coordinates

تانسور انرژی–تکانه برای $\gamma > \frac{4}{3}$

شرط نسبیتی بودن برای معادله حرکت برای شاره سخت[1] با معادله حالت با $\gamma > \frac{4}{3}$ در $t \to 0$ بدین ترتیب بدست می آید:

$$u^0 \gg 1$$

$$u_0 \propto u_3 t^{-p_3} = t^{(p_1+p_2)(\gamma-1)/(2-\gamma)} t^{-p_3} \gg 1$$

$$t^{[(p_1+p_2)(\gamma-1)-p_3(2-\gamma)]/(2-\gamma)} \gg 1$$

$$(p_1 + p_2)(\gamma - 1) - p_3(2 - \gamma) << 0$$

در اینجا از $\gamma - 1 \propto 1$ استفاده می کنیم و سرانجام شرط به صورت زیر بدست می آید:

$$\gamma - 1 + 2\delta < p_3 \tag{$0$-$٣٤$}$$

در شرایطی که شرط ($٣٤$-$٥$) برقرار نباشد در مجاورت $t \to 0$ شرط زیر برقرار خواهد بود:

$$u^\alpha u_\alpha \to 0 \quad , \quad u_0 \to 1 \tag{$0$-$٣٥$}$$

یعنی تقریب سرعت غالب دیگر برقرار نیست و سرعت ها به سمت صفر میل می کنند. در واقع اینرسی فوق العاده بالای شاره سخت باعث می شود که حرکت به سمت سکون پیش رود. در این شرایط روابط ($٢٨$-$٥$) و ($٢٩$-$٥$) دیگر برقرار نیستند و برای حالت حرکت غیر نسبیتی باید دوباره بدست آیند.

$$\frac{\partial}{\partial t}\left(t^{p_1+p_2+p_3}\rho^{\frac{1}{\gamma}}u^0\right) = \frac{\partial}{\partial t}\left(t^{p_1+p_2+p_3}\rho^{\frac{1}{\gamma}}\right) = 0$$

$$t^{p_1+p_2+p_3}\rho^{\frac{1}{\gamma}} = constant \tag{$0$-$٣٦$}$$

شرط فوق با در نظر گرفتن شرط ($٢٧$-$٥$) عبارت زیر را برای چگالی ماده و مؤلفه های فضایی چهاربردار سرعت بدست می دهد:

$$\rho \propto t^{-\gamma(1+2\delta)} \tag{$0$-$٣٧$}$$

---

[1] stiff fluid

$$u_\alpha \propto t^{(\gamma-1)(1+2\delta)} \qquad\qquad (٣٨-٥)$$

شرط $u^\alpha u_\alpha \to 0$ می دهد:

$$u_3 u^3 > u^1 u_1 \sim u^2 u_2$$

$$u^\alpha u_\alpha = u^1 u_1 + u^2 u_2 + u_3 u^3$$

$$t^{2(\gamma-1)(1+2\delta)} t^{-2p_3} \ll 0$$

$$2(\gamma-1)(1+2\delta) - 2p_3 \gg 0$$

$$(\gamma-1)(1+2\delta) > p_3 \qquad\qquad (٣٩-٥)$$

در واقع این شرط باید برای حرکات غیر نسبیتی صادق باشد.

برای بررسی رفتار تانسور انرژی-تکانه در این حالت چون در مجاورت $t \to 0$ داریم پس همیشه تقریب زیر بر قرار خواهد بود:

$$\rho \gg \rho u_\alpha u^\alpha \qquad\qquad (٤٠-٥)$$

یعنی برای این که حل کسر برای متریک تحت تاثیر شاره مختل نشود، فقط کافی است که $\rho$ با سرعت کمتری نسبت به جملات خلاء $t^{-2(1+\delta)}$ واگرا شود.

برای بدست آوردن شرط مربوط به این حالت می نویسیم:

$$\rho \propto t^{-\gamma(1+2\delta)} < t^{-2(1+\delta)}$$

$$-\gamma(1+2\delta) > -2(1+\delta)$$

$$(\gamma-2)(1+2\delta) < -2\delta$$

از آنجایی که $(\gamma-2) < 0$ و $1+2\delta > 0$ پس باید داشته باشیم:

$$\delta > 0 \qquad\qquad (٤١-٥)$$

در نتیجه داریم:

$$(\gamma-2)(1+2\delta) < 0$$

نتیجه ای که می توان گرفت این است که برای ماده سخت با $\gamma > \frac{4}{3}$ برای $\delta > 0$ حل کسنری در شرایطی که ماده وجود داشته باشد حل معتبر خواهد بود. بدین معنا که به ماده می توان به شکل یک ذره آزمون نگاه کرد که اثری روی متریک فضا–زمان نداشته و فقط به شکل یک اختلال غیر مؤثر حضور دارد. یعنی در این شرایط متریک کسنر که حل خلاء معادلات اینشتین برای متریک نوع یک بیانکی است می تواند توصیف کننده فضا–زمان با وجود ماده به شکل ذره آزمون باشد. یعنی همین که قید های (۲۰–۵) بر قرار باشند کافی است که ماده اثری ناچیز داشته باشد. برای $\delta < 0$ نمی توان شرط (۲۲–۵) را به دست آورد. در این حالت رفتار عالم با توجه به ضرایب جدید کسنر تغییر می کند بدین شکل که عالم وارد یک فاز جدید از بینهایت نوسان انبساطی و انقباطی آشوبناک از نوع mixmaster شده و از حالت پایدار خود دور می شود.

۵–۳–۳ بررسی شرایط انرژی و انرژی تاریک

در این قسمت قید هایی را که از ملاحضات انرژی می توان بدست آورد را مورد مطالعه قرار می دهیم.

برای این کار از تعریف تانسور انرژی–تکانه موثر و فشار و چگالی موثر که در رابطه (۷–۳) معرفی شـدند استفاده می کنیم. در فصل سوم به این نکته اشاره شد که روابط (۷–۳) بـرای یـک مـدل $f(R)$ بـا متریک فریدمان رابرتسون واکر بدست آمده اند، ولی ما اینجا فضا ناهمسانگرد زمانی در واقع متریـک زمینـه متفاوت است. بنا براین باید روابط (۷–۳) را دوباره برای این حالت خـاص بـاز نویسـی کنـیم. در واقـع در روابط مذکور جملاتی که دارای ثابت هابل $H$ هستند بایسد تغییر کنند.

با اعمال تغییرات روابط جدید عبارت ان از:

$$\rho_{\text{eff}} = -\frac{1}{f'(R)}\left\{\rho + \frac{1}{2}\left[f(R) - Rf'(R)\right] + \mathbf{P}t^{-1}\dot{R}f''(R)\right\} \qquad (۲۲–۵)$$

$$(۲۳–۵)$$

$$P_{\text{eff}} = \frac{1}{3f'(R)}\left\{T - \rho + \frac{3}{2}\left[f(R) - Rf'(R)\right] + 3\ddot{R}f''(R) + 3\dot{R}^2f'''(R) + 2\,\mathbf{P}t^{-1}\dot{R}f''(R)\right\}$$

که تعریف کرده ایم $T = T_\alpha{}^\alpha$ و $\mathbf{P} = \sum_{i=1}^{3} p_i$

۷۱

در اینجا برای نشان دادن ویژگی های خود هندسه از حضور ماده صرف نظر می کنیم یعنی فرض می کنیم $P = \rho = T = 0$ باشد.در نتیجه روابط به شکل زیر ساده می شوند:

$$\rho_{eff} = \frac{1}{f'(R)}\left\{-\frac{1}{2}[f(R) - Rf'(R)] - \boldsymbol{P}t^{-1}\dot{R}f''(R)\right\} \qquad (٤٤-٥)$$

$$\mathrm{P}_{eff} = \frac{1}{f'(R)}\left\{\frac{1}{2}[f(R) - Rf'(R)] + \ddot{R}f''(R) + \dot{R}^2f'''(R) + \frac{2}{3}\boldsymbol{P}t^{-1}\dot{R}f''(R)\right\} (٤٥-٥)$$

در ادامه با داشتن فشار و چگالی انرژی موثر از روابط (۳-۱) تـا (۴-۳) اسـتفاده کـرده و شـرایط انـرژی را بدست می آوریم.

شرط انرژی تهی

این شرط به شکل $\rho_{eff} + P_{eff} \geq 0$ است که با جاگذاری از روابط بالا داریم:

$$(٤٦-٥)$$

$$\rho_{eff} + P_{eff} = \frac{1}{f'(R)}\left\{\ddot{R}f''(R) + \dot{R}^2f'''(R) - \frac{1}{3}\boldsymbol{P}t^{-1}\dot{R}f''(R)\right\} \geq 0$$

با محاسبه اسکالر ریچی برای متریک کسنر و نیز $f(R) = R^{1+\delta}$ رابطه فوق به شکل زیر ساده می شود:

$$4\delta^2 + 2\delta + \frac{2}{3}\delta\boldsymbol{P} \geq 0$$

چون فرض کرده ایم که خلاء داریم از رابطه (۱۸-۵) برای $\boldsymbol{P}$ جاگذاری کرده و نتیجه زیر را بـه دسـت مـی آوریم:

$$\frac{8}{3}\delta(1 + 2\delta) \geq 0 \qquad (٤٧-٥)$$

شرط انرژی ضعیف

برای $\rho_{eff} \geq 0$ از (۴۴-۵) داریم:

$$\frac{1}{1+\delta}\left[\frac{\delta}{2}\boldsymbol{Q} + \delta(1 + 2\delta) + \frac{\delta}{2}\boldsymbol{P}^2\right] \geq 0$$

که تعریف کرده ایم $\boldsymbol{Q} = \sum_{i=1}^{3} p_i^2$. با استفاده از (۱۸-۵) و (۱۹-۵) شرط زیر بدست می آید:

$$2\delta(1 + 2\delta) \geq 0 \qquad (٤٨-٥)$$

پس برای شرط انرژی ضعیف داریم:

$$\begin{cases} \rho_{eff} + P_{eff} \geq 0 & \rightarrow & \frac{8}{3}\delta(1 + 2\delta) \geq 0 \\ \rho_{eff} \geq 0 & \rightarrow & 2\delta(1 + 2\delta) \geq 0 \end{cases} \qquad (٤٩-٥)$$

شرط انرژی قوی

برای $\rho_{eff} + 3P_{eff} \geq 0$ داریم:

$$(٥٠-٥)$$

$$\rho_{eff} + 3P_{eff} =$$
$$\frac{1}{f'(R)}\{f(R) - Rf'(R) + 3\ddot{R}f''(R) + 3\dot{R}^2f'''(R) + Pt^{-1}\dot{R}f''(R)\} \geq 0$$

که بعد از ساده سازی برابر می شود با:

$$\frac{1}{1+\delta}[-\mathbf{Q} - 2\delta^2\mathbf{P} - \delta\mathbf{P}^2 + 6\delta(1 + 2\delta)] \geq 0$$

با جاگذاری برای $\boldsymbol{P}$ و $\boldsymbol{Q}$ نتیجه زیر بدست می آید:

$$4\delta(1 + 2\delta) \geq 0 \qquad (٥١-٥)$$

که در نتیجه برای شرط انرژی قوی خواهیم داشت:

$$\begin{cases} \rho_{eff} + P_{eff} \geq 0 & \rightarrow & \frac{8}{3}\delta(1 + 2\delta) \geq 0 \\ \rho_{eff} + 3P_{eff} \geq 0 & \rightarrow & 4\delta(1 + 2\delta) \geq 0 \end{cases} \qquad (٥٢-٥)$$

شرط انرژی غالب

برای $\rho_{eff} - P_{eff} \geq 0$ داریم:

$$(٥٣-٥)$$

$$\rho_{eff} - P_{eff} =$$

$$\frac{1}{f'(R)}\left\{-f(R) + Rf'(R) - \ddot{R}f''(R) - \dot{R}^2 f'''(R) - \frac{5}{3}\,\mathbf{P}t^{-1}\dot{R}f''(R)\right\} \geq 0$$

که ساده سازی عبارت زیر را بدست می دهد:

$$\frac{1}{3(1+\delta)}\left[3\delta\mathbf{Q} + (10\delta^2 + 4\delta)\mathbf{P} + 3\delta\mathbf{P}^2\right] - 4\delta^2 - 2\delta \geq 0$$

که در نهایت زیر بدست می آید:

$$\frac{4}{3}\delta(1 + 2\delta) \geq 0 \qquad\qquad (\text{۵۴–۵})$$

که در نتیجه برای شرط انرژی غالب خواهیم داشت:

$$\begin{cases} \rho_{eff} \geq 0 & \rightarrow \quad 2\delta(1+2\delta) \geq 0 \\ \rho_{eff} + P_{eff} \geq 0 & \rightarrow \quad \frac{8}{3}\delta(1+2\delta) \geq 0 \\ \rho_{eff} - P_{eff} \geq 0 & \rightarrow \quad \frac{4}{3}\delta(1+2\delta) \geq 0 \end{cases} \qquad (\text{۵۵–۵})$$

بحث

با مرور هر چهار شرط انرژی می بینیم که این شرط ها به صورت شگفت آوری همه به یک قیـد بـه شـکل زیر منجر می شوند:

$$\delta(1 + 2\delta) \geq 0 \qquad\qquad (\text{۵۶–۵})$$

برای ارضاء شدن این قید باید $\frac{1}{2} - \leq \delta$ یا $0 \geq \delta$ باشد. جواب اول مورد توجه نیست زیرا در بخش های قبلی دریافتیم که مدل $R^{1+\delta}$ زمانی حل دقیق خلاء است که $\frac{1}{4} < \delta < \frac{1}{2} -$ باشـد پـس $0 \geq \delta$ نتیجـه نهایی است که از ارضاء شدن شرایط انرژی بدست می آید و به ازاء تمام مقادیر $\delta$ در این بـازه هـر چهـار شرط انرژی ارضاء می شوند.

از سویی دیگر گفته می شود برای داشتن انبساطی شتاب دار لازم است شرط انرژی قوی نقـض شـود، کـه این در مدل ما به معنای وجود قید زیر است:

$$\delta(1 + 2\delta) \leq 0 \qquad\qquad (\text{۵۷–۵})$$

که این شرط فقط به ازاء تمام مقادیر $\delta$ در بازه $0 < \delta < -\frac{1}{2}$ بر قرار است. از طرفی در (۲۰-۵) و توضیحات بعد از آن گفته شد که در این مدل انبساط شتاب دار فقط به ازاء $\delta > -\frac{1}{2}$ به وجود می آید که نتیجه ای که در اینجا بدست آوردیم با این موضوع همخوانی خوبی نشان می دهد.

از طرفی این نتیجه باعث می شود که هیچ کدام از سه شرط انرژی دیگر، ارضاء نشوند. اجازه بدهید مفاهیم چهار شرط انرژی را با فرض ارضاء نشدن این شرط ها مرور کنیم.

ارضاء نشدن دو شرط انرژی ضعیف و تهی به این معنی است که چگالی انرژی ماده باید مقداری نا مثبت باشد. از آنجایی که ما در بررسی مان فقط اثرات گرانشی هندسه را بررسی کردیم، این موضوع بدین معناست که اگر بخواهیم هندسه را عامل انبساط شتابدار عالم کنونی بدانیم باید بپذیریم که هندسه از این حیث نقش یک ماده غیر معمولی را ایفا می کند. ارضاء شدن شرط انرژی قوی به معنای وجود میدان های گرانشی همگرا به عبارتی میدان های گرانشی جاذب است. در اینجا این شرط نقض می شود که به این معناست که میدان گرانشی بجای جاذبه باید دافعه باشند. و در آخر نقض شرط انرژی غالب به معنای این است که انرژی ماده منتسب به هندسه با سرعتی بیشتر از سرعت نور شارش می یابد.

در نتیجه می توانیم فرض کنیم در زمان های اولیه تحول عالم یعنی زمان هایی از تحول عالم که میدان ماده تاثیر غیر قابل توجهی روی میدان گرانشی ناشی از هندسه داشته است، هندسه در نقش یک ماده غیر معمول، یعنی ماده ای با چگالی انرژی منفی، دارای میدان های گرانشی دافعه بجای جاذبه و سرعت انتشار شارش انرژی بیشتر از سرعت نور، عامل ایجاد یک انبساط شتاب دار در عالم بوده است. در قسمت بعدی می خواهیم این نتیجه گیری را رد کنیم.

اگر واقعاً هندسه می تواند چنین نقشی ایفا کند پس باید این موضوع را در معادله حالت خود نشان بدهد. در فصل چهار در رابطه (۱۶-۴) $w_{curv}$ را معرفی کردیم. در این قسمت می خواهیم این کمیت را در مدل $R^{1+\delta}$ و برای متریک کسنر بدست آوریم.

عبارت باز نویسی شده برای $w_{curv}$ عبارت است از:

$$w_{curv} = \frac{\frac{1}{2}[f(R)-Rf'(R)]+\dot{R}f''(R)+\dot{R}^2f'''(R)+\frac{2}{3}Pt^{-1}\dot{R}f''(R)}{-\frac{1}{2}[f(R)-Rf'(R)]-Pt^{-1}\dot{R}f''(R)} \qquad (۵۸-۵)$$

در این قسمت $P_{curv}$ و $\rho_{curv}$ را بطور جداگانه محاسبه می کنیم، نتیجه به شکل زیر است:

$$\rho_{curv} = 2\delta(1 + 2\delta) \; , P_{curv} = \frac{2}{3}\delta(1 + 2\delta) \qquad (\text{۵۹-۵})$$

نتیجه بالا شگفت انگیز است زیرا با نقض شدن شرط انرژی قوی فشار و چگالی انرژی هندسه مـی شوند. نتیجه جالب تر به صورت زیر است:

$$w_{curv} = \frac{1}{3} \qquad (\text{۶۰-۵})$$

که این معادله حرکت ذره ای به جرم صفر است که با سرعت نور حرکت می کند یعنی تابش. منفی شـدن چگالی انرژی یک نتیجه غیر فیزیکی است و در مدل ما به این معنی است که شرط انرژی قوی نباید نقـض شود. این نتیجه دور از انتظار نیست زیرا مدل ناهمسانگرد ما نمی تواند یک انبساط شتابدار داشته باشد زیرا مشاهدات حال حاضر گویای یک همسانگردی با درجه بالا هستند. این یعنی مدل ناهمسانگرد مـا در آینـده خود باید به یک مدل همسانگرد تغییر کند. مدل های ناهمسانگردی که چنین ویژگی را ندارند خیلی مـورد توجه نیستند.

۵-۴ مدل بعدی که به بررسی آن علاقه مندیم چگالی لاگرانژی به شکل زیر دارد[۳۵]:

$$L = \frac{1}{\chi}(R_{\alpha\beta} R^{\alpha\beta})^n \qquad (\text{۶۱-۵})$$

۵-۴-۱ حل معادلات در شرایط خلاء

با حل معادله میدان برای تابع فوق به دو معادله زیر می رسیم:

$$(\text{۶۲-۵})$$

$$Y^{n-1}\big((P^2 + Q - 4PQ + P^2Q + Q^2) -$$
$$2(3P^2 + P^3 + 3Q - 9PQ + 2Q^2)n + (4P^2 + 4Q - 8PQ)\big) = 0$$

$$(\text{۶۳-۵})$$

$$Y^{n-1}\big((24P - 2P^2 - 2P^3 - 30Q +$$
$$10PQ)n + (8P^2 - 32P + 24Q)n^2 + (P^2 - 4P + 2P^3 + 9Q - 10PQ - P^2Q +$$
$$3Q^2)\big) = 0$$

که P و Q را به شکل زیر تعریف کرده ایم:

$$P = \sum_{i=1}^{3} p_i$$

$$Q = \sum_{i=1}^{3} p_i^2$$

که $p_i$ ها ضرایب کسنر هستند.

این معادلات پنج دسته جواب به شکل زیر را دارند:

دسته اول:

$$P = 0 = Q \qquad\qquad (٦٤-٥)$$

این جواب فقط به ازاء $p_i = 0$ بدست می آید. که در واقع بـا ایـن ضـرایب متریـک فضـای مینکوسـکی بدست می آید و در فضا زمان مینکوسکی تانسور ریچی صـفر اسـت در نتیجـه ایـن جـواب بـه ازاء همـه $n \geq 0$ برقرار است.

دسته دوم:

$$P = 1 = Q \qquad\qquad (٦٥-٥)$$

متریکی که این جواب بدست می دهد همان متریک کسنر است و  $p_i$ ها در بـازه هـای زیـر تعریـف مـی شوند:

$$-\frac{1}{3} \leq p_1 \leq 0$$

$$0 \leq p_2 \leq \frac{2}{3}$$

$$\frac{2}{3} \leq p_3 \leq 1$$

از آنجایی که تانسور ریچی برای حل کسری صفر است مجددا این حل فقط به ازاء $n \geq 0$ برقرار است.

دسته سوم:

$$P = \frac{3(1-3n+4n^2) \pm \sqrt{3(-1+10n-5n^2-40n^3+48n^4)}}{2(1-n)} \qquad (66-5)$$

$$Q = \frac{P^2}{3} \qquad (67-5)$$

مشابه مدل قبلی اگر این دو معادله را با هم حل کنیم فقط به ازاء مقادیر برابر $p_i$ ها جواب حقیقی خـواهیم داشت. این دسته از جوابها فقط متریک همسانگرد با خمش صفر خلاء فریدمان-رابرتسون-واکر را بدست می دهند. مقادیر $p_i$ ها همه برابر با $\frac{P}{3}$ می باشد. به این علت که در رابطـه (66-5) بـه ازاء $n = 1$ مخـرج صفر می شود این دسته به ازاء همه $n \neq 1$ معتبر است.

دسته چهارم:

$$P = 4n - 1 \qquad (68-5)$$

$$Q = -3 + 12n - 8n^2 \pm 2(1-2n)\sqrt{2(1-4n+2n^2)} \qquad (69-5)$$

این دسته جواب به ازاء همه $n \neq \frac{1}{4}$ (به ازاء این مقدار هر دو رابطه بالا صفر شده و دوباره حـل دسـته اول بدست می آید) مقادیری موهومی برای ضرایب کسر بدست می دهد. در واقع $Q$ همواره مقادیری منفی را اختیار می کند.

دسته پنجم:

$$P = (1-2n)^2 \qquad (70-5)$$

$$Q = 1 - 8n(n-1)^2 \qquad (71-5)$$

این تنها جوابی است که متریکی نا همسانگرد را معرفی مـی کنـد. زیبـایی ایـن جـواب در ایـن اسـت کـه در $n = 1$ همان حل کسر در خلاء نسبیت عام اینشتین بدست می آید. بـه روش توصـیف شـده در مـدل قبلی بازه هایی که ضرایب کسر در آنها تعریف می شوند عبارت اند از:

$$(1-2n)^2 - 2A \leq 3p_1 \leq (1-2n)^2 - A \qquad \text{(الف –۷۲–۵)..}$$

$$(1-2n)^2 - A \leq 3p_2 \leq (1-2n)^2 + A \qquad \text{(ب –۷۲–۵)}$$

$$(1-2n)^2 + A \leq 3p_3 \leq (1-2n)^2 + 2A \qquad \text{(ج –۷۲–۵)}$$

که A را به شکل زیر تعریف کرده ایم:

$$A \equiv \sqrt{(1-2n)(1-6n+4n^3)}$$

مثل قبل بدون از دست دادن عمومیت مساله فرض کرده ایم که رابطه (۲۱–۵) همچنان بر قـرار اسـت. ایـن سه بازه به ازاء n های مختلف برای ضرایب $p_i$ رسم شده است.

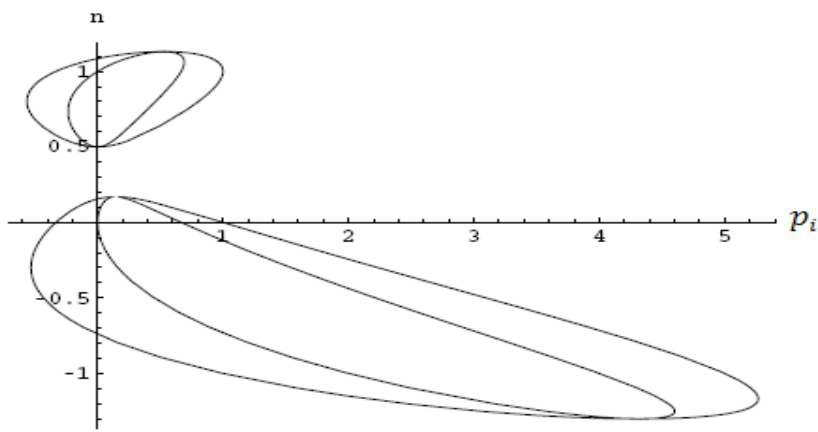

منحنی ۵–۲ منحنی جواب های مربوط به ضرایب تعمیم یافته کسری برای مدل $L = \frac{1}{\chi}(\mathbf{R}_{\alpha\beta}\mathbf{R}^{\alpha\beta})^n$

که مثل مدل قبلی با رسم خط ثابت $n =$ می توانیم به ازاء این n سه بازه عددی برای ضرایب $p_i$ بدسـت آوریم.

یک تحلیل کوتاه از ضریب A جهت درک کیفی بهتر از نمودار فوق شایسته است. اگر توابـع زیـر را تعریـف کنیم:

$$f \equiv 1 - 2n$$

$$g \equiv 1 - 6n + 4n^3$$

دامنه توابع فوق عبارتند از:

$$f \rightarrow \begin{cases} n \leq \dfrac{1}{2} & \text{مثبت} \\ n > \dfrac{1}{2} & \text{منفی} \end{cases}$$

$$g \rightarrow \begin{cases} n_1 \leq n \leq n_2 \quad , \quad n \geq n_3 & \text{مثبت} \\ n_2 \leq n \leq n_3 \, , \, n \leq n_1 & \text{منفی} \end{cases}$$

با توجه به اینکه برای معنی دار بودن $A$ باید زیر رادیکال مثبت باشد باید همواره $fg \geq 0$ یعنی یـا هـر دو به ازاء دامنه ای معین منفی یا هر دو باید مثبت باشند. با توجه بـه ایـن توضیحات مقـادیر مجـاز بـرای $n$ عبارت است از:

$$n_1 \leq n \leq n_2 \qquad\qquad (\text{۵}-\text{۷۳})$$

$$\frac{1}{2} \leq n \leq n_3 \qquad\qquad (\text{۵}-\text{۷٤})$$

($n_1 < n_2 < n_3$). با توجه به رابطه (۷۰-۵) $P$ همواره مثبت است محدوده های تعیین شده در (۷۳-۵) و (۷٤-۵) همواره یک عالم انبساط یابنده را معرفی می کنند.

با بررسی این پنج دسته جواب برای مدل فوق برای حالت خلاء به این نتیجه رسیدیم که مـدل فـوق دارای یک جواب دقیق نا همسانگرد می باشد که به ازاء $n$ های یافت شده در(۷۳-۵) و (۷٤-۵) معتبر است.

## ۵-٤-۲ حل در مجاورت ماده

حال این مدل را در حضور ماده با تحلیلی مشابه با مدل قبلی بررسی می کنیم. با توجه به ایـن کـه در مـدل قبلی این حالت به صورت مشروح توضیح داده شد از شرح مجدد جزئیات خودداری کرده وتا حـد امکـان نتایج را مطالعه می نماییم.

در این مدل تابعیت جملات میدان در حالت خلاء به صورت زیر می باشد:

جملات میدان بدون حضور ماده$\sim O(t^{-4})$

۸۰

با توجه به توضیحات داده شده در قبل برای اینکه جملات میدان بر جملات تانسور انرژی-تکانه غلبه کنـد از مولفه $T_3^3$ استفاده می کنیم:

برای مقادیر $\gamma < \frac{4}{3}$ در سیستم غیر همراه داریم:

$$-P - p_3 > -4n$$

$$-(1 - 2n)^2 - p_3 > -4n$$

$$p_3 < -1 + 8n - 4n^2 \qquad (\text{۵}-\text{۷۵})$$

و در سیستم همراه برای $\gamma = \frac{4}{3}$ داریم:

$$\rho \propto t^{-2(p_1 + p_2)} = t^{-2(P - p_3)} < t^{-4n}$$

$$-2(P - p_3) > -4n$$

$$p_3 > 1 - 6n + 4n^2 \qquad (\text{۵}-\text{۷۶})$$

حال با توجه به این موضوع که برای این که پاسخ ها (یعنی بازه هـایی کـه بـرای n مجـاز هسـتند) بـدون حضور ماده یعنی (۵-۷۳) و (۵-۷۴) زیر مجموعه ای از پاسخ ها در حضور ماده باشند ما در جستجوی این جواب ها هستیم. در واقع این بدین معنی است که با توجه به توضیحات داده شـده در ابتـدای ایـن فصـل حضور ماده که فرض می شود تاثیر مهمی روی تحول متریک فضازمان ندارد نبایـد جـواب هـای جدیـد نادرستی بدهد. نادرست بدین معنی که جواب ها بدون حضور ماده بایـد حـالات خاصـی از جـواب هـای جدید باشند. یعنی برای این که جواب قدیم وجواب جدید دارای حداقل n های مشترک باشند باید بیشینه $p_3$ در جواب جدید طبق رابطه (۵-۷۵) از بیشینه جواب در جواب قـدیم طبـق رابطـه (۵-۷۲-ج) بزرگتـر باشد یعنی:

$$-1 + 8n - 4n^2 > \frac{(1-2n)^2 + 2\sqrt{(1-2n)(1-6n+4n^3)}}{3}$$

ما به دنبال n هایی هستیم که شرط فوق را ارضاء کنند. با ساده سازی عبارت بالا به رابطه زیر می رسیم:

$$-2 + 14n - 8n^2 - \sqrt{(1-2n)(1-6n+4n^3)} > 0$$

در حالت خلاء ما دو بازه مجاز یعنی جواب های (۵-۷۳) و (۵-۷۴) را برای $n$ بدست آوردیم. اگر مقادیر تعریف شده $n$ در (۵-۷۴) را در (۵-۷۵) و (۵-۷۶) قرار دهیم مقادیری برای $p_3$ بدست می آوریم که با مقایسه با مقادیر خوانده شده از نمودار ۵-۲ همواره درست هستند. مثلاً برای $n \sim \frac{1}{2}$ حدودا داریم:$2 < p_3 < 1-$ ولی با مقایسه با نمودار می بینیم که جواب خلاء برای $n \sim \frac{1}{2}$ همواره بین صفر و یک است. یعنی جواب با حضور ماده جوابی منطقی است.

از سویی دیگر چنین اعتباری همواره برای جواب دیگر خلاء یعنی (۵-۷۳) وجود ندارد. می بینیم که با اضافه شدن ماده به مساله این بازه محدود تر می شود. به نامعادله ای که بدست آوردیم بر می گردیم. برای ارضاء نامعادله فوق باید داشته باشیم:

$$n > \frac{1}{6}$$

در نتیجه بازه $n_1 \le n \le n_2$ برای ماده های با $\gamma < \frac{1}{4}$ به شکل زیر تصحیح می شود:

$$\frac{1}{6} < n < n_2$$

و برای ماده هایی با $\gamma = \frac{1}{4}$ برای این که جواب قدیم و جواب جدید دارای حداقل $n$ های مشترک باشند باید کمینه $p_3$ در جواب جدید طبق رابطه (۵-۷۶) از کمینه جواب در جواب قدیم طبق رابطه (۵-۷۲-ج) کوچکتر باشد یعنی:

$$1 - 6n + 4n^2 < \frac{(1-2n)^2 + \sqrt{(1-2n)(1-6n+4n^3)}}{3}$$

که بعد از ساده سازی داریم:

$$2 - 14n + 8n^2 - \sqrt{(1-2n)(1-6n+4n^3)} < 0$$

که این نامساوی برای بازه زیر ارضاء می شود:

$$n > n_4$$

($n_4 \cong 0.115$) در نتیجه بازه $n_1 \le n \le n_2$ برای ماده های با $\gamma = \frac{1}{4}$ به شکل زیر تصحیح می شود:

$$n_4 < n < n_2$$

بنا براین بازه های مجاز برای $n$ برای اینکه ماده با $\gamma < \frac{4}{3}$ روی متریک توصیف کننده فضا زمان تاثیر ناچیزی داشته باشد عبارت اند از:

$$\frac{1}{2} \leq n \leq n_3 \qquad (٦٥-٥)$$

$$\frac{1}{6} < n < n_2 \qquad (٧٧-٥)$$

و بازه های مجاز $n$ برای $\gamma = \frac{4}{3}$ عبارت اند از:

$$\frac{1}{2} \leq n \leq n_3 \qquad (٦٥-٥)$$

$$n_4 < n < n_2 \qquad (٧٨-٥)$$

حال به تحلیلی مشابه برای ماده سخت یعنی برای $\gamma > \frac{4}{3}$ می پردازیم. همانطور که گفته شد برای این ماده $u_0 \to 1$ و برای مولفه های فضایی چار بردار $u_a u^a \to 0$ یعنی ما باید دنبال $\rho$ و $u_a$ جدید باشیم. این مقادیر با جا گذاری $P$ بجای $1 + 2\delta$ عبارت اند از:

$$\rho \sim t^{-\gamma P}$$

$$u_a \sim t^{(\gamma-1)P}$$

با استدلالی مشابه در مدل قبلی برای ماده سخت شرطی که در آن تقریب سرعت غالب با شکست مواجه می شود عبارت است از:

$$p_3 < (\gamma - 1)P$$

تا اینجا این روابط همان روابط (٥-٣٧), (٥-٣٨) و (٥-٣٩) هستند که برای مدل جدید باز نویسی شده اند. برای اینکه در این حالت مولفه های تانسور انرژی تکانه با سرعت کمتری نسبت به جملات میدان واگرا شوند همان طور که گفتیم باید شرط زیر برقرار باشد:

مرتبه بزرگی جملات میدان در حالت خلاء $< \rho$

$$t^{-\gamma P} < t^{-4n}$$

$$-\gamma P < -4n$$

٨٣

$$\gamma < \frac{4n}{(1-2n)^2} \qquad (٥-٧٩)$$

در بازه $\frac{1}{2} < n < n_3$ تمام مقادیر $n$ وقتی که در (٥-٧٩) قرار داده شوند همواره جواب هایی بـرای $\gamma$ بدست می دهند که تمام بازه $2 < \gamma$ در آن جواب ها قرار دارد. یعنی جواب خلاء (٥-٧٤) با حضور مـاده سخت دست نخورده باقی می ماند. ولی دوباره جواب (٥-٧٣) مشکل سـاز اسـت. یعنی همـه $n$ هـای موجود در (٥-٧٣) با (٥-٧٩) سازگاری ندارند. مثلاً اگر $n_1 \sim 1.3 -$ را در (٥-٧٩) قـرار دهـیم جواب تقریبی $\gamma < -0.4$ را بدست می آوریم که با فرض اولیـه $1 < \gamma < 2$ در تضاد مـی باشـد. پـس بایـد جواب (٥-٧٣) تصحیح شود. برای تصحیح این بازه ما از این شرط که برای مـاده سـخت همـواره $\gamma > \frac{4}{3}$ است استفاده می کنیم. با توجه به (٥-٧٩) داریم:

$$\gamma_{max} \sim \frac{4n}{(1-2n)^2}$$

$$\gamma_{max} > \frac{4}{3}$$

با حل این نامعادله می بینیم که رابطه فوق فقط به ازاء دو مقدار $r_{1,2} = \frac{7 \pm \sqrt{33}}{8}$ که ریشه های نامعادله زیر هستند صادق است:

$$1 - 7n + 4n^2 = 0$$

یعنی حد پایین جواب (٥-٧٣) در حالتی که ماده سخت حضـور دارد یکـی از دو ریشـه فـوق اسـت. بـازه بزرگ تر برای (٥-٧٣) برای ریشه کوچک تر به شکل زیر تصحیح می شود:

$$\frac{7-\sqrt{33}}{8} < n < n_2 \qquad (٥-٨٠)$$

در نهایت بازه های مجاز برای $n$ برای اینکه ماده سخت روی تحول متریک توصیف کننده فضا زمان تـاثیر ناچیزی داشته باشد عبارت اند از:

$$\frac{1}{2} \le n \le n_3 \qquad (٥-٦٥)$$

$$\frac{7-\sqrt{33}}{8} < n < n_2 \qquad (٥-٨٠)$$

۵-۵ آخرین مدل مورد بررسی ما چگالی لاگرانژی زیر است[۳۵]:

$$L_G = \frac{1}{\chi} \sqrt{-g} (R_{\alpha\beta\mu\nu} R^{\alpha\beta\mu\nu})^n \qquad (۵-۸۱)$$

بعد از حل معادلات میدان و با جاگذاری عناصر متریک تعمیم یافته کسر به دو معادله قیدی زیر می رسیم:

$$(۵-۸۲)$$

$$Z^{n-1}\left(\left(\frac{P^2Q}{2} - \frac{P^4}{12} - Q - \frac{3Q^2}{4} + 2S - \frac{2PS}{3}\right) + \left(\frac{P^4}{3} + 6Q + 2PQ - 2P^2Q + 3Q^2 - 10S + \frac{2PS}{3}\right)n + 8(S - Q)n^2\right) = 0$$

$$(۵-۸۳)$$

$$Z^{n-1}\left(\left(\frac{P^4}{4} - P - \frac{3P^2}{2} - \frac{P^3}{2} + \frac{13Q}{2} + \frac{7PQ}{2} - \frac{3P^2Q}{2} + \frac{9Q^2}{4} - 8S\right) + (6P + 4P^2 - 16Q - 2PQ + 8S)n + 8(Q - P)n^2\right) = 0$$

که $Z \equiv R_{\alpha\beta\mu\nu} R^{\alpha\beta\mu\nu}$ می باشد.

که در دو قید بالا $P$ و $Q$ تعریف قبل را دارند ولی $S$ به شکل زیر تعریف می شود:

$$S \equiv \sum_{i=1}^{3} p_i^3$$

دو معادله بالا چهار دسته جواب به شکل زیر دارند:

دسته اول:

این جواب متعلق است به:

$$P = 0 = Q = S \qquad (۵-۸٤)$$

این حل متریک فضا زمان مینکوسکی را بدست می دهد. چون برای فضا زمان مینکوسکی مولفه های متریک ثابتند پس همه ستارها صفر و در نتیجه تانسور ریمان صفر می باشد. یعنی در این حل $Z = 0$. و در نتیجه برای واگرا نشدن چگالی لاگرانژی این جواب برای $n > 0$ معتبراست.

دسته دوم:

جواب های متعلق به این دسته حل با

$$P = 1 = Q = S \qquad (\Lambda\Delta-\Delta)$$

داده می شود. برای اینکه نتیجه فوق همیشه برقرار باشد باید همواره یکی از ضرایب متریک تعمیم یافته متریک مقدار یک و دو ضریب دیگر مقدار صفر داشته باشد. این بدین معنی است که متریک ما فقط در راستای یک بعد تحول زمانی داشته و در دو راستای دیگر دارای ضرایب ثابت یک می باشد. فرض می کنیم

$$p_1 = 0 = p_2, p_3 = 1$$

باشد. در این صورت متریک به شکل زیر در می آید:

$$dS^2 = -dt^2 + dx^2 + dy^2 + t^2 dz^2$$

تبدیل مختصات به شکل زیر را استفاده می کنیم:

$$\bar{z} = t \sinh z$$

$$\bar{t} = t \cosh z$$

در این صورت با تبدیلات بالا شکل نهایی متریک عبارت است از:

$$dS^2 = -d\bar{t}^2 + dx^2 + dy^2 + d\bar{z}^2$$

این متریک مربوط به فضای مینکوسکی است. بنابراین این حل به دسته اول تعلق دارد و به ازاء $n > 0$ معتبراست.

دسته سوم:

در این دسته داریم:

$$P = \frac{3(1-2n+4n^2 \pm \sqrt{-1+10n-16n^2+16n^4}}{2(1-n)} \qquad (\Lambda\mathcal{F}-\Delta)$$

$$Q = \frac{P^2}{3}$$

$$S = \frac{P^3}{9}$$

ما قبلا با این گونه حل آشنا شدیم. این دسته فقط به ازاء مقادیر برابر ضرایب $p_i$ جواب های غیر موهومی

برای $p_i$ ها بدست می دهد. یعنی اگر این سه معادله را با هم حل کنیم، برای همه $p_i = \frac{P}{3}$ ، $n \neq 1$ برای خواهند بود.

دسته چهارم:

$$P = 4n - 1 \qquad\qquad (\text{۸۷–۵})$$

$$Q = \frac{1}{3}\{16n^2 - 8n - 1 \pm 4\sqrt{2[n(1-2n) + S(1-n)]}\}$$

در این دسته ما مقادیر موهومی برای $p_i$ ها بدست می آوریم و در نتیجه این جواب مورد توجه ما نیست.

جمع بندی

در این پایان نامه وجود برخی حل های دقیق کیهان شناسی ناهمسانگرد در نظریه گرانش مرتبه بالا مورد بررسی قرار گرفت. در حالی که نظریه استاندارد نسبیت عام که توسط کنش اینشتین-هیلبرت تعریف می شود با آزمایشات میدان ضعیف همخوانی دارد، با وجود این دلیل کمتری وجود دارد که چنین رفتاری در موقعیت هایی با خمش بالا، مثلاً در مجاورت با تکینگی اولیه کیهانی وجود داشته باشد. در واقع در حد خمش های بالاست که اثرات کوانتومی مهم می شوند و باید انحرافی را در نظریه استاندارد نسبیت عام انتظار داشت.

بدون دانشی در مورد شکل دقیق این انحراف ها، ما این مساله را با بررسی یک رده عمومی از نظریه هایی که می توانند از یک تابع تحلیلی اختیاری از سه کمیت اسکالر $R$، $R_{\mu\nu}R^{\mu\nu}$ و $R_{\alpha\beta\mu\nu}R^{\alpha\beta\mu\nu}$ مشتق شوند مورد تحقیق قرار داده ایم. با بسط این تابع در حوالی تکینگی بر حسب توان هایی از این سه متغیر، انتظار می رود جملات غالب در بسط به شکل $R^n$، $(R_{\mu\nu}R^{\mu\nu})^n$ یا $(R_{\alpha\beta\mu\nu}R^{\alpha\beta\mu\nu})^n$ باشد. در این کار تمام حل های کسری موجود مربوط به این نظریه ها زیر استخراج شده اند. در مدل هایی از نوع $R^n$ و $(R_{\mu\nu}R^{\mu\nu})^n$ حل ناهمسانگرد از نوع حل کسنر در خلاء یافت شد. ولی برای مدل هایی از نوع $(R_{\alpha\beta\mu\nu}R^{\alpha\beta\mu\nu})^n$ چنین حل هایی یافت نشد.

از سویی دیگر هر حل مربوط به خلاء ممکن است با اضافه شدن ماده به مساله تغییر کند. یعنی مثلاً در مدل هایی که در حل خلاء دارای تکینگی هستند، ممکن است بعد از اضافه شدن ماده به مساله ویژگی های این تکینگی تغییر کند. در بعضی مواقع ممکن است تکینگی از بین برود. وقتی از تکینگی در حل خلاء صحبت می شود در واقع از تکینگی در هندسه مدل صحبت می کنیم. این در حالی است که مدلی خوب بشمار می آید که این تکینگی بعد از اضافه شدن ماده همچنان ویژگی های خود را حفظ کند یعنی در واقع در این صورت تکینگی یک مفهوم فیزیکی بشمار می آید.

در این کار به این موضوع پرداخته شده است که چه شرایطی لازم است ارضاء شود تا تکینگی موجود در این مدل ها در حضور ماده همچنان ویژگی های خود را حفظ کند. شرط های لازم با فرض آنکه در مجاورت تکینگی سرعت واگرایی جملات تانسور انرژی تکانه از سرعت جملات میدان خلاء کمتر باشد به دست آمده اند. در این شرایط با توجه به قید هایی که بدست می آید، ماده رفتاری شبیه ذره آزمون داشته و در

واقع تاثیری روی میدان مساله نخواهد داشت. عملاً چنین تقریبی را در فیزیک کلاسیک نیز داریم. مثلاً در بررسی مساله مربوط به رفتار ذره آزمون در حضور میدان الکتریکی فرض می شود بار ذره اغتشاشی روی میدان الکتریکی خارجی ایجاد نمی کند. همچنین وقتی رفتار یک جرم نقطه ای در میدان گرانشی خارجی مورد نظر است فرض می شود میدان گرانشی جرم نقطه ای در مقابل میدان خارجی ناچیز است.

همه جواب های یافت شده در حضور ماده نسبیتی و تابش جواب های خوش رفتاری هستند، ولی برخی از آن ها در حضور ماده غیر نسبیتی خوش رفتاری از خود نشان نمی دهند. در جواب های خوش رفتار، متریک کسنر همراه با قیدهایی بدست می آید یک حل مناسب حتی در حضور ماده است. به یاد داشته باشیم که این متریک حل دقیق خلاء است.

همچنین در مورد مدل $R^{1+\delta}$، شرایط انرژی را مورد بررسی قرار داده ایم. مشاهده می شود که اگر شرایط انرژی در مورد این مدل از گرانش های مرتبه بالاتر با متریک زمینه ناهمسانگرد کسنر بکار رود تمام چهار شرط انرژی به یک قید $\delta(2\delta + 1) \geq 0$ برای $\delta$ منجر می شوند. که در نتیجه اگر $\delta \geq 0$ باشد هندسه نقش یک ماده معمولی با چگالی انرژی مثبت، میدان گرانشی جاذب و سرعت انتشار انرژی کمتر از سرعت نور را دارد.

از سویی دیگر می دانیم که برای داشتن انبساط شتاب دار در مدل فریدمان-رابرتسون-واکر باید شرط انرژی قوی نقض شود. در نسبیت عام نیز این شرط برای یک مدل ناهمسانگرد از نوع بیانلی نوع اول برقرار است. در نتیجه با فرض درست بودن این موضوع برای مدل گرانشی ما، یعنی با فرض اینکه در اینجا با نقض شرط انرژی قوی عالمی با انبساط شتاب دار خواهیم داشت، یک جواب غیر فیزیکی بدست می آوریم. در اینجا چگالی انرژی باید منفی باشد. در نتیجه ما نقض شرط انرژی را رها کرده و به ارضاء شدن آن اکتفا کردیم. ارضاء شدن این شرط به معنای این است که انبساط ناهمسانگرد در این مدل کند شونده است. تعبیر فیزیکی که برای توجیه این *انبساط کند شونده ناهمسانگرد* می توان یافت این است که این نتیجه قابل انتظار است زیرا این خود به معنای از بین رفتن ناهمسانگردی و احتمالاً تغییر سیمای عالم ناهمسانگرد مساله به یک عالم همسانگرد می تواند باشد.

ولی در فصل چهارم ذکر کردیم که برای ماده ای که می تواند عامل انبساط شتاب دار باشد باید داشته باشیم $w < -\frac{1}{3}$ ولی بعد از محاسبه دیدیم که در مدل ما $w = \frac{1}{3}$ بدست می آید که این نتیجه در بازه $\delta \geq 0$ معتبر است. به عنوان نتیجه نهایی می توانیم بگوییم که مدل $R^{1+\delta}$ با متریک زمینه کسنر در بازه $\delta \geq 0$ چند ویژگی را دارد که عبارت اند از:

الف) در این بازه یک حل خلاء است.

ب) در حضور تمام ماده های تقریبا نسبیتی و بسیار نسبیتی و نیز ماده های غیر نسبیتی برای دوره هایی از تحول عالم یک حل تقریبی خوب در اطراف تکینگی اولیه است.

ج) در این بازه یک عالم با انبساط شتاب دار ولی با شتاب منفی را معرفی می کند.

د) به ازاء تمام مقادیر $\delta$ در این بازه، از معادله حالت تابش تبعیت می کند.

ضمیمه الف

ورش کنش $S = \int d^4x \mathcal{L}$ با $\mathcal{L} = \sqrt{-g}L = \sqrt{-g}R^2$ به روش متریکی

کنش به شکل زیر است:

$$S = \int d^4x \mathcal{L} = \int d^4x \sqrt{-g}R^2 \qquad (1)$$

$$\delta S = \int d^4x \delta \mathcal{L}$$

$(2)$

$$\delta \mathcal{L} = \delta \left\{ \sqrt{-g}R^2 \right\} = R^2 \delta(\sqrt{-g}) + 2\sqrt{-g}R\delta R = -\frac{1}{2}\sqrt{-g}R^2 g_{\alpha\beta}\delta g^{\alpha\beta} + 2\sqrt{-g}R\delta R$$

حالا به محاسبه $\delta R$ به طور جداگانه می پردازیم.

$$\sqrt{-g}R\delta R = \sqrt{-g}R\delta(g^{\mu\nu}R_{\mu\nu}) = \sqrt{-g}RR_{\mu\nu}\delta g^{\mu\nu} + \sqrt{-g}g^{\mu\nu}R\delta R_{\mu\nu} =$$

$$\sqrt{-g}RR_{\mu\nu}\delta g^{\mu\nu} + \frac{1}{2}\sqrt{-g}g^{\mu\nu}g^{\alpha\beta}R[\delta g_{\mu\alpha;\nu\beta} + \delta g_{\nu\beta;\mu\alpha} - \delta g_{\mu\nu;\alpha\beta} - \delta g_{\alpha\beta;\mu\nu}] =$$

$$\sqrt{-g}RR_{\mu\nu}\delta g^{\mu\nu} + \sqrt{-g}g^{\mu\nu}g^{\alpha\beta}R[\delta g_{\mu\alpha;\nu\beta} - \delta g_{\alpha\beta;\mu\nu}]$$

دوباره رابطه (۲) را می نویسیم:

$(3)$

$$\delta \mathcal{L} = \delta \left\{ \sqrt{-g}R^2 \right\} = R^2 \delta(\sqrt{-g}) + 2\sqrt{-g}R\delta R = -\frac{1}{2}\sqrt{-g}R^2 g_{\alpha\beta}\delta g^{\alpha\beta} + 2\sqrt{-g}RR_{\mu\nu}\delta g^{\mu\nu} + 2\sqrt{-g}g^{\mu\nu}g^{\alpha\beta}R[\delta g_{\mu\alpha;\nu\beta} - \delta g_{\alpha\beta;\mu\nu}]$$

حالا جمله آخر را ساده می کنیم:

$$\sqrt{-g}g^{\mu\nu}g^{\alpha\beta}R[\delta g_{\mu\alpha;\nu\beta} - \delta g_{\alpha\beta;\mu\nu}] = \sqrt{-g}[g^{\mu\nu}g^{\alpha\beta}R_{;\nu\beta}\delta g_{\mu\alpha} - g^{\mu\nu}g^{\alpha\beta}R_{;\mu\nu}\delta g_{\alpha\beta}]$$

۹۱

در ضمیمه مربوط به وردش متریکی مدل $\mathcal{L} = \sqrt{-g}R^{\mu\nu}R_{\mu\nu}$ روشی که در خط بالا استفاده کردیم توضیح داده شده است و در اینجا از توضیح مجدد آن خودداری می شود. حالا در ادامه روابط خط بالا را برحسب $\delta g^{\alpha\beta}$ باز نویسی می کنیم:

$$\sqrt{-g}[g^{\mu\nu}g^{\alpha\beta}R_{;\nu\beta}\delta g_{\mu\alpha} - g^{\mu\nu}g^{\alpha\beta}R_{;\mu\nu}\delta g_{\alpha\beta}] = \sqrt{-g}g^{\mu d}g^{bc}R_{;dc}\delta g_{\mu b} - $$
$$\sqrt{-g}g^{d\nu}g^{bc}R_{;d\nu}\delta g_{bc} = $$
$$-\sqrt{-g}g^{\mu d}g^{bc}R_{;dc}g_{\beta b}g_{\alpha\mu}\delta g^{\alpha\beta} + \sqrt{-g}g^{d\nu}g^{bc}R_{;d\nu}g_{\beta c}g_{\alpha b}\delta g^{\alpha\beta} = $$
$$-\sqrt{-g}R_{;dc}\delta_\alpha{}^d\delta_\beta{}^c\delta g^{\alpha\beta} + \sqrt{-g}g^{d\nu}g_{\alpha\beta}R_{;d\nu}\delta^b{}_\beta\delta g^{\alpha\beta} = \sqrt{-g}[g^{d\nu}g_{\alpha\beta}R_{;d\nu} - $$
$$R_{;\alpha\beta}]\delta g^{\alpha\beta} = \sqrt{-g}[g_{\alpha\beta}\Box R - R_{;\alpha\beta}]\delta g^{\alpha\beta}$$

حال عبارت فوق را در رابطه (۳) جاگذاری می کنیم:

$$\delta\mathcal{L} = \delta\left\{\sqrt{-g}R^2\right\} = R^2\delta\left(\sqrt{-g}\right) + 2\sqrt{-g}R\delta R = -\frac{1}{2}\sqrt{-g}R^2 g_{\alpha\beta}\delta g^{\alpha\beta} + $$
$$2\sqrt{-g}RR_{\mu\nu}\delta g^{\mu\nu} + 2\sqrt{-g}[g_{\alpha\beta}\Box R - R_{;\alpha\beta}]\delta g^{\alpha\beta} = \sqrt{-g}\left\{-\frac{1}{2}R^2 g_{\alpha\beta} + 2RR_{\mu\nu} + \right.$$
$$\left. 2g_{\alpha\beta}\Box R - 2R_{;\alpha\beta}\right\}\delta g^{\alpha\beta}$$

$$\delta S = \int d^4x \delta\mathcal{L} = \int d^4x\sqrt{-g}\left\{-\frac{1}{2}R^2 g_{\alpha\beta} + 2RR_{\mu\nu} + 2g_{\alpha\beta}\Box R - 2R_{;\alpha\beta}\right\}\delta g^{\alpha\beta} = $$
$$0$$

$$-\frac{1}{2}R^2 g_{\alpha\beta} + 2RR_{\mu\nu} + 2g_{\alpha\beta}\Box R - 2R_{;\alpha\beta} = 0 \qquad (\text{٤})$$

معادله فوق معدله میدان مدل معرفی شده در روش وردشی متریکی است.

ضمیمه ب

ورودش کنش $S = \int d^4x \mathcal{L}$ با $\mathcal{L} = \frac{1}{\chi}\sqrt{-g}L = \sqrt{-g}R^{\mu\nu}R_{\mu\nu}$ به روش متریکی

برای شروع ابتدا چند رابطه را که بعدا از آنها استفاده خواهیم کرد را معرفی می کنیم.

با استفاده از اینکه مشتق همموردای تانسور اینشتین باید صفر باشد دو رابطه را به شکل زیر بدست خواهیم

آورد:

$$\nabla_\mu G^{\mu\nu} = \nabla_\mu \left(R^{\mu\nu} - \frac{1}{2}Rg^{\mu\nu}\right) = 0$$

$$\nabla_\mu R^{\mu\nu} = \frac{1}{2}g^{\mu\nu}\nabla_\mu R = \frac{1}{2}\nabla^\nu R$$

$$\nabla_\nu \nabla_\mu R^{\mu\nu} = \frac{1}{2}\nabla_\nu \nabla^\nu R \equiv \frac{1}{2}\Box R \qquad (1)$$

$$\nabla^\mu R_{\mu\nu} = \frac{1}{2}\nabla_\nu R$$

$$\nabla_\alpha \nabla^\mu R_{\mu\nu} = \frac{1}{2}\nabla_\alpha \nabla_\nu R \qquad (2)$$

وآخرین رابطه ای که معرفی می کنم رابطه زیر است که از اتحاد دوم بیانکی[1] بدست می آید:

$$R_{\alpha\mu}R^\mu_{\ \beta} = R_{\alpha c;\beta}^{\quad c} - \frac{1}{2}R_{;\alpha\beta} + R_{\alpha\rho\beta\sigma}R^{\rho\sigma} \qquad (3)$$

$$\delta R_{\mu\nu} = \frac{1}{2}g^{\alpha\beta}[\delta g_{\mu\alpha;\nu\beta} + \delta g_{\nu\beta;\mu\alpha} - \delta g_{\mu\nu;\alpha\beta} - \delta g_{\alpha\beta;\mu\nu}] \qquad (4)$$

که در روابط بالا علامت ; به معنی مشتق همموردا است یعنی $A_{;\alpha} \equiv \nabla_\alpha A$ و برای شاخص های بالا

$$A_{;}^{\ \alpha} = \nabla^\alpha A$$

---

[1] second Bianchi identity

دو رابطه زیر را نیز داریم:

$$\delta g_{\alpha\beta} = -g_{\alpha\mu} g_{\beta\nu} \delta g^{\mu\nu} \qquad (\text{٥})$$

$$\delta\left(\sqrt{-g}\right) = -\frac{1}{2}\sqrt{-g} g_{\alpha\beta} \delta g^{\alpha\beta} \qquad (\text{٦})$$

حالا به وردش کنش می پردازیم. کنش به شکل زیر است:

$$S = \int d^4x \mathcal{L} = \frac{1}{\chi} \int d^4x \sqrt{-g} R^{\mu\nu} R_{\mu\nu} \qquad (\text{٧})$$

$$\delta S = \int d^4x \delta\mathcal{L}$$

$$\delta\mathcal{L} = \delta\left\{\sqrt{-g} R^{\mu\nu} R_{\mu\nu}\right\} = R^{\mu\nu} R_{\mu\nu} \delta\left(\sqrt{-g}\right) + \sqrt{-g} R_{\mu\nu} \delta(R^{\mu\nu}) + \sqrt{-g} R^{\mu\nu} \delta(R_{\mu\nu})$$

(فرض کرده ایم $\frac{1}{\chi}$ برابر یک است) حال برای طولانی نشدن مساله هریک از جمله ها را جداگانه محاسبه

می کنیم. ابتدا جمله سوم راحساب می کنیم:

$$\sqrt{-g} R^{\mu\nu} \delta(R_{\mu\nu}) = \frac{1}{2}\sqrt{-g} R^{\mu\nu} g^{\alpha\beta}\left[\delta g_{\mu\alpha;\nu\beta} + \delta g_{\nu\beta;\mu\alpha} - \delta g_{\mu\nu;\alpha\beta} - \delta g_{\alpha\beta;\mu\nu}\right]$$

$$(\text{٨})$$

$$= \frac{1}{2}\sqrt{-g}\left\{R^{\mu\nu} g^{\alpha\beta} \delta g_{\mu\alpha;\nu\beta} + R^{\mu\nu} g^{\alpha\beta} \delta g_{\nu\beta;\mu\alpha} - R^{\mu\nu} g^{\alpha\beta} \delta g_{\mu\nu;\alpha\beta} - R^{\mu\nu} g^{\alpha\beta} \delta g_{\alpha\beta;\mu\nu}\right\}$$

حالا هریک از جمله های عبارت فوق به شکل زیر ساده می شوند، مثلاً:

$$R^{\mu\nu} g^{\alpha\beta} \delta g_{\mu\alpha;\nu\beta} = \left[(R^{\mu\nu} g^{\alpha\beta} \delta g_{\mu\alpha;\nu})_{;\beta} - R^{\mu\nu}_{;\beta} g^{\alpha\beta} \delta g_{\mu\alpha;\nu} - R^{\mu\nu} g^{\alpha\beta}_{;\beta} \delta g_{\mu\alpha;\nu}\right]$$

جمله اول دیورژانس کامل است و در انتگرال گیری کنش کنار گذاشته می شود. جمله آخر به علت قید

متریک سازگاری یعنی $g^{\alpha\beta}_{;\mu} = 0$ صفر می شود در نتیجه داریم:

$$R^{\mu\nu}g^{\alpha\beta}\delta g_{\mu\alpha;\nu\beta} = -R^{\mu\nu}_{\phantom{\mu\nu};\beta}g^{\alpha\beta}\delta g_{\mu\alpha;\nu}$$

اگر یک بار دیگر همین عملیات را انجام دهیم عبارت زیر را بدست می آوریم

$$R^{\mu\nu}g^{\alpha\beta}\delta g_{\mu\alpha;\nu\beta} = R^{\mu\nu}_{\phantom{\mu\nu};\beta\nu}g^{\alpha\beta}\delta g_{\mu\alpha}$$

برمی گردیم به مساله. هریک از جمله های رابطه (۸) را دو باره باز نویسی می کنیم:

$$\delta\mathcal{L}_3 = \frac{1}{2}\sqrt{-g}\{R^{\mu\nu}_{\phantom{\mu\nu};\nu\beta}g^{\alpha\beta}\delta g_{\mu\alpha} + R^{\mu\nu}_{\phantom{\mu\nu};\mu\alpha}g^{\alpha\beta}\delta g_{\nu\beta} - R^{\mu\nu}_{\phantom{\mu\nu};\alpha\beta}g^{\alpha\beta}\delta g_{\mu\nu} - \\ R^{\mu\nu}_{\phantom{\mu\nu};\mu\nu}g^{\alpha\beta}\delta g_{\alpha\beta}\}$$

منظور از شاخص ۳ در چگالی لاگرانژی جمله سوم است. بالا در این مرحله از رابطه (۵) استفاده می کنیم:

$$\delta\mathcal{L}_3 = \\ -\frac{1}{2}\sqrt{-g}\{R^{\mu\nu}_{\phantom{\mu\nu};\nu\beta}g^{\alpha\beta}g_{\mu i}g_{\alpha j}\delta g^{ij} + R^{\mu\nu}_{\phantom{\mu\nu};\mu\alpha}g^{\alpha\beta}g_{\nu m}g_{\beta n}\delta g^{mn} - \\ R^{\mu\nu}_{\phantom{\mu\nu};\alpha\beta}g^{\alpha\beta}g_{\mu p}g_{\nu q}\delta g^{pq} - R^{\mu\nu}_{\phantom{\mu\nu};\mu\nu}g^{\alpha\beta}g_{\alpha x}g_{\beta y}\delta g^{xy}\}$$

$$\delta\mathcal{L}_3 = -\frac{1}{2}\sqrt{-g}\{R^{\mu\nu}_{\phantom{\mu\nu};\nu\beta}\delta_j^{\phantom{j}\beta}g_{\mu i}\delta g^{ij} + R^{\mu\nu}_{\phantom{\mu\nu};\mu\alpha}\delta_n^{\phantom{n}\alpha}g_{\nu m}\delta g^{mn} - R_{pq;\alpha\beta}g^{\alpha\beta}\delta g^{pq} - \\ R^{\mu\nu}_{\phantom{\mu\nu};\mu\nu}g_{\beta y}\delta_x^{\phantom{x}\beta}\delta g^{xy}\}$$

$$\delta\mathcal{L}_3 = -\frac{1}{2}\sqrt{-g}\{R^{\mu\nu}_{\phantom{\mu\nu};\nu j}g_{\mu i}\delta g^{ij} + R^{\mu\nu}_{\phantom{\mu\nu};\mu n}g_{\nu m}\delta g^{mn} - \square R_{pq}\delta g^{pq} - R^{\mu\nu}_{\phantom{\mu\nu};\mu\nu}g_{xy}\delta g^{xy}\}$$

در این مرحله باید از تمام $\delta g$ ها فاکتور بگیریم. برای این کار باید با تغییر شاخص آنها را هم شاخص کنیم. در چهار جمله تغییر شاخص ها به ترتیب شکل $ij \to \alpha\beta$، $mn \to \alpha\beta$، $pq \to \alpha\beta$ و $xy \to \alpha\beta$ هستند در نتیجه بعد از انجام این کار خواهیم داشت:

$$\delta\mathcal{L}_3 = -\frac{1}{2}\sqrt{-g}\left\{R_\alpha^{\phantom{\alpha}\nu}_{\phantom{\nu};\nu\beta}g_{\mu\alpha} + R^\mu_{\phantom{\mu}\alpha;\mu\beta} - \square R_{\alpha\beta} - R^{\mu\nu}_{\phantom{\mu\nu};\mu\nu}g_{\alpha\beta}\right\}\delta g^{\alpha\beta} \qquad (۹)$$

در مرحله بعدی جمله دوم را حساب می کنیم:

(۱۰)

$$\delta\mathcal{L}_2 = \sqrt{-g}R_{\mu\nu}\delta(R^{\mu\nu}) = \sqrt{-g}R_{\mu\nu}\delta(g^{\alpha\mu}g^{\beta\nu}R_{\alpha\beta}) = \sqrt{-g}R_{\mu\nu}R_{\alpha\beta}g^{\beta\nu}\delta g^{\alpha\mu} +$$
$$\sqrt{-g}R_{\mu\nu}R_{\alpha\beta}g^{\alpha\mu}\delta g^{\beta\nu} + \sqrt{-g}R_{\mu\nu}g^{\beta\nu}g^{\alpha\mu}\delta R_{\alpha\beta}$$

اگر در دو جمله اول تغییر شاخص دهیم خواهیم داشت:

$$\sqrt{-g}R_{\beta\nu}R_\alpha{}^\nu\delta g^{\alpha\beta} + \sqrt{-g}R_{\mu\alpha}R^\mu{}_\beta\delta g^{\beta\alpha} = 2\sqrt{-g}R_{\alpha c}R^c{}_\beta = 2\sqrt{-g}\left[R_{\alpha c;\beta}{}^c - \right.$$
$$\left.\frac{1}{2}R_{;\alpha\beta} + R_{\alpha\rho\beta\sigma}R^{\rho\sigma}\right]\delta g^{\beta\alpha}$$

که از رابطه (۳) استفاده کردیم. با محاسبه خواهیم دید جمله سوم عبارت (۱۰) دقیقا برابر رابطه (۹) می شود. که برای جلو گیری از محاسبه تکراری از این کار خودداری شده است. پس داریم:

(۱۱)

$$\delta\mathcal{L}_2 = \sqrt{-g}\left\{2\left[R_{\alpha c;\beta}{}^c - \frac{1}{2}R_{;\alpha\beta} + R_{\alpha\rho\beta\sigma}R^{\rho\sigma}\right] - \frac{1}{2}\left[R_\alpha{}^\nu{}_{;\nu\beta}g_{\mu\alpha} + R^\mu{}_{\alpha;\mu\beta} - R_{\alpha\beta} - \right.\right.$$
$$\left.\left.R^{\mu\nu}{}_{;\mu\nu}g_{\alpha\beta}\right]\right\}\delta g^{\alpha\beta}$$

و در انتها جمله اول لاگرانژی (۷) برابر است با:

$$\delta\mathcal{L}_1 = R^{\mu\nu}R_{\mu\nu}\delta(\sqrt{-g}) = -\frac{1}{2}R^{\mu\nu}R_{\mu\nu}g_{\alpha\beta}\delta g^{\alpha\beta}$$

چگالی لاگرانژی نهایی برابر است با جمع سه چگالی لاگرانژی قبل به شکل زیر:

$$\delta\mathcal{L} = \delta\mathcal{L}_1 + \delta\mathcal{L}_2 + \delta\mathcal{L}_3 = -\sqrt{-g}\frac{1}{2}R^{\mu\nu}R_{\mu\nu}g_{\alpha\beta}\delta g^{\alpha\beta} + \sqrt{-g}\left\{2\left[R_{\alpha c;\beta}{}^c - \right.\right.$$

$$\left.\frac{1}{2}R_{;\alpha\beta} + R_{\alpha\rho\beta\sigma}R^{\rho\sigma}\right] - \frac{1}{2}\left[R_{\alpha}{}^{\nu}{}_{;\nu\beta}g_{\mu\alpha} + R^{\mu}{}_{\alpha;\mu\beta} - \Box R_{\alpha\beta} - R^{\mu\nu}{}_{;\mu\nu}g_{\alpha\beta}\right]\right\}\delta g^{\alpha\beta} -$$

$$\frac{1}{2}\sqrt{-g}\left[R_{\alpha}{}^{\nu}{}_{;\nu\beta} + R^{\mu}{}_{\alpha;\mu\beta} - \Box R_{\alpha\beta} - R^{\mu\nu}{}_{;\mu\nu}g_{\alpha\beta}\right]\delta g^{\alpha\beta}$$

$$\delta\mathcal{L} = \sqrt{-g}\left\{-\frac{1}{2}R^{\mu\nu}R_{\mu\nu}g_{\alpha\beta} + 2R_{\alpha c;\beta}{}^c - R_{;\alpha\beta} + 2R_{\alpha\rho\beta\sigma}R^{\rho\sigma} - R_{\alpha}{}^{\nu}{}_{;\nu\beta} - \right.$$

$$\left. R^{\mu}{}_{\alpha;\mu\beta} + \Box R_{\alpha\beta} + R^{\mu\nu}{}_{;\mu\nu}g_{\alpha\beta}\right\}\delta g^{\alpha\beta}$$

در عبارت بالا مجموع جمله های دوم و پنجم و ششم صفر است. برای جمله آخر نیز از رابطه (۱) استفاده می کنیم. در نتیجه خواهیم داشت:

<div dir="rtl">(۱۲)</div>

$$\delta\mathcal{L} = \sqrt{-g}\left\{-\frac{1}{2}R^{\mu\nu}R_{\mu\nu}g_{\alpha\beta} - R_{;\alpha\beta} + 2R_{\alpha\rho\beta\sigma}R^{\rho\sigma} + \Box R_{\alpha\beta} + \frac{1}{2}g_{\alpha\beta}\Box R\right\}\delta g^{\alpha\beta}$$

$$\delta S = \int d^4x\delta\mathcal{L} = \int d^4x\sqrt{-g}\left\{-\frac{1}{2}R^{\mu\nu}R_{\mu\nu}g_{\alpha\beta} - R_{;\alpha\beta} + 2R_{\alpha\rho\beta\sigma}R^{\rho\sigma} + \Box R_{\alpha\beta} + \right.$$

$$\left. \frac{1}{2}g_{\alpha\beta}\Box R\right\}\delta g^{\alpha\beta} = 0$$

$$-\frac{1}{2}R^{\mu\nu}R_{\mu\nu}g_{\alpha\beta} - R_{;\alpha\beta} + 2R_{\alpha\rho\beta\sigma}R^{\rho\sigma} + R_{\alpha\beta} + \frac{1}{2}g_{\alpha\beta}\Box R = 0 \qquad (۱۳)$$

توجه کنید که اگر ازبجای وردش نسبت $\delta g^{\alpha\beta}$ از وردش گیری نسبت به $\delta g_{\alpha\beta}$ و در ادامه به ابتدا بجای (۵) از $\delta g^{\mu\nu} = -g^{\mu\alpha}g^{\nu\beta}\delta g_{\alpha\beta}$ استفاده می کردیم، رابطه (۱۳) علامتش تغییر می کرد و شاخص های $\beta$ و $\alpha$ در بالای متغیر ها قرار می گرفتند.

ورش کنش $S = \int d^4x\mathcal{L}$ با $\mathcal{L} = \sqrt{-g}f(R)$ به روش متریکی

$$A = \int d^4x \sqrt{-g}f(R) \qquad (1)$$

$$\delta A = \int d^4x \, \delta\left(\sqrt{-g}f(R)\right) = \int d^4x\{\delta(\sqrt{-g})f(R) + \sqrt{-g}\delta f(R)\}$$

$$\int d^4x\left\{-\tfrac{1}{2}\sqrt{-g}g_{\alpha\beta}f(R)\delta g^{\alpha\beta} + \sqrt{-g}\tfrac{\delta f(R)}{\delta R}\delta R\right\} =$$

$$\int d^4x\{-\tfrac{1}{2}\sqrt{-g}g_{\alpha\beta}f(R)\delta g^{\alpha\beta}$$

$$+\sqrt{-g}f'(R)\delta(g^{\alpha\beta}R_{\alpha\beta})\}$$

$$= \int d^4x\{-\tfrac{1}{2}\sqrt{-g}g_{\alpha\beta}f(R)\delta g^{\alpha\beta} + \sqrt{-g}f'(R)R_{\alpha\beta}\delta g^{\alpha\beta} + \sqrt{-g}f'(R)\delta(R_{\alpha\beta})\}$$

$$= \int d^4x\sqrt{-g}\{-\tfrac{1}{2}f(R)g_{\alpha\beta}\delta g^{\alpha\beta} + f'(R)R_{\alpha\beta}\delta g^{\alpha\beta} + f'(R)g^{\alpha\beta}g^{\rho\sigma}(\delta g_{\alpha\rho;\beta\sigma} - \delta g_{\alpha\beta;\rho\sigma})\}$$

در روابط بالا از نکات زیر استفاده کرده ایم:

$$\delta(\sqrt{-g}) = -\tfrac{1}{2}\sqrt{-g}g_{\alpha\beta}\delta g^{\alpha\beta} \qquad (2)$$

و نیز برای ورش اسکالر ریچی داریم:

$$\delta(R_{\alpha\beta}) = g^{\alpha\beta}g^{\rho\sigma}(\delta g_{\alpha\rho;\beta\sigma} - \delta g_{\alpha\beta;\rho\sigma}) \qquad (3)$$

در ادامه داریم:

$$(4)$$

$$\delta A =$$

$$\int d^4x \sqrt{-g} \{ -\frac{1}{2} f(R) g_{\alpha\beta} \delta g^{\alpha\beta} + f'(R) R_{\alpha\beta} \delta g^{\alpha\beta} + [f'(R) g^{\alpha\beta} g^{\rho\sigma} \delta g_{\alpha\rho;\beta\sigma} +$$
$$f'(R) g^{\alpha\beta} g^{\rho\sigma} \delta g_{\alpha\beta;\rho\sigma}]$$

حالا عبارت $f'(R) g^{\alpha\beta} g^{\rho\sigma} \delta g_{\alpha\rho;\beta\sigma}$ را جداگانه حساب می کنیم:

$$f'(R) g^{\alpha\beta} g^{\rho\sigma} \delta g_{\alpha\rho;\beta\sigma} = (f'(R) g^{\alpha\beta} g^{\rho\sigma} \delta g_{\alpha\rho;\beta})_{;\sigma} - g^{\alpha\beta} g^{\rho\sigma} f'(R)_{;\sigma} \delta g_{\alpha\rho;\beta} -$$
$$f'(R) g^{\rho\sigma} g^{\alpha\beta}_{\ \ ;\sigma} \delta g_{\alpha\rho;\beta} - f'(R) g^{\alpha\beta} g^{\rho\sigma}_{\ \ ;\sigma} \delta g_{\alpha\rho;\beta} = (f'(R) g^{\alpha\beta} g^{\rho\sigma} \delta g_{\alpha\rho;\beta})_{;\sigma} -$$
$$g^{\alpha\beta} g^{\rho\sigma} f'(R)_{;\sigma} \delta g_{\alpha\rho;\beta}$$

که به علت شرط سازگاری متریک $g^{\alpha\beta}_{\ \ ;\sigma} = g^{\rho\sigma}_{\ \ ;\sigma} = 0$ است. از آنجایی که عبارت $(f'(R) g^{\alpha\beta} g^{\rho\sigma} \delta g_{\alpha\rho;\beta})_{;\sigma}$ یک واگرایی کامل است یعنی می توان آن را به شکل $A^{\alpha}_{\ ;\alpha}$ در نتیجه انتگرال آن به دلیل شرایط مرزی صفر می شود، در نتیجه خواهیم داشت:

$$f'(R) g^{\alpha\beta} g^{\rho\sigma} \delta g_{\alpha\rho;\beta\sigma} = -g^{\alpha\beta} g^{\rho\sigma} f'(R)_{;\sigma} \delta g_{\alpha\rho;\beta}$$

اگر همین کار را یک بار دیگرانجام دهیم خواهیم داشت:

$$f'(R) g^{\alpha\beta} g^{\rho\sigma} \delta g_{\alpha\rho;\beta\sigma} = g^{\alpha\beta} g^{\rho\sigma} f'(R)_{;\beta\sigma} \delta g_{\alpha\rho} \qquad (٥)$$

به طور مشابه برای جمله آخر رابطه (٤) داریم:

$$f'(R) g^{\alpha\beta} g^{\rho\sigma} \delta g_{\alpha\beta;\rho\sigma} = g^{\alpha\beta} g^{\rho\sigma} f'(R)_{;\rho\sigma} \delta g_{\alpha\beta} \qquad (٦)$$

اگر (٥) و (٦) را در (٤) جاگذاری کنیم داریم:

$$\delta A =$$

$$\int d^4x \sqrt{-g} \{-\frac{1}{2} f(R) g_{\alpha\beta} \delta g^{\alpha\beta} + f'(R) R_{\alpha\beta} \delta g^{\alpha\beta} + [g^{\alpha\beta} g^{\rho\sigma} f'(R)_{;\beta\sigma} \delta g_{\alpha\rho} -$$

$$g^{\alpha\beta} g^{\rho\sigma} f'(R)_{;\rho\sigma} \delta g_{\alpha\beta}]$$

حالا برای برای دو جمله آخر از فرمول زیر استفاده می کنیم:

$$\delta g_{\lambda\phi} = -g_{\alpha\lambda} g_{\beta\phi} \delta g^{\alpha\beta} \qquad (\Lambda)$$

$$\delta A =$$

$$\int d^4x \sqrt{-g} \{-\frac{1}{2} f(R) g_{\alpha\beta} \delta g^{\alpha\beta} + f'(R) R_{\alpha\beta} \delta g^{\alpha\beta} +$$

$$[g^{\alpha\beta} g^{\rho\sigma} f'(R)_{;\beta\sigma} (-g_{\alpha\lambda} g_{\rho\phi} \delta g^{\lambda\phi}) - g^{\alpha\beta} g^{\rho\sigma} f'(R)_{;\rho\sigma} (-g_{\alpha\lambda} g_{\beta\phi} \delta g^{\lambda\phi})]\}$$

$$\delta A =$$

$$\int d^4x \sqrt{-g} \{-\frac{1}{2} f(R) g_{\alpha\beta} \delta g^{\alpha\beta} + f'(R) R_{\alpha\beta} \delta g^{\alpha\beta} - [\delta_\lambda{}^\beta \delta_\phi{}^\sigma f'(R)_{;\beta\sigma} \delta g^{\lambda\phi} -$$

$$\delta_\lambda{}^\beta f'(R)_{;\rho\sigma} g^{\rho\sigma} g_{\beta\phi} \delta g^{\lambda\phi}]\}$$

$$\delta A = \int d^4x \sqrt{-g} \{-\frac{1}{2} f(R) g_{\alpha\beta} \delta g^{\alpha\beta} + f'(R) R_{\alpha\beta} \delta g^{\alpha\beta} - [f'(R)_{;\lambda\phi} \delta g^{\lambda\phi} -$$

$$f'(R)_{;\rho\sigma} g^{\rho\sigma} g_{\lambda\phi} \delta g^{\lambda\phi}]\}$$

جمله سوم را در (۹) با تغییر $\lambda \rightarrow \alpha$, $\phi \rightarrow \beta$ به شکل زیر می نویسیم:

$$f'(R)_{;\lambda\phi} \delta g^{\lambda\phi} = f'(R)_{;\alpha\beta} \delta g^{\alpha\beta} = f'(R)^{;\epsilon\theta} g_{\epsilon\alpha} g_{\theta\beta} \delta g^{\alpha\beta}$$

برای جمله چهارم نیز به طور مشابه داریم:

$$f'(R)_{;\rho\sigma} = g_{\rho i} g_{\sigma j} f'(R)^{;ij}$$

حالا این دو را جاگذاری می کنیم:

$$\delta A =$$

$$\int d^4 x \sqrt{-g} \{ -\frac{1}{2} f(R) g_{\alpha\beta} \delta g^{\alpha\beta} + f'(R) R_{\alpha\beta} \delta g^{\alpha\beta} - [f'(R)^{;\epsilon\theta} g_{\epsilon\alpha} g_{\theta\beta} \delta g^{\alpha\beta} -$$

$$g_{\rho i} g_{\sigma j} f'(R)^{;ij} g^{\rho\sigma} g_{\lambda\phi} \delta g^{\lambda\phi}] \}$$

جمله چهارم را ساده می کنیم:

$$g_{\rho i} g_{\sigma j} f'(R)^{;ij} g^{\rho\sigma} g_{\lambda\phi} \delta g^{\lambda\phi} = f'(R)^{;ij} \delta_i{}^\sigma g_{\sigma j} g_{\lambda\phi} \delta g^{\lambda\phi} = f'(R)^{;ij} g_{ij} g_{\lambda\phi} \delta g^{\lambda\phi} =$$

$$f'(R)^{;\epsilon\theta} g_{\epsilon\theta} g_{\alpha\beta} \delta g^{\alpha\beta}$$

که درنهایت در جمله فوق اندیس ها را به شکل $\beta \to \phi$ ، $\alpha \to \lambda$ و $i \to \epsilon$ و $j \to \theta$ تغییر داده ایم. حالا

جمله سوم و چهارم را در (۹) جاگذاری می کنیم:

$$\delta A =$$

$$\int d^4 x \sqrt{-g} \{ -\frac{1}{2} f(R) g_{\alpha\beta} \delta g^{\alpha\beta} + f'(R) R_{\alpha\beta} \delta g^{\alpha\beta} - [f'(R)^{;\epsilon\theta} g_{\epsilon\alpha} g_{\theta\beta} \delta g^{\alpha\beta} -$$

$$f'(R)^{;\epsilon\theta} g_{\epsilon\theta} g_{\alpha\beta} \delta g^{\alpha\beta}] \}$$

$$\delta A = \int d^4 x \sqrt{-g} \delta g^{\alpha\beta} \{ -\frac{1}{2} f(R) g_{\alpha\beta} + f'(R) R_{\alpha\beta} - f'(R)^{;\epsilon\theta} (g_{\epsilon\alpha} g_{\theta\beta} -$$

$$g_{\epsilon\theta} g_{\alpha\beta}) \}$$

چون باید وردش کنش کمینه باشد داریم:

$$\delta A = 0 \qquad\qquad (\text{۱۰})$$

که در نهایت  با تغییر اندیس $\rho \to \epsilon$ و $\sigma \to \theta$ رابطه ای که می خواستیم به دست آمد:

$$-\frac{1}{2} f(R) g_{\alpha\beta} + f'(R) R_{\alpha\beta} - f'(R)^{;\rho\sigma} (g_{\rho\alpha} g_{\sigma\beta} - g_{\rho\sigma} g_{\alpha\beta}) = 0 \qquad (\text{۱۱})$$

ضمیمه ت

وردش چگالی لاگرانژی $R^2$ $S = \int d^4x \sqrt{-g}R^2$ به روش پالاتینی

در محاسبات زیر از رابطه زیر موسوم به رابطه پالاتینی [۱] استفاده می کنیم:

$$\delta R^a{}_{bcd} = \nabla_c(\delta\Gamma^a{}_{bd}) - \nabla_d(\delta\Gamma^a{}_{bc}) \qquad (\text{۱})$$

$$\delta R^c{}_{bcd} = \delta R_{bd} = \nabla_c(\delta\Gamma^c{}_{bd}) - \nabla_d(\delta\Gamma^c{}_{bc}) \qquad (\text{۲})$$

ابتدا معادله **g** را بدست می آوریم.

$$\delta S = \int d^4x \delta\mathcal{L} = \int d^4x \delta(\sqrt{-g}R^2) \qquad (\text{۳})$$

$$\delta\mathcal{L} = \delta(\sqrt{-g}R^2) = R^2\delta(\sqrt{-g}) + \sqrt{-g}\delta R^2$$

جمله اول عبارت است از:

$$R^2\delta(\sqrt{-g}) = -\frac{1}{2}\sqrt{-g}R^2 g_{\alpha\beta}\delta g^{\alpha\beta}$$

و جمله دوم به شکل زیر محاسبه می شود:

$$\sqrt{-g}\delta R^2 = 2\sqrt{-g}R\delta R = 2\sqrt{-g}R\delta(g^{\alpha\beta}R_{\alpha\beta}) = 2\sqrt{-g}RR_{\alpha\beta}\delta g^{\alpha\beta} + 2\sqrt{-g}Rg^{\alpha\beta}\delta R_{\alpha\beta}$$

در روش پالاتینی وردش تانسور ریچی نسبت به متریک صفر است زیرا این تانسور تابع هموستار ها هستند که آنها هم مستقل از متریک هستند. پس برای چگالی لاگرانژی داریم:

$$\delta\mathcal{L} = \delta\left(\sqrt{-g}R^2\right) = R^2\delta\left(\sqrt{-g}\right) + \sqrt{-g}\delta R^2 = -\frac{1}{2}\sqrt{-g}R^2 g_{\alpha\beta}\delta g^{\alpha\beta} +$$

$$2\sqrt{-g}RR_{\alpha\beta}\delta g^{\alpha\beta} = \sqrt{-g}\left[-\frac{1}{2}R^2 g_{\alpha\beta} + 2RR_{\alpha\beta}\right]\delta g^{\alpha\beta}$$

$$\delta S = \int d^4x\,\delta\left(\sqrt{-g}R^2\right) = \int d^4x\sqrt{-g}\left[-\frac{1}{2}R^2 g_{\alpha\beta} + 2RR_{\alpha\beta}\right]\delta g^{\alpha\beta} = 0$$

عبارت فوق برای هر $\delta g^{\alpha\beta}$ دلخواه باید بر قرار باشد حتی برای $\delta g^{\beta\alpha}$ . از آنجایی که متریک متقارن است

نتیجه می شود که ضریب آن در انتگرال کنش باید متقارن باشد. در نتیجه با یک ساده سازی داریم:

$$R\left(R_{(\alpha\beta)} - \frac{1}{4}g_{\alpha\beta}R\right) = 0 \qquad\qquad (\text{٤})$$

در مرحله بعد معادله $\Gamma$ را بدست می آوریم. در این قسمت باید نسبت به همبوستار ها وردش بگیریم.

$$\delta\mathcal{L} = \delta\left(\sqrt{-g}R^2\right) = R^2\delta\left(\sqrt{-g}\right) + \sqrt{-g}\delta R^2 = 0 + 2\sqrt{-g}R\delta\left(g^{\alpha\beta}R_{\alpha\beta}\right) =$$

$$2\sqrt{-g}Rg^{\alpha\beta}\delta(R_{\alpha\beta}) = 2\sqrt{-g}Rg^{\alpha\beta}\left[\nabla_c(\delta\Gamma^c_{\alpha\beta}) - \nabla_\beta(\delta\Gamma^c_{\alpha c})\right]$$

$$\delta S =$$
$$2\int d^4x\sqrt{-g}Rg^{\alpha\beta}\left[\nabla_c(\delta\Gamma^c_{\alpha\beta}) - \nabla_\beta(\delta\Gamma^c_{\alpha c})\right] = 2\int d^4x\sqrt{-g}Rg^{\alpha\beta}\nabla_c(\delta\Gamma^c_{\alpha\beta}) -$$
$$\sqrt{-g}Rg^{\alpha\beta}\nabla_\beta(\delta\Gamma^c_{\alpha c})$$

$$\delta S =$$
$$2\int d^4x[\nabla_c[\sqrt{-g}Rg^{\alpha\beta}\delta\Gamma^c_{\alpha\beta}] - \nabla_c[R\sqrt{-g}\,g^{\alpha\beta}]\delta\Gamma^c_{\alpha\beta} - \nabla_\beta[\sqrt{-g}Rg^{\alpha\beta}\delta\Gamma^c_{\alpha c}] +$$
$$\nabla_\beta[R\sqrt{-g}g^{\alpha\beta}]\delta\Gamma^c_{\alpha c}\,]$$

جمله اول و سوم دیورژانس کامل است و صفر می شود

$$\delta S = 2\int d^4x\{-\nabla_c[Rg^{\alpha\beta}\sqrt{-g}]\,\delta\Gamma^c_{\alpha\beta} + \nabla_\beta[R\sqrt{-g}g^{\alpha\beta}]\delta\Gamma^c_{\alpha c}\,\}$$

$$\delta S = 2\int d^4x\{-\nabla_c[g^{\alpha\beta}R\sqrt{-g}] + \delta_c{}^\beta\nabla_d[R\sqrt{-g}g^{\alpha d}]\}\delta\Gamma^c_{\alpha\beta} = 0$$

$$-\nabla_c[g^{\alpha\beta}R\sqrt{-g}] + \delta_c{}^\beta\nabla_d[R\sqrt{-g}g^{\alpha d}] = 0$$

$$\nabla_c[\sqrt{-g}\,g^{\alpha\beta}R] = 0 \qquad (\text{۵})$$

معادلات (۴) و (۵) معادلات میدان مورد نظر در روش وردشی پالاتینی هستند.

ضمیمه ث

وردش چگالی لاگرانژی ($R_{\mu\nu} R^{\mu\nu}$) $S = \int d^4x \sqrt{-g}(R_{\mu\nu} R^{\mu\nu})$ به روش پالاتینی

قبل از شروع روابط (۱) و (۲) را از ضمیمه ت در ذهن داشته باشید. ابتدا معادلہ **g** را بدست می آوریم.

$$\delta S = \int d^4x \delta\mathcal{L} = \int d^4x \delta(\sqrt{-g} R_{\mu\nu} R^{\mu\nu}) \qquad (\text{۱})$$

$$\delta\mathcal{L} = \delta(\sqrt{-g} R_{\mu\nu} R^{\mu\nu}) = \delta(\sqrt{-g} g^{\beta\nu} g^{\alpha\mu} R_{\mu\nu} R_{\alpha\beta}) = R_{\mu\nu} R^{\mu\nu} \delta(\sqrt{-g}) +$$

$$\sqrt{-g} g^{\beta\nu} g^{\alpha\mu} R_{\alpha\beta} \delta R_{\mu\nu} + \sqrt{-g} g^{\beta\nu} g^{\alpha\mu} R_{\mu\nu} \delta R_{\alpha\beta} + \sqrt{-g} g^{\beta\nu} R_{\alpha\beta} R_{\mu\nu} \delta g^{\alpha\mu} +$$

$$\sqrt{-g} g^{\alpha\mu} R_{\mu\nu} R_{\alpha\beta} \delta g^{\beta\nu}$$

جمله دوم و سوم به علت مستقل بودن تانسور ریچی از متریک صفر می باشند.

$$\delta\mathcal{L} = \delta(\sqrt{-g} R_{\mu\nu} R^{\mu\nu}) = \delta(\sqrt{-g} g^{\beta\nu} g^{\alpha\mu} R_{\mu\nu} R_{\alpha\beta}) = R_{\mu\nu} R^{\mu\nu} \delta(\sqrt{-g}) +$$

$$\sqrt{-g} g^{\beta\nu} R_{\alpha\beta} R_{\mu\nu} \delta g^{\alpha\mu} + \sqrt{-g} g^{\alpha\mu} R_{\mu\nu} R_{\alpha\beta} \delta g^{\beta\nu} = -\frac{1}{2}\sqrt{-g} R_{\mu\nu} R^{\mu\nu} g_{\alpha\beta} \delta g^{\alpha\beta} +$$

$$\sqrt{-g} R_\alpha{}^\nu R_{\mu\nu} \delta g^{\alpha\mu} + \sqrt{-g} R^\alpha{}_\nu R_{\alpha\beta} \delta g^{\beta\nu} = -\frac{1}{2}\sqrt{-g} R_{\mu\nu} R^{\mu\nu} g_{\alpha\beta} \delta g^{\alpha\beta} +$$

$$\sqrt{-g} R_\alpha{}^\nu R_{\beta\nu} \delta g^{\alpha\beta} + \sqrt{-g} R^\nu{}_\alpha R_{\nu\beta} \delta g^{\beta\alpha}$$

که در مرحله آخر شاخص های سه جمله را بر حسب $\alpha\beta$ مرتب کرده ایم.

$$\delta\mathcal{L} = \sqrt{-g}\left\{-\frac{1}{2} R_{\mu\nu} R^{\mu\nu} g_{\alpha\beta} + R_\alpha{}^\nu R_{\beta\nu} + R^\nu{}_\alpha R_{\nu\beta}\right\} \delta g^{\beta\alpha} \qquad (\text{۲})$$

$$\delta S = \int d^4x \delta\mathcal{L} = \int d^4x \sqrt{-g}\left\{-\frac{1}{2} R_{\mu\nu} R^{\mu\nu} g_{\alpha\beta} + R_\alpha{}^\nu R_{\beta\nu} + R^\nu{}_\alpha R_{\nu\beta}\right\} \delta g^{\beta\alpha} = 0$$

که بدین ترتیب معادله **g** بدست می آید:

$$-\frac{1}{2} R_{\mu\nu} R^{\mu\nu} g_{\alpha\beta} + R_\alpha{}^\nu R_{\beta\nu} + R^\nu{}_\alpha R_{\nu\beta} = 0 \qquad (\text{۳})$$

در مرحله بعد معادله $\Gamma$ را بدست می آوریم. در معادله (۲)به علت مستقل بودن متریک وهموستار جمله های اول، چهارو و پنجم صفر می باشند.

$$\delta\mathcal{L} = \delta(\sqrt{-g}R_{\mu\nu}R^{\mu\nu}) = \delta(\sqrt{-g}g^{\beta\nu}g^{\alpha\mu}R_{\mu\nu}R_{\alpha\beta}) = \sqrt{-g}g^{\beta\nu}g^{\alpha\mu}R_{\alpha\beta}\delta R_{\mu\nu} +$$
$$\sqrt{-g}g^{\beta\nu}g^{\alpha\mu}R_{\mu\nu}\delta R_{\alpha\beta} = 2\sqrt{-g}R^{\mu\nu}\delta R_{\mu\nu}$$

$$\delta S = \int d^4x \delta\mathcal{L} = 2\int d^4x \sqrt{-g}R^{\mu\nu}\delta R_{\mu\nu} = 2\int d^4x \sqrt{-g}R^{\mu\nu}[\nabla_c(\delta\Gamma^c{}_{\mu\nu}) - \nabla_\nu(\delta\Gamma^c{}_{\mu c})] =$$

در این مرحله مشابه کاری که در ضمیمه ت انجام دادیم انتگرال جزء به جزء می گیریم و سپس جمله های دیورژانس کامل را کنار می گذاریم.نتیجه برابر است با:

$$\delta S =$$
$$2\int d^4x[\nabla_\nu(\sqrt{-g}R^{\mu\nu})\delta\Gamma^c{}_{\mu c} - \nabla_c(\sqrt{-g}R^{\mu\nu})\delta\Gamma^c{}_{\mu\nu}] =$$
$$2\int d^4x\left[\delta_c{}^t\nabla_\nu(\sqrt{-g}R^{\mu\nu})\delta\Gamma^c{}_{\mu t} - \nabla_c(\sqrt{-g}R^{\mu t})\delta\Gamma^c{}_{\mu t}\right]$$

$$\delta S = 2\int d^4x\left[\delta_c{}^t\nabla_\nu(\sqrt{-g}R^{\mu\nu}) - \nabla_c(\sqrt{-g}R^{\mu t})\right]\delta\Gamma^c{}_{\mu t} = 0$$

که در نتیجه داریم:

$$\delta_c{}^t\nabla_\nu(\sqrt{-g}R^{\mu\nu}) - \nabla_c(\sqrt{-g}R^{\mu t}) = 0$$

$$\nabla_c(\sqrt{-g}R^{\mu t}) = 0 \qquad (\text{٤})$$

معادلات (۳) و(٤) معادلات میدان مورد نظر در روش وردشی پالاتینی هستند.

ضمیمه ج

ورذش چگالی لاگرانژی $(R_{abcd}R^{abcd})$ $S = \int d^4x\sqrt{-g}$ به روش پالاتینی

قبل از شروع روابط (۱) و (۲) را از ضمیمه ت در ذهن داشته باشید. ابتدا معادله **g** را بدست می آوریم.

$$\delta S = \int d^4x \delta\mathcal{L} = \int d^4x \delta(\sqrt{-g}R_{abcd}R^{abcd}) \tag{۱}$$

$$\tag{۲}$$

$$\begin{aligned}
\delta\mathcal{L} = \delta\left(\sqrt{-g}R_{abcd}R^{abcd}\right) = \delta\left(\sqrt{-g}g^{a\rho}g^{b\sigma}g^{c\mu}g^{d\nu}R_{abcd}R_{\rho\sigma\mu\nu}\right) = \\
R_{abcd}R^{abcd}\delta(\sqrt{-g}) + \sqrt{-g}R_{abcd}R_{\rho\sigma\mu\nu}g^{b\sigma}g^{c\mu}g^{d\nu}\delta g^{a\rho} + \\
\sqrt{-g}R_{abcd}R_{\rho\sigma\mu\nu}g^{a\rho}g^{c\mu}g^{d\nu}\delta g^{b\sigma} + \sqrt{-g}R_{abcd}R_{\rho\sigma\mu\nu}g^{a\rho}g^{b\sigma}g^{d\nu}\delta g^{c\mu} + \\
\sqrt{-g}R_{abcd}R_{\rho\sigma\mu\nu}g^{a\rho}g^{b\sigma}g^{c\mu}\delta g^{d\nu}
\end{aligned}$$

$$\begin{aligned}
\delta\mathcal{L} = -\frac{1}{2}\sqrt{-g}g_{\alpha\beta}R_{abcd}R^{abcd}\delta g^{\alpha\beta} + \sqrt{-g}R_{abcd}R_{\beta\sigma\mu\nu}g^{b\sigma}g^{c\mu}g^{d\nu}\delta g^{\alpha\beta} + \\
\sqrt{-g}R_{a\alpha cd}R_{\rho\beta\mu\nu}g^{a\rho}g^{c\mu}g^{d\nu}\delta g^{\alpha\beta} + \sqrt{-g}R_{ab\alpha d}R_{\rho\sigma\beta\nu}g^{a\rho}g^{b\sigma}g^{d\nu}\delta g^{\alpha\beta} + \\
\sqrt{-g}R_{abc\alpha}R_{\rho\sigma\mu\beta}g^{a\rho}g^{b\sigma}g^{c\mu}\delta g^{\alpha\beta}
\end{aligned}$$

$$\delta\mathcal{L} =$$

با استفاده از خواص تانسور ریمان در جابه جایی شاخص ها جمله چهارم و پنجم را به شکل زیر ساده می کنیم

$$\begin{aligned}
R_{ab\alpha d}R_{\rho\sigma\beta\nu}g^{a\rho}g^{b\sigma}g^{d\nu} + R_{abc\alpha}R_{\rho\sigma\mu\beta}g^{a\rho}g^{b\sigma}g^{c\mu} = R_{ab\alpha d}R^{ab}{}_{\beta}{}^{d} + \\
R_{abc\alpha}R^{abc}{}_{\beta} = 2R_{ab\alpha d}R^{ab}{}_{\beta}{}^{d} = 2R_{ab\alpha}{}^{d}R^{ab}{}_{\beta d}
\end{aligned}$$

برای جمله دوم داریم

$$R_{\alpha bcd}R_{\beta\sigma\mu\nu}g^{b\sigma}g^{c\mu}g^{d\nu} = R_{\alpha bcd}R_{\beta}{}^{bcd}$$

<div dir="rtl">برای جمله سوم نیز داریم</div>

$$R_{a\alpha cd}R_{\rho\beta\mu\nu}g^{a\rho}g^{c\mu}g^{d\nu} = R_{a\alpha}{}^{cd}R^{a}{}_{\beta cd}$$

<div dir="rtl">در نتیجه داریم:</div>

$$\delta\mathcal{L} = \sqrt{-g}\Big[-\frac{1}{2}g_{\alpha\beta}R_{abcd}R^{abcd} + 2R_{ab\alpha}{}^{d}R^{ab}{}_{\beta d} + R_{\alpha bcd}R_{\beta}{}^{bcd} +$$
$$R_{a\alpha}{}^{cd}R^{a}{}_{\beta cd}\Big]\delta g^{\alpha\beta}$$

$$\delta S = \int d^4x\,\delta\mathcal{L} = \int d^4x\sqrt{-g}\Big[-\frac{1}{2}g_{\alpha\beta}R_{abcd}R^{abcd} + 2R_{ab\alpha}{}^{d}R^{ab}{}_{\beta d} +$$
$$R_{\alpha bcd}R_{\beta}{}^{bcd} + R_{a\alpha}{}^{cd}R^{a}{}_{\beta cd}\Big]\delta g^{\alpha\beta} = 0$$

<div dir="rtl">بنا بر این برای معادله <strong>g</strong> خواهیم داشت:</div>

<div dir="rtl">(۳)</div>

$$-\frac{1}{2}g_{\alpha\beta}R_{abcd}R^{abcd} + 2R_{ab\alpha}{}^{d}R^{ab}{}_{\beta d} + R_{\alpha bcd}R_{\beta}{}^{bcd} + R_{a\alpha}{}^{cd}R^{a}{}_{\beta cd} =$$
$$0$$

<div dir="rtl">در مرحله بعد معادله <strong>Γ</strong> را بدست می آوریم. در این قسمت باید نسبت به همبستار ها وردش بگیریم.</div>

<div dir="rtl">در رابطه (۲) تمام جملات مربوط به وردش متریک صفر می شود در نتیجه داریم</div>

$$\delta\mathcal{L} = \delta\big(\sqrt{-g}R_{abcd}R^{abcd}\big) = \sqrt{-g}\{R_{abcd}\delta(R^{abcd}) + R^{abcd}\delta(R_{abcd})\} =$$
$$\sqrt{-g}\Big\{R^{abcd}\delta\big(g_{a\rho}R^{\rho}{}_{abcd}\big) + R_{abcd}\delta(g^{\sigma b}g^{\beta c}\delta g^{\mu d}R^{a}{}_{\sigma\beta\mu})\Big\} =$$
$$\sqrt{-g}\{R^{abcd}g_{a\rho}\delta(R^{\rho}{}_{abcd}) + R_{abcd}g^{\sigma b}g^{\beta c}g^{\mu d}\delta(R^{a}{}_{\sigma\beta\mu})\} =$$
$$\sqrt{-g}\{R_{\rho}{}^{bcd}\delta(R^{\rho}{}_{abcd}) + R_{a}{}^{\sigma\beta\mu}\delta(R^{a}{}_{\sigma\beta\mu})\} = 2\sqrt{-g}R_{\rho}{}^{bcd}\delta(R^{\rho}{}_{abcd})$$

$$\delta\mathcal{L} = 2R_{\rho}{}^{bcd}\delta(R^{\rho}{}_{abcd}) = 2\sqrt{-g}R_{\rho}{}^{bcd}\{\nabla_c(\delta\Gamma^{\rho}{}_{bd}) - \nabla_d(\delta\Gamma^{\rho}{}_{bc})\}$$

<div dir="rtl">۱۰۸</div>

$$\delta S = \int d^4x \delta \mathcal{L} = 2 \int d^4x \sqrt{-g} R_\rho{}^{bcd} \{\nabla_c(\delta \Gamma^\rho{}_{bd}) - \nabla_d(\delta \Gamma^\rho{}_{bc})\}$$

در این مرحله مشابه کاری که در ضمیمه ت انجام دادیم انتگرال جزء به جزء می گیریم و سپس جمله های

دیورژانس کامل را کنار می گذاریم. نتیجه برابر است با:

$$\delta S =$$

$$2 \int d^4x \delta \mathcal{L} = 2 \int d^4x \left\{ \nabla_d \left( \sqrt{-g} R_\rho{}^{bcd} \right) \delta \Gamma^\rho{}_{bc} - \nabla_c \left( \sqrt{-g} R_\rho{}^{bcd} \right) \delta \Gamma^\rho{}_{bd} \right\} =$$

$$2 \int d^4x \left\{ \nabla_m \left( \sqrt{-g} R_\rho{}^{bcm} \right) \delta \Gamma^\rho{}_{bc} - \nabla_c \left( \sqrt{-g} R_\rho{}^{bcd} \right) \delta \Gamma^\rho{}_{bd} \right\} =$$

$$2 \int d^4x \left\{ \delta_c{}^d \nabla_m \left( \sqrt{-g} R_\rho{}^{bcm} \right) \delta \Gamma^\rho{}_{bd} - \nabla_c \left( \sqrt{-g} R_\rho{}^{bcd} \right) \delta \Gamma^\rho{}_{bd} \right\} =$$

$$2 \int d^4x \left\{ \delta_c{}^d \nabla_m \left( \sqrt{-g} R_\rho{}^{bcm} \right) - \nabla_c \left( \sqrt{-g} R_\rho{}^{bcd} \right) \right\} \delta \Gamma^\rho{}_{bd} = 0$$

$$\delta_c{}^d \nabla_m \left( \sqrt{-g} R_\rho{}^{bcm} \right) - \nabla_c \left( \sqrt{-g} R_\rho{}^{bcd} \right) = 0$$

که در خط دوم تنها تغییری که دادیم، تغییر شاخص $d$ به $m$ در جمله اول است. در خط

سوم نیز $\delta \Gamma^\rho{}_{bd}$ ها را هم شاخص کردیم تا بتوانیم از آن فاکتور بگیریم. در نتیجه داریم:

$$\nabla_c \left( \sqrt{-g} R_\rho{}^{bcd} \right) = 0 \qquad (٤)$$

معادلات (٣) و (٤) معادلات میدان در روش وردشی پالاتینی هستند.

وردش کنش $S = \int d^4x \sqrt{-g}R^2$ به روش ساختار مقید مرتبه اول

در این روش کنش مساله به علت وجود قید $\Gamma^\mu_{\alpha\beta} = \left\{ \begin{matrix} \mu \\ \alpha\beta \end{matrix} \right\}$ به شکل زیر تغییر می کند:

$$S = \int d^4x \sqrt{-g} \left\{ R^2 + \Lambda_\mu{}^{\alpha\beta} \left( \Gamma^\mu_{\alpha\beta} - \left\{ \begin{matrix} \mu \\ \alpha\beta \end{matrix} \right\} \right) \right\} \tag{1}$$

معادلات میدان عبارت اند از معادله $\Gamma$ و معادلات $g$ به شکل زیر:

$$\frac{\delta \mathcal{L}}{\delta \Gamma^\mu_{\alpha\beta}}|_g + \Lambda_\mu{}^{\alpha\beta} = 0 \tag{2}$$

$$\frac{\delta \mathcal{L}}{\delta g^{\alpha\beta}}|_\Gamma + \sqrt{-g}B_{\alpha\beta} = 0 \tag{3}$$

الگوی محاسبات به شکلی است که در متن ارائه شد. ابتدا با وردش نسبت به $\Gamma$ ضرایب لاگرانژ را بدست می آوریم. این کار در قسمت های قبلی ضمیمه انجام شده است. یعنی داریم:

$$\tag{4}$$

$$\frac{\delta \mathcal{L}}{\delta \Gamma^\mu_{\alpha\beta}}|_g = 2\{\delta_c{}^\beta \nabla_d [R\sqrt{-g}g^{\alpha d}] - \nabla_c[g^{\alpha\beta}R\sqrt{-g}]\}|_g = 2\delta_c{}^\beta \sqrt{-g}g^{\alpha d}\nabla_d R - $$
$$2R\sqrt{-g}g^{\alpha\beta}\nabla_c R = 2\sqrt{-g}(\delta_c{}^\beta g^{\alpha d}\nabla_d R - \delta_c{}^d g^{\alpha\beta}\nabla_d R)$$

$$2\sqrt{-g}(\delta_c{}^\beta g^{\alpha d}\nabla_d R - \delta_c{}^d g^{\alpha\beta}\nabla_d R) + \sqrt{-g}\Lambda_c{}^{\alpha\beta} = 0 \tag{5}$$

از آنجا که فرض کرده ایم هموستار ها متقارن اند نتیجه می شود که ضرایب لاگرانژ نیز متقارن اند. معادله (۵) را برای $\Lambda_c{}^{\beta\alpha}$ به شکل زیر باز نویسی می کنیم:

$$2\sqrt{-g}(\delta_c{}^\alpha g^{\beta d}\nabla_d R - \delta_c{}^d g^{\beta\alpha}\nabla_d R) + \sqrt{-g}\Lambda_c{}^{\beta\alpha} = 0 \tag{6}$$

روابط (۵) و (۶) را جمع می کنیم

$$2\delta_c{}^\beta g^{\alpha d}\nabla_d R - 2\delta_c{}^d g^{\alpha\beta}\nabla_d R + 2\delta_c{}^\alpha g^{\beta d}\nabla_d R - 2\delta_c{}^d g^{\beta\alpha}\nabla_d R + \Lambda_c{}^{\beta\alpha} + \Lambda_c{}^{\alpha\beta} = $$
$$0$$

چون داریم $\Lambda_c{}^{\alpha\beta} = \Lambda_c{}^{\beta\alpha}$ و متریک نیز متقارن است خواهیم داشت:

$$2\delta_c{}^\beta g^{\alpha d}\nabla_d R - 4\delta_c{}^d g^{\alpha\beta}\nabla_d R + 2\delta_c{}^\alpha g^{\beta d}\nabla_d R + 2\Lambda_c{}^{\alpha\beta} = 0$$

$$\Lambda_c{}^{\alpha\beta} = (2\delta_c{}^d g^{\alpha\beta} - \delta_c{}^\beta g^{\alpha d} - \delta_c{}^\alpha g^{\beta d})\nabla_d R \qquad (7)$$

در مرحله بعدی معادله **g** را بدست می آوریم. این کار را قبلا در ضمیمه ت انجام دادیم

نتیجه عبارت است از:

$$\frac{\delta \mathcal{L}}{\delta g^{\alpha\beta}}\big|_\Gamma = \sqrt{-g}\left[-\frac{1}{2}R^2 g_{\alpha\beta} + 2RR_{\alpha\beta}\right]$$

طبق معادله (۳) داریم:

$$\sqrt{-g}\left[-\frac{1}{2}R^2 g_{\alpha\beta} + 2RR_{\alpha\beta}\right] + \sqrt{-g}B_{\alpha\beta} = 0 \qquad (8)$$

حالا $B_{\alpha\beta}$ را محاسبه می کنیم:

$$\sqrt{-g}B_{\alpha\beta} \equiv -\frac{1}{2}\sqrt{-g}\nabla^c(\Lambda_{\beta\alpha c} + \Lambda_{\alpha c\beta} - \Lambda_{c\alpha\beta})$$

حالا از رابطه (۷) سه ضریب لاگرانژ فوق را محاسبه می کنیم.

$$\Lambda_{c\alpha\beta} = (2\delta_c{}^d g_{\alpha\beta} - g_{c\beta}\delta_\alpha{}^d - g_{c\beta}\delta_\alpha{}^d)\nabla_d R$$

$$\Lambda_{\alpha c\beta} = (2\delta_\alpha{}^d g_{c\beta} - g_{\alpha\beta}\delta_c{}^d - g_{\alpha\beta}\delta_c{}^d)\nabla_d R$$

$$\Lambda_{\beta\alpha c} = (2\delta_\beta{}^d g_{\alpha c} - g_{c\beta}\delta_\alpha{}^d - g_{\beta c}\delta_\alpha{}^d)\nabla_d R$$

$$\Lambda_{\beta\alpha c} + \Lambda_{\alpha c\beta} - \Lambda_{c\alpha\beta} = \{2\delta_\beta{}^d g_{\alpha c} - g_{c\beta}\delta_\alpha{}^d - g_{\beta c}\delta_\alpha{}^d + 2\delta_\alpha{}^d g_{c\beta} - g_{\alpha\beta}\delta_c{}^d - $$
$$g_{\alpha\beta}\delta_c{}^d - 2\delta_c{}^d g_{\alpha\beta} + g_{c\beta}\delta_\alpha{}^d + g_{c\beta}\delta_\alpha{}^d\}\nabla_d R$$

$$\Lambda_{\beta\alpha c} + \Lambda_{\alpha c\beta} - \Lambda_{c\alpha\beta} = \{2\delta_\beta{}^d g_{\alpha c} + 2\delta_\alpha{}^d g_{c\beta} - 2g_{\alpha\beta}\delta_c{}^d - 2\delta_c{}^d g_{\alpha\beta}\}\nabla_d R$$

$$\nabla^c[\Lambda_{\beta\alpha c} + \Lambda_{\alpha c\beta} - \Lambda_{c\alpha\beta}] =$$

$$\{2\delta_\beta{}^d g_{\alpha c} + 2\delta_\alpha{}^d g_{c\beta} - 2g_{\alpha\beta}\delta_c{}^d - 2\delta_c{}^d g_{\alpha\beta}\}\nabla_d\nabla^c R = \{2\delta_\beta{}^d\nabla_d\nabla_\alpha R +$$

$$2\delta_\alpha{}^d\nabla_d\nabla_\beta R - 2g_{\alpha\beta}\nabla_d\nabla^d R - 2g_{\alpha\beta}\nabla_d\nabla^d R\} = 4\nabla_d\nabla_\beta R - 4g_{\alpha\beta}\nabla_d\nabla^d R$$

$$B_{\alpha\beta} = 2g_{\alpha\beta}\nabla_d\nabla^d R - 2\nabla_d\nabla_\beta R \qquad (\text{٩})$$

اگر (٩) را در (٨) قرار دهیم خواهیم داشت:

$$\sqrt{-g}\left[-\frac{1}{2}R^2 g_{\alpha\beta} + 2RR_{\alpha\beta}\right] + \sqrt{-g}\left[2g_{\alpha\beta}\nabla_d\nabla^d R - 2\nabla_d\nabla_\beta R\right] = 0$$

$$-\frac{1}{2}R^2 g_{\alpha\beta} + 2RR_{\alpha\beta} + 2g_{\alpha\beta}\nabla_d\nabla^d R - 2\nabla_d\nabla_\beta R = 0 \qquad (\text{١٠})$$

که این معادله را قبلا توسط وردش متریکی بدست آوردیم.

ضمیمه ح

وردش کنش $S = \frac{1}{\chi} \int d^4x \sqrt{-g} R_{\mu\nu} R^{\mu\nu}$ به روش ساختار مقید مرتبه اول

در این روش کنش مساله به علت وجود قید $\Gamma^{\mu}{}_{\alpha\beta} = \begin{Bmatrix} \mu \\ \alpha\beta \end{Bmatrix}$ به شکل زیر تغییر می کند:

$$S = \int d^4x \sqrt{-g} \left\{ R_{\mu\nu} R^{\mu\nu} + \Lambda_\mu{}^{\alpha\beta} \left( \Gamma^{\mu}{}_{\alpha\beta} - \begin{Bmatrix} \mu \\ \alpha\beta \end{Bmatrix} \right) \right\} \qquad (\text{۱})$$

برای معادله $\Gamma$ داریم (جمله اول کنش زیر را در قبلا حساب کردیم):

$$\delta S = \int d^4x + \left\{ 2 \left[ \delta_c{}^\beta \nabla_\nu (\sqrt{-g} R^{\alpha\nu}) - \nabla_c (\sqrt{-g} R^{\alpha\beta}) \right] + \Lambda_\mu{}^{\alpha\beta} \right\} \delta \Gamma^{\mu}{}_{\alpha\beta} = 0$$

$$2 \left[ \delta_c{}^\beta \nabla_\nu (\sqrt{-g} R^{\alpha\nu}) - \nabla_c (\sqrt{-g} R^{\alpha\beta}) \right] + \sqrt{-g} \Lambda_\mu{}^{\alpha\beta} = 0 \qquad (\text{۲})$$

برای $\Gamma^{\mu}{}_{\beta\alpha}$ نیز می نویسیم

$$2 \left[ \delta_c{}^\alpha \nabla_\nu (\sqrt{-g} R^{\beta\nu}) - \nabla_c (\sqrt{-g} R^{\beta\alpha}) \right] + \sqrt{-g} \Lambda_\mu{}^{\beta\alpha} = 0 \qquad (\text{۳})$$

اگر این دو را جمع کنیم خواهیم داشت:

$$\sqrt{-g} \Lambda_\mu{}^{\alpha\beta} = 2 \nabla_c (\sqrt{-g} R^{\alpha\beta}) - \delta_c{}^\beta \nabla_\nu (\sqrt{-g} R^{\alpha\nu}) - \delta_c{}^\alpha \nabla_\nu (\sqrt{-g} R^{\beta\nu})$$

که البته چون در وردش گیری نسبت به هموستار متریک ثابت فرض می شود، $\sqrt{-g}$ از داخل پرانتز ها بیرون می آید:

$$\Lambda_\mu{}^{\alpha\beta} = 2 \nabla_c (R^{\alpha\beta}) - \delta_c{}^\beta \nabla_\nu (R^{\alpha\nu}) - \delta_c{}^\alpha \nabla_\nu (R^{\beta\nu}) \qquad (\text{٤})$$

برای معادله $\mathbf{g}$ داریم ($\frac{\delta \mathcal{L}}{\delta g^{\alpha\beta}}|_\Gamma$ را در قبلا محاسبه کردیم):

$$\frac{\delta \mathcal{L}}{\delta \mathrm{g}^{\alpha\beta}}|_{\Gamma} + \sqrt{-g}B_{\alpha\beta} = 0$$

$$-\frac{1}{2}R_{\mu\nu}R^{\mu\nu}g_{\alpha\beta} + R_{\alpha}{}^{\nu}R_{\beta\nu} + R^{\nu}{}_{\alpha}R_{\nu\beta} + \sqrt{-g}B_{\alpha\beta} = 0 \qquad (\text{٥})$$

حال به محاسبه $B_{\alpha\beta}$ از طریق ضرایب لاگرانژ طبق معادله زیر می پردازیم:

$$\sqrt{-g}B_{\alpha\beta} \equiv -\frac{1}{2}\sqrt{-g}\nabla^c(\Lambda_{\beta\alpha c} + \Lambda_{\alpha c\beta} - \Lambda_{c\alpha\beta})$$

$$\Lambda_{c\alpha\beta} = 2\nabla_c(\mathrm{R}_{\alpha\beta}) - \mathrm{g}_{c\beta}\nabla_\nu(R_\alpha{}^\nu) - \mathrm{g}_{c\alpha}\nabla_\nu(R_\beta{}^\nu)$$

$$\Lambda_{\alpha c\beta} = 2\nabla_\alpha(\mathrm{R}_{c\beta}) - \mathrm{g}_{\alpha\beta}\nabla_\nu(R_c{}^\nu) - \mathrm{g}_{\alpha c}\nabla_\nu(R_\beta{}^\nu)$$

$$\Lambda_{\beta\alpha c} = 2\nabla_\beta(\mathrm{R}_{\alpha c}) - \mathrm{g}_{\beta c}\nabla_\nu(R_\alpha{}^\nu) - \mathrm{g}_{\beta\alpha}\nabla_\nu(R_c{}^\nu)$$

$$\Lambda_{\beta\alpha c} + \Lambda_{\alpha c\beta} - \Lambda_{c\alpha\beta} = 2\nabla_\beta(\mathrm{R}_{\alpha c}) - \mathrm{g}_{\beta c}\nabla_\nu(R_\alpha{}^\nu) - \mathrm{g}_{\beta\alpha}\nabla_\nu(R_c{}^\nu) + 2\nabla_\alpha(\mathrm{R}_{c\beta}) -$$

$$\mathrm{g}_{\alpha\beta}\nabla_\nu(R_c{}^\nu) - \mathrm{g}_{\alpha c}\nabla_\nu(R_\beta{}^\nu) - 2\nabla_c(\mathrm{R}_{\alpha\beta}) + \mathrm{g}_{c\beta}\nabla_\nu(R_\alpha{}^\nu) + \mathrm{g}_{c\alpha}\nabla_\nu(R_\beta{}^\nu) =$$

$$2\nabla_\beta(\mathrm{R}_{\alpha c}) + 2\nabla_\alpha(\mathrm{R}_{c\beta}) - 2\mathrm{g}_{\alpha\beta}\nabla_\nu(R_c{}^\nu) - 2\nabla_c(\mathrm{R}_{\alpha\beta}) =$$

$$-\frac{1}{2}\sqrt{-g}\nabla^c(\Lambda_{\beta\alpha c} + \Lambda_{\alpha c\beta} - \Lambda_{c\alpha\beta}) =$$

$$-\frac{1}{2}\sqrt{-g}\{4\nabla_\beta\nabla^c(\mathrm{R}_{\alpha c}) - 2\mathrm{g}_{\alpha\beta}\nabla_\nu\nabla^c(R_c{}^\nu) - 2\nabla_c\nabla^c(\mathrm{R}_{\alpha\beta})\} =$$

$$\sqrt{-g}\{\nabla_c\nabla^c(\mathrm{R}_{\alpha\beta}) + \mathrm{g}_{\alpha\beta}\nabla_\nu\nabla^c(R_c{}^\nu) - 2\nabla_\beta\nabla^c(\mathrm{R}_{\alpha c})\}$$

$$\sqrt{-g}B_{\alpha\beta} = \sqrt{-g}\{\nabla_c\nabla^c(\mathrm{R}_{\alpha\beta}) + \mathrm{g}_{\alpha\beta}\nabla_\nu\nabla^c(R_c{}^\nu) - 2\nabla_\beta\nabla^c(\mathrm{R}_{\alpha c})\} \qquad (\text{٦})$$

حال اگر (٦) را در (٥) قرار دهیم داریم

$$(\text{٧})$$

$$-\frac{1}{2}R_{\mu\nu}R^{\mu\nu}g_{\alpha\beta} + R_{\alpha}{}^{\nu}R_{\beta\nu} + R^{\nu}{}_{\alpha}R_{\nu\beta} + \nabla_c\nabla^c(\mathrm{R}_{\alpha\beta}) + \mathrm{g}_{\alpha\beta}\nabla_\nu\nabla^c(R_c{}^\nu) -$$

$$2\nabla_\beta\nabla^c(\mathrm{R}_{\alpha c}) = 0$$

طبق نکته (۱) که در ضمیمه ت گفته شد داریم

$$g_{\alpha\beta}\nabla_\nu\nabla^c(R_c{}^\nu) = g_{\alpha\beta}\nabla^\nu\nabla^c(R_{c\nu}) = \frac{1}{2}g_{\alpha\beta}\Box R$$

و طبق نکته (۳) همان ضمیمه

$$R_\alpha{}^\nu R_{\beta\nu} + R^\nu{}_\alpha R_{\nu\beta} = 2R^\nu{}_\alpha R_{\nu\beta} = 2R_{\alpha c;\beta}{}^c - R_{;\alpha\beta} + 2R_{\alpha\rho\beta\sigma}R^{\rho\sigma}$$

می باشد. حالا آنها را در (۷) جاگذاری می کنیم:

$$-\frac{1}{2}R_{\mu\nu}R^{\mu\nu}g_{\alpha\beta} + 2R_{\alpha c;\beta}{}^c - R_{;\alpha\beta} + 2R_{\alpha\rho\beta\sigma}R^{\rho\sigma} + \Box R_{\alpha\beta} + \frac{1}{2}g_{\alpha\beta}\Box R -$$
$$2\nabla_\beta\nabla^c(R_{\alpha c}) = 0$$

جمله دوم و جمله آخر با هم ساده می شوند.رابطه بالا از نکته (۲) ضمیمه ت جاگذاری می کنیم

$$-\frac{1}{2}R_{\mu\nu}R^{\mu\nu}g_{\alpha\beta} - R_{;\alpha\beta} + 2R_{\alpha\rho\beta\sigma}R^{\rho\sigma} + \Box R_{\alpha\beta} + \frac{1}{2}g_{\alpha\beta}\Box R = 0$$

این معادله ای است که قبلا در روش متریکی بدست آوردیم.

ضمیمه خ

توضیح در مورد روابط (۳-۱۷) و (۳-۱۹)

با توجه به معادله میدان (۲-۶) و این که برای ماده کامل $\rho = -T^t{}_t = T_{tt}$ و $P = T^i{}_i$ هستند و استفاده از رد معادله این دو معادله بدست می آیند:

$$f'(R)R_{\alpha\beta} - \frac{f(R)}{2}g_{\alpha\beta} - \left(\nabla_\alpha\nabla_\beta - g_{\alpha\beta}g^{\alpha\beta}\nabla_\alpha\nabla_\beta\right)f'(R) = T_{\alpha\beta}$$

(در بند الف، این معادله میدان برای خلاء بدست آمد. محاسبه برای حالتی که تانسور ماده داریم سر راست است).

$$\rho = T_{00} = f'(R)R_{00} - \frac{f(R)}{2}g_{00} - \left(\nabla_0\nabla_0 - g_{00}\Box\right)f'(R)$$

با استفاده از رایانه و یا به طور دستی می توان $R_{00}$ را حساب کرد. نتیجه عبارت است از:

$$R_{00} = 3\frac{\ddot{a}}{a} = -3qH^2$$

و طبق قرار دادی که در قبل ذکر شد $g_{00} = -1$ و در بند چ گفتیم که نتیجه عبارت داخل پرانتز $-3(\frac{\dot{a}}{a})f''(R)\dot{R}$ است. در نتیجه داریم:

$$\rho = T_{00} = -3f'(R)qH^2 - \frac{f(R)}{2} + 3(\frac{\dot{a}}{a})f''(R)\dot{R} \tag{۱}$$

با استفاده از رد معادله میدان و معادله چگالی که بدست آوردیم رابطه فشار نیز بدست  می آید.

<div dir="rtl">

استخراج معادلات مربوط به (٣-٢٤)

در فصل مربوط به شرایط انرژی چهار مدل بررسی شده روش اسخراج معادلات مربوط به آن ها مشابه است. جهت اطلاع یکی از آن ها مورد بررسی قرار می گیرد.

</div>

$$f(R) = R + \alpha R^n \tag{١}$$

$$f'(R) = 1 + n\alpha R^{n-1}$$

$$f''(R) = n(n-1)\alpha R^{n-2}$$

<div dir="rtl">

با استفاده از (٣-١٤) داریم:

</div>

$$R_0 + \alpha R_0{}^n + 6H_0{}^2(1-q_0)\left(1 + n\alpha R_0{}^{n-1}\right) + 36H_0{}^4(j_0 - q_0 - 2)n(n-1)\alpha R_0{}^{n-2} \geq 0$$

<div dir="rtl">

می دانیم:
</div>

$$R_0 = -6H_0{}^2(1-q_0) = -|R_0|$$

$$1 - q_0 = -\frac{R_0}{6H_0{}^2}$$

$$36H_0{}^4 = \frac{R_0{}^2}{(1-q_0)^2}$$

$$A \equiv \frac{j_0 - q_0 - 2}{(1-q_0)^2}$$

<div dir="rtl">

حال چهار عبارت فوق را در عبارت اول جاگذاری می کنیم

</div>

$$-6H_0{}^2(1-q_0) + \alpha(-|R_0|)^n + 6H_0{}^2(1-q_0) -$$

$$\underbrace{6H_0{}^2(1-q_0)}_{-|R_0|} n\alpha(-|R_0|)^{n-1} + \left(\frac{R_0{}^2}{(1-q_0)^2}\right)(j_0 - q_0 - 2)n^2\alpha(-|R_0|)^{n-2} -$$

$$\left(\frac{R_0{}^2}{(1-q_0)^2}\right)(j_0 - q_0 - 2)\alpha n(-|R_0|)^{n-2} \geq 0$$

<div dir="rtl" align="center">

١١٧

</div>

$$\alpha(-|R_0|)^n - n\alpha(-|R_0|)^n + An^2\alpha(-|R_0|)^n - A\alpha n(-|R_0|)^n \geq 0$$

$$\alpha(-1)^n - n\alpha(-1)^n + An^2\alpha(-1)^n - A\alpha n(-1)^n \geq 0$$

$$(-1)^n\{\alpha - n\alpha + An^2\alpha - A\alpha n\} \geq 0$$

$$\alpha(-1)^n(An^2 - (A+1)n + 1) \geq 0 \qquad\qquad (\text{۲})$$

شرط بعدی (۳-۲۶) می باشد که مشابه روش بالا بدست می آید.

اثبات رابطه (٤-٤)

با استفاده از رابطه زیر که در ضمیمه ث بدست آمد داریم:

$$R_{\alpha\beta} = \frac{1}{f'(R)}\{\frac{1}{2}f(R)g_{\alpha\beta} + f'(R)^{;\rho\sigma}(g_{\rho\alpha}g_{\sigma\beta} - g_{\rho\sigma}g_{\alpha\beta})\}$$

$$G_{\alpha\beta} = R_{\alpha\beta} - \frac{1}{2}g_{\alpha\beta}R = \frac{1}{f'(R)}\{\frac{1}{2}f(R)g_{\alpha\beta} + f'(R)^{;\rho\sigma}(g_{\rho\alpha}g_{\sigma\beta} - g_{\rho\sigma}g_{\alpha\beta})\} -$$

$$\frac{1}{2}g_{\alpha\beta}R = \frac{1}{f'(R)}\{\frac{1}{2}f(R)g_{\alpha\beta} + f'(R)^{;\rho\sigma}(g_{\rho\alpha}g_{\sigma\beta} - g_{\rho\sigma}g_{\alpha\beta}) - \frac{1}{2}g_{\alpha\beta}Rf'(R)\} =$$

$$\frac{1}{f'(R)}\{\frac{1}{2}g_{\alpha\beta}[f(R) - Rf'(R)] + f'(R)^{;\rho\sigma}(g_{\rho\alpha}g_{\sigma\beta} - g_{\rho\sigma}g_{\alpha\beta})\}$$

در نتیجه داریم:

$$T^{curve}_{\alpha\beta} \equiv \frac{1}{f'(R)}\{\frac{1}{2}g_{\alpha\beta}[f(R) - Rf'(R)] + f'(R)^{;\rho\sigma}(g_{\rho\alpha}g_{\sigma\beta} - g_{\rho\sigma}g_{\alpha\beta})\} \qquad (1)$$

پ) توضیحاتی در مورد رابطه (٤-٧) و روابط منتج از آن

روشی که برای استخراج معادلات میدان در اینجا استفاده شده است، روش کانونیکال کردن کنش (٤-١) است. در این روش ضریب مقیاس $a$ و اسکالر ریچی $R$ به عنوان متغیر های مستقل در نظر گرفته می شوند. حال اگر متریک مساله را فریدمان – رابرتسون – واکر انتخاب کنیم بعد از محاسبه می بینیم که رابطه $R = -6\left[\frac{\ddot{a}}{a} + \left(\frac{\dot{a}}{a}\right)^2 + \frac{k}{a^2}\right]$ به عنوان یک قید بدست می آید. با استفاده از روش ضرایب لاگرانژ که در فصل قبل توضیح داده شد جمله قیدی به شکل زیر به منش مساله اضافه می شود:

$$S = \int d^4x\sqrt{-g}\left\{f(R) - \lambda\left(R + 6\left[\frac{\ddot{a}}{a} + \left(\frac{\dot{a}}{a}\right)^2 + \frac{k}{a^2}\right]\right)\right\} \qquad (2)$$

حال اگر از کنش فوق نسبت به $R$ وردش بگیریم بدست می آوریم $\lambda = f'(R)$ با جاگذاری این نتیجه،

کنش به صورت زیر تغییر می کند:

$$S = \int d^4\sqrt{-g}\left\{f(R) - f'(R)\left(R + 6\left[\frac{\ddot{a}}{a} + \left(\frac{\dot{a}}{a}\right)^2 + \frac{k}{a^2}\right]\right)\right\}$$

حالا کنش به شکل شبهه–نقطه ای[1] می نویسیم یعنی:

$$S = \int d^3x \int dt \sqrt{-g}\left\{f(R) - f'(R)\left(R + 6\left[\frac{\ddot{a}}{a} + \left(\frac{\dot{a}}{a}\right)^2 + \frac{k}{a^2}\right]\right)\right\} \qquad (٣)$$

از آنجایی که برای متریک فریدمان – رابرتسون – واکر داریم $\sqrt{-g} = a^3(t)\frac{r^2}{\sqrt{1-kr^2}}$، وقتی کنش را

به شکل شبهه–نقطه ای می نویسیم در واقع این کار بدین معنی است که ما علاقه مند به محاسبه یک

قسمت یک بعدی کنش هستیم. چون کنش ما به طور غیر صریح تابع زمان است و هدف ما استخراج رفتار

دینامیکی معادلات است، در اینجا ما تنها علاقه مند به کار با قسمت زمانی کنش هستیم:

$$S = \int \frac{r^2}{\sqrt{1-kr^2}}d^3x \int dt\, a^3\left\{f(R) + f'(R)\left(R + 6\left[\frac{\ddot{a}}{a} + \left(\frac{\dot{a}}{a}\right)^2 + \frac{k}{a^2}\right]\right)\right\}$$

پس تعریف می کنیم:

$$A_{(curv)} \equiv \int dt\, a^3(t)\left\{f(R) - f'(R)\left(R + 6\left[\frac{\ddot{a}}{a} + \left(\frac{\dot{a}}{a}\right)^2 + \frac{k}{a^2}\right]\right)\right\} \qquad (٤)$$

حال کنش فوق را ساده کرده و سپس از روش انتگرال گیری جزء به جزء استفاده می کنیم:

$$A_{(curv)} = \int dt\{a^3 f(R) - a^3 f'(R)R - 6\, a^3 f'(R)\frac{\ddot{a}}{a} - 6\, a^3 f'(R)\left(\frac{\dot{a}}{a}\right)^2 -$$
$$6\, a^3 f'(R)\frac{k}{a^2}\}$$

---

[1] point-like

$$A_{(curv)} = \int dt\{ a^3 f(R) - a^3 f'(R)R - 6\ddot{a}\, a^2 f'(R) - 6\dot{a}^2 a f'(R) - 6a f'(R)k\}$$

جمله سوم را به صورت زیر ساده می کنیم:

$$6\ddot{a}\, a^2 f'(R) = 6\frac{\mathrm{d}}{\mathrm{d}t}(a^2 f'(R)\dot{a}) - 12a\dot{a}^2 f'(R) - 6a^2 \dot{a}\frac{\mathrm{d}}{\mathrm{d}t} f'(R)$$

اگر از عبارت فوق نسبت به زمان انتگرال بگیریم جمله اول سمت راست یک مشتق کامل است و به علت شرایط مرزی انتگرال زمانی آن صفر می شود. برای جمله آخر نیز از مشتق گیری زنجیره ای استفاده می کنیم:

$$\frac{d}{dt} f'(R) = \frac{df'(R)}{dR}\frac{dR}{dt} = f''(R)\dot{R}$$

حال در (۵) به جای $6\ddot{a}\, a^2 f'(R)$ طبق توضیحات بالا جاگذاری می کنیم:

$$A_{(curv)} = \int dt\{ a^3 f(R) - a^3 f'(R)R + 12a\dot{a}^2 f'(R) + 6a^2 \dot{a}\dot{R}f''(R) - 6\dot{a}^2 a f'(R) - 6a f'(R)k\}$$

بعد از ساده سازی داریم:

$$A_{(curv)} = \int dt\{ a^3 (f(R) - f'(R)R) + 6a\dot{a}^2 f'(R) + 6a^2 \dot{a}\dot{R}f''(R) - 6a f'(R)k\}$$

حال کنش را به شکل زیر می نویسیم:

$$A_{(curv)} = \int dt\, \mathcal{L}\big(a, \dot{a}; R, \dot{R}\big)$$

که تعریف کرده ایم:

$$\mathcal{L}(a, \dot{a}; R, \dot{R}) = a^3\big(f(R) - Rf'(R)\big) + 6a\dot{a}^2 f'(R) + 6a^2\dot{a}\dot{R}f''(R) - 6af'(R)k$$

ج) استخراج روابط (٤-٨) تا (٤-١٠)

معادلات حرکت از معادلات اویلر – لاگرانژ برای متغیر های مستقل مساله به شکل معمول و شرط هامیلتونی بدست می آیند. در اینجا چون ماده از مساله حذف شده است یعنی مساله را در حالت خلاء بررسی می کنیم هامیلتونی کل صفر است.

معادله اویلر – لاگرانژ برای ضریب مقیاس عبارت است از:

$$\frac{\partial}{\partial t}\left(\frac{\partial \mathcal{L}}{\partial \dot{a}}\right) - \frac{\partial \mathcal{L}}{\partial a} = 0 \tag{٧}$$

$$\frac{\partial}{\partial t}\Big(12a\dot{a}f'(R) + 6a^2\dot{R}f''(R)\Big) - 3a^2[f(R) - Rf'(R)] - 6\dot{a}^2 f'(R) - 12a\dot{a}\dot{R}f''(R) + 6kf' = 0$$

$$12\dot{a}^2 f'(R) + 12a\ddot{a}f'(R) + 12a\dot{a}\dot{R}f''(R) + 12a\dot{a}\dot{R}f''(R) + 6a^2\ddot{R}f''(R) + 6a^2\dot{R}^2 f'''(R) - 3a^2 f(R) + 3a^2 Rf'(R) - \dot{a}^2 f'(R) - 12a\dot{a}\dot{R}f''(R) + 6kf'(R) = 0$$

با کمی ساده سازی به راحتی به معادله (٤-٨) می رسیم:

$$\tag{٨}$$

$$2\left(\frac{\ddot{a}}{a}\right) + \left(\frac{\dot{a}}{a}\right)^2 + \frac{k}{a^2} = -\frac{1}{f'(R)}\{2\left(\frac{\dot{a}}{a}\right)\dot{R}f''(R) + \ddot{R}f''(R) + \dot{R}^2 f'''(R) - \frac{1}{2}[f(R) - Rf'(R)]\} = -p_{(curv)}$$

معادله اولر – لاگرانژ برای اسکالر ریچی عبارت است از:

$$\frac{\partial}{\partial t}\left(\frac{\partial \mathcal{L}}{\partial \dot{R}}\right) - \frac{\partial \mathcal{L}}{\partial R} = 0 \qquad (۹)$$

$$12a\dot{a}^2 f''(R) + 6a^2\ddot{a}f''(R) + 6a^2\dot{a}\frac{\partial}{\partial t}f''(R) + a^3 R f''(R) - 6a\dot{a}^2 f''(R) -$$
$$6a^2\dot{a}\frac{\partial}{\partial t}f''(R) + 6kaf''(R) = 0$$

$$6a\dot{a}^2 f''(R) + 6a^2\ddot{a}f''(R) + a^3 R f''(R) + 6kaf''(R) = 0$$

$$f''(R)\left\{R + 6\left[\frac{\ddot{a}}{a} + \left(\frac{\dot{a}}{a}\right)^2 + \frac{k}{a^2}\right]\right\} = 0 \qquad (۱۰)$$

شرط صفر بودن هامیلتونی به شکل زیر است:

$$\mathcal{H} = \dot{q}\frac{\partial \mathcal{L}}{\partial \dot{q}} - \mathcal{L} = 0 \qquad (۱۱)$$

که جمله اول جمع روی کلیه متغیر های مستقل مساله است.

$$\mathcal{H} = \dot{a}\left[12a\dot{a}\ddot{a}f'(R) + 6a^2\dot{R}f''(R)\right] + \dot{R}[6a^2\dot{a}f''(R)] - \left[a^3\big(f(R) - Rf'(R)\big) + 6a\dot{a}^2 f'(R) + 6a^2\dot{a}\dot{R}f''(R) - 6af'(R)k\right] = 0$$

$$\mathcal{H} = 12a\dot{a}^2 f'(R) + 6a^2\dot{a}\dot{R}f''(R) + 6\dot{R}a^2\dot{a}f''(R) - a^3\big(f(R) - Rf'(R)\big) - 6a\dot{a}^2 f'(R) - 6a^2\dot{a}\dot{R}f''(R) + 6af'(R)k = 0$$

$$\mathcal{H} = 6a\dot{a}^2 f'(R) + 6a^2\dot{a}\dot{R}f''(R) - a^3[f(R) - Rf'(R)] + 6af'(R)k = 0$$

$$6a\dot{a}^2 f'(R) + 6af'(R)k = -6a^2\dot{a}\dot{R}f''(R) + a^3[f(R) - Rf'(R)]$$

$$(۱۲)$$

$$\left(\frac{\dot{a}}{a}\right)^2 + \frac{k}{a^2} = \frac{1}{3}\frac{1}{f'(R)}\left\{\frac{1}{2}[f(R) - Rf'(R)] - 3\left(\frac{\dot{a}}{a}\right)\dot{R}f''(R)\right\} = \frac{1}{3}\rho_{(curv)}$$

می دانیم برای یک ماده کامل تانسور انرژی – تکانه قطری است. اگر از همین الگو برای تعریف فشار و چگالی انرژی خلاء استفاده کنیم داریم:

$$\rho_{curv} \equiv -T_0^{\ 0\,curv} = T_{00}^{\ curv} \tag{١٣}$$

$$p_{curv} \equiv T_i^{\ i\,curv} \tag{١٤}$$

که $T_i^{\ i}{}_{curv}$ هریک از سه مولفه فضایی این تانسور است. اگر از علامت (+ + + -) برای متریک فریدمان – رابرتسون – واکر استفاده کنیم داریم:

$$T_{\alpha\beta}^{curve} = \frac{1}{f'(R)}\{\tfrac{1}{2}g_{\alpha\beta}[f(R) - Rf'(R)] + f'(R)^{;\rho\sigma}(g_{\rho\alpha}g_{\sigma\beta} - g_{\rho\sigma}g_{\alpha\beta})\} =$$

$$\frac{1}{f'(R)}\{\tfrac{1}{2}g_{\alpha\beta}[f(R) - Rf'(R)] + f'(R)_{;\alpha\beta} - g_{\alpha\beta}\Box R\}$$

که تعریف کرده ایم $\Box R = g_{\rho\sigma}f'(R)^{;\rho\sigma} = f'(R)_{;\rho}^{\ ;\rho}$

$$T_{00}^{curve} = \frac{1}{f'(R)}\left\{\tfrac{1}{2}g_{00}[f(R) - Rf'(R)] + f'(R)_{;00} - g_{00}\Box R\right\} =$$

$$\frac{1}{f'(R)}\left\{\tfrac{1}{2}[f(R) - Rf'(R)] + f'(R)_{;00} - \Box R\right\}$$

به طور دستی یا با راینه می توان جمله آخر را محاسبه کرد:

$$\Box R = f'''(R)\dot{R} + f''(R)\dot{R}^2 + 3(\tfrac{\dot{a}}{a})f''(R)\dot{R}$$

جمله دوم نیز به شکل زیر ساده می شود:

$$f'(R)_{;00} = \frac{\partial}{\partial t}\left(\frac{\partial}{\partial t}f'(R)\right) = \frac{\partial}{\partial t}\left(\frac{\partial f'(R)}{\partial R}\frac{\partial R}{\partial t}\right) = \frac{\partial}{\partial t}\left(f''(R)\dot{R}\right) = f'''(R)\dot{R} +$$
$$f''(R)\dot{R}^2$$

اگر عبارات فوق را جاگذاری کنیم خواهیم داشت:

$$\rho_{curv} = \mathrm{T}_{00}^{\mathrm{curve}} = -\frac{1}{f'(R)}\left\{\frac{1}{2}[f(R) - Rf'(R)] - 3\left(\frac{\dot{a}}{a}\right)f''(R)\dot{R}\right\} \qquad (\backslash\delta)$$

برای فشار از رد تانسور انرژی – تکانه خمش استفاده می کنیم:

$$T \equiv trace\left(T_{\alpha\beta}^{curve}\right) = -\rho + 3p = \frac{1}{f'(R)}\mathrm{g}^{\alpha\beta}\left\{\frac{1}{2}\mathrm{g}_{\alpha\beta}[f(R) - Rf'(R)] +\right.$$
$$\left. f'(R)_{;\alpha\beta} - \mathrm{g}_{\alpha\beta}\Box R\right\}$$

$$T = \left\{2[f(R) - Rf'(R)] + f'(R)_{;\alpha}{}^{;\alpha} - 4R\right\} = \left\{2[f(R) - Rf'(R)] - 3\Box R\right\}$$

$$-\rho + 3p = 2[f(R) - Rf'(R)] - 3\Box R$$

$$p = \frac{1}{3}\{\rho + (2[f(R) - Rf'(R)] - 3\Box R)\}$$

با جاگذاری $\Box R$ و چگالی و یک ساده سازی مختصر به عبارت مورد نظر برای فشار دست خواهیم یافت:

$$(\backslash\mathcal{F})$$

$$p_{(curv)} = -\frac{1}{f'(R)}\left\{2\left(\frac{\dot{a}}{a}\right)\dot{R}f''(R) + \ddot{R}f''(R) + \dot{R}^2f'''(R) - \frac{1}{2}[f(R) - Rf'(R)]\right\}$$

وردش رابطه (۵-۱۰-ب)

وردش کنش $R^{1+\delta}$

$$S = \int \mathscr{L}_G \, d^4 x + S_m$$

با در نظر گرفتن چگالی لاگرانژی $\mathscr{L}_G = \dfrac{1}{\chi} \sqrt{-g} \, R^{1+\delta}$ وردش کنش را به صورت زیر محاسبه می‌کنیم:

$$\delta S = \delta S_m + \int \delta \mathscr{L}_G \, d^4 x \quad , \quad \delta S_m = -\frac{1}{2} \int T_{\mu\nu} \sqrt{-g} \, \delta g^{\mu\nu} d^4 x$$

که $T_{ab}$ تانسور انرژی-تکانه است.

$$\delta \mathscr{L}_G = \delta (\frac{1}{\chi} \sqrt{-g} \, R^{1+\delta}) = \frac{1}{\chi} \left[ \delta(\sqrt{-g}) R^{1+\delta} + \sqrt{-g} \, (1+\delta) R^{\delta} \delta R \right]$$

$$\delta \sqrt{-g} = -\frac{1}{2} \sqrt{-g} \, g_{\mu\nu} \delta g^{\mu\nu} \quad , \quad \delta R = g^{\alpha\beta} g^{\mu\nu} (\delta g_{\mu\alpha;\nu\beta} - \delta g_{\mu\nu;\alpha\beta}) + R_{\mu\nu} \delta g^{\mu\nu}$$

که عبارت سمت راست نتیجه محاسبه‌ای است کـه جداگانـه آورده خواهـد شـد. (جـزوه تکمیلـی دکتـر فرهودی)

$$\delta \mathscr{L}_G = \frac{\sqrt{-g}}{\chi} \left[ -\frac{1}{2} R^{1+\delta} g_{\mu\nu} \delta g^{\mu\nu} + (1+\delta) \left( {\color{red} R^{\delta} g^{\mu\nu} g^{\alpha\beta} \delta g_{\mu\alpha;\nu\beta}} - R^{\delta} g^{\mu\nu} g^{\alpha\beta} \delta g_{\mu\nu;\alpha\beta} \right) + R_{\mu\nu} \delta g^{\mu\nu} \right]$$

جمله مشخص شده در عبارت فوق را به صورت زیر بازنویسی می‌کنیم:

$$R^{\delta} g^{\mu\nu} g^{\alpha\beta} \delta g_{\mu\alpha;\nu\beta} = \left( R^{\delta} g^{\mu\nu} g^{\alpha\beta} \delta g_{\mu\alpha;\nu} \right)_{;\beta} - R^{\delta}_{\ ;\beta} g^{\mu\nu} g^{\alpha\beta} \delta g_{\mu\alpha;\nu}$$

حاصل انتگرال جمله سمت چپ عبارت فوق به علت اینکه یک دیورژانس کامل است و با استفاده از قضیه استوکس برابر صفر خواهد بود. در نتیجه:

$$R^{\delta} g^{\mu\nu} g^{\alpha\beta} \delta g_{\mu\alpha;\nu\beta} = -R^{\delta}_{\ ;\beta} g^{\mu\nu} g^{\alpha\beta} \delta g_{\mu\alpha;\nu}$$

با تکرار تکنیک مشابه خواهیم داشت:

$$R^{\delta}{}_{;\gamma\beta}g^{\mu\nu}g^{\alpha\beta}\delta g_{\mu\alpha;\gamma\beta} = -R^{\delta}{}_{;\beta}g^{\mu\nu}g^{\alpha\beta}\delta g_{\mu\alpha;\nu} = R^{\delta}{}_{;\beta\nu}g^{\mu\nu}g^{\alpha\beta}\delta g_{\mu\alpha}$$

برای جمله مشابه نیز به همین طریق می‌توانیم بنویسیم:

$$R^{\delta}{}_{;\gamma}g^{\mu\nu}g^{\alpha\beta}\delta g_{\mu\nu;\alpha\beta} = -R^{\delta}{}_{;\beta}g^{\mu\nu}g^{\alpha\beta}\delta g_{\mu\nu;\alpha} = R^{\delta}{}_{;\beta\alpha}g^{\mu\nu}g^{\alpha\beta}\delta g_{\mu\nu}$$

در نتیجه:

$$\delta\mathcal{L}_G = \frac{\sqrt{-g}}{\chi}[-\frac{1}{2}g_{\mu\nu}R^{1+\delta}\delta g^{\mu\nu} + (1+\delta)R^{\delta}R_{\mu\nu}\delta g^{\mu\nu}$$
$$+ (1+\delta)R^{\delta}{}_{;\beta\nu}g^{\mu\nu}g^{\alpha\beta}\delta g_{\mu\alpha} - (1+\delta)R^{\delta}{}_{;\beta\alpha}g^{\mu\nu}g^{\alpha\beta}\delta g_{\mu\nu}]$$

با استفاده از این واقعیت:

$$R^{\delta}{}_{;\gamma\beta} = (R^{\delta}{}_{;\gamma})_{;\beta} = (\delta R_{,\nu}R^{\delta-1})_{;\beta} = \delta R^{\delta}\frac{R_{;\beta\nu}}{R} + \delta(\delta-1)R^{\delta}\frac{R_{,\beta}R_{,\nu}}{R^2}$$

جملات سوم و چهارم داخل براکت را می‌توانیم به صورت زیر بنویسیم:

$$\delta(1+\delta)R^{\delta}\frac{R_{;\beta\nu}}{R}g^{\mu\nu}g^{\alpha\beta}\delta g_{\mu\alpha} - \delta(1-\delta^2)R^{\delta}\frac{R_{,\beta}R_{,\nu}}{R^2}g^{\mu\nu}g^{\alpha\beta}\delta g_{\mu\alpha}$$

$$-\delta(1+\delta)R^{\delta}\frac{R_{;\beta\alpha}}{R}g^{\mu\nu}g^{\alpha\beta}\delta g_{\mu\nu} + \delta(1-\delta^2)R^{\delta}\frac{R_{,\beta}R_{,\alpha}}{R^2}g^{\mu\nu}g^{\alpha\beta}\delta g_{\mu\nu}$$

همچنین می‌دانیم: $\delta g_{\mu\nu} = -g_{\mu\mu'}g_{\nu\nu'}\delta g^{\mu'\nu'}$، در نتیجه خواهیم داشت:

$$\delta\mathcal{L}_G = \frac{\sqrt{-g}}{\chi}[-\frac{1}{2}g_{\mu\nu}R^{1+\delta}\delta g^{\mu\nu} + (1+\delta)R^{\delta}R_{\mu\nu}\delta g^{\mu\nu}$$
$$-\delta(1+\delta)R^{\delta}\frac{R_{;\beta\nu}}{R}\delta^{\nu}{}_{\mu'}\delta^{\beta}{}_{\alpha'}\delta g^{\mu'\alpha'} + \delta(1-\delta^2)R^{\delta}\frac{R_{,\beta}R_{,\nu}}{R^2}\delta^{\nu}{}_{\mu'}\delta^{\beta}{}_{\alpha'}\delta g^{\mu'\alpha'}$$
$$+\delta(1+\delta)R^{\delta}\frac{R_{;\beta\alpha}}{R}\delta^{\nu}{}_{\mu'}g_{\nu\nu'}g^{\alpha\beta}\delta g^{\mu'\nu'} - \delta(1-\delta^2)R^{\delta}\frac{R_{,\beta}R_{,\alpha}}{R^2}\delta^{\nu}{}_{\mu'}g_{\nu\nu'}g^{\alpha\beta}\delta g^{\mu'\nu'}]$$

$$\delta \mathcal{L}_G = \frac{\sqrt{-g}}{\chi} \left[ -\frac{1}{2} g_{\mu\nu} R^{1+\delta} + (1+\delta) R^{\delta} R_{\mu\nu} - \delta(1+\delta) R^{\delta} \frac{R_{;\nu\mu}}{R} + \delta(1-\delta^2) R^{\delta} \frac{R_{,\nu} R_{,\mu}}{R^2} \right.$$

$$\left. + \delta(1+\delta) R^{\delta} \frac{\Box R}{R} g_{\mu\nu} - \delta(1-\delta^2) R^{\delta} \frac{R_{,\beta} R_{,}^{\ \beta}}{R^2} g_{\mu\nu} \right] \delta g^{\mu\nu}$$

<div dir="rtl">

$\delta S = 0$ معادلات میدان را نتیجه می‌دهد:

</div>

$$-\frac{1}{2} g_{\mu\nu} R^{1+\delta} + (1+\delta) R^{\delta} R_{\mu\nu} - \delta(1+\delta) R^{\delta} \frac{R_{;\nu\mu}}{R} + \delta(1-\delta^2) R^{\delta} \frac{R_{,\nu} R_{,\mu}}{R^2}$$

$$+ \delta(1+\delta) R^{\delta} \frac{\Box R}{R} g_{\mu\nu} - \delta(1-\delta^2) R^{\delta} \frac{R_{,\beta} R_{,}^{\ \beta}}{R^2} g_{\mu\nu} = \frac{\chi}{2} T_{\mu\nu}$$

**Abstract**

This thesis has considered the existence of anisotropic exact vacuum solutions in the context of higher order gravities. The investigated models generally are a function of three scalars $R$, $R_{\alpha\beta}R^{\alpha\beta}$ and $R_{\mu\nu\alpha\beta}R^{\mu\nu\alpha\beta}$. Near singularity, dominant terms in the expansion of analytic type of these functions are in terms of $R^n$ (practically $R^{1+\delta}$, for indicating deviation from Einstein-Hilbert action), $(R_{\alpha\beta}R^{\alpha\beta})^n$ or $(R_{\mu\nu\alpha\beta}R^{\mu\nu\alpha\beta})^n$. Investigation shows that there always exists Kasner type solutions in $R^n$ and $(R_{\alpha\beta}R^{\alpha\beta})^n$ models. But, in the third type model, anisotropic Kasner type solution has not been found. Furthermore, the behavior of these models in the presence of matter has been investigated, and it has been revealed that these solutions are always valid for relativistic matter, however this is not true for some non relativistic matter. Moreover, the energy condition for $R^{1+\delta}$ model in vacuum has been investigated and has been shown that the application of four energy conditions (weak, null, strong and dominant) lead to one constraint on $\delta$. Usually in an accelerating expansion, the strong energy condition is violated. In our particular model, the violation of this condition leads to a negative energy density. However this result means that the strong energy condition must be satisfied in this model which in turn means an anisotropic expansion follows with a deceleration. In addition in a vacuum state, defining energy and pressure density for gravity leads to an equation of state of radiation. Hence, investigating energy conditions shows that the Kasner metric can be approximately a good solution even in the presence of relativistic and ultra relativistic matters.

Keywords: higher order gravity, singularity, Kasner type solution, anisotropic solution, relativistic mattet, non relativistic matter, ultra relativistic matter, energy condition, accelerating expansion, deceleration

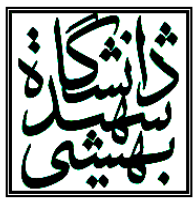

Shahid Beheshti University

Faculty of Sciences

Department of Physics

M. Sc. Thesis

Title:

# On Exact Anisotropic Solutions of Kasner type in Higher Order Gravities

Supervisor:

Dr. Mehrdad Farhoudi

Advisor:

Dr. Hamidreza Sepanji

Student:

Hamid Shabani

Summer 2009